\documentclass[]{aa}
\usepackage[varg]{txfonts}
\usepackage{graphicx}
\usepackage{hyperref}
\usepackage{natbib}
\def\simle{\mathrel{\hbox{\rlap{\hbox{\lower4pt\hbox{$\sim$}}}\hbox{$<$}}}}
\def\simgr{\mathrel{\hbox{\rlap{\hbox{\lower4pt\hbox{$\sim$}}}\hbox{$>$}}}}

\usepackage{xcolor}

\begin{document}

\title{A constant upper luminosity limit of cool supergiant stars down to the extremely low metallicity of I Zw 18}

\titlerunning{A constant $L_\mathrm{max}$ 
of cool supergiants down to the extremely low metallicity of I Zw 18}

\author{Abel Schootemeijer\inst{1} \and Ylva G\"{o}tberg\inst{2} \and Norbert Langer\inst{1,3} \and Giacomo Bortolini\inst{4} \and Alec S.\ Hirschauer\inst{5}\and Lee Patrick\inst{6}
}

\institute{Argelander-Institut f\"{u}r Astronomie, Universit\"{a}t Bonn, Auf dem H\"{u}gel 71, 53121 Bonn, Germany\\  \email{aschoot@astro.uni-bonn.de} 
\and Institute of Science and Technology Austria (ISTA), Am Campus 1, 3400 Klosterneuburg, Austria
\and Max-Planck-Institut für Radioastronomie, Auf dem H\"{u}gel 69, 53121 Bonn, Germany 
\and Department of Astronomy, The Oskar Klein Centre, Stockholm University, AlbaNova, 10691 Stockholm, Sweden
\and Department of Physics \& Engineering Physics, Morgan State University, 1700 East Cold Spring Lane, Baltimore, MD 21251, USA
\and Centro de Astrobiolog\'{i}a (CSIC-INTA), Ctra. Torrej\'{o}n a Ajalvir km 4, 28850 Torrej\'{o}n de Ardoz, Spain
}


\abstract{Stellar wind mass loss of massive stars is often assumed to depend on their metallicity $Z$. Therefore, evolutionary models predict that massive stars in lower-$Z$ environments are able to retain more of their hydrogen-rich layers and evolve into brighter cool supergiants (cool SGs; $T_\mathrm{eff} < 7$\,kK). Surprisingly, in galaxies in the metallicity range $0.2 \lesssim Z / Z_\odot \lesssim 1.5$, previous studies have not found a metallicity dependence on the upper luminosity limit $L_\mathrm{max}$ of cool SGs. Here, we add four galaxies to the sample studied for this purpose with data from the Hubble Space Telescope and the James Webb Space Telescope (JWST). Observations of the extremely metal-poor dwarf galaxy I\,Zw\,18 from JWST allow us to extend the studied metallicity range down to $Z / Z_\odot \approx 1/40$. For cool SGs in all studied galaxies, including I\,Zw\,18, we find a constant value of $L_\mathrm{max} \approx 10^{5.6}$\,L$_\odot$, similar to literature results for $0.2 \lesssim Z / Z_\odot \lesssim 1.5$. 
In I\,Zw\,18 and the other studied galaxies, the presence of Wolf-Rayet stars has been previously inferred. Although we cannot rule out that some of them become intermediate-temperature objects, this paints a picture in which evolved stars with $L>10^{5.6}$\,L$_\odot$ burn helium as hot, helium-rich stars down to extremely low metallicity. 
We argue that metallicity-independent late-phase mass loss would be the most likely mechanism responsible for this. Regardless of the exact stripping mechanism (winds or, for example, binary interaction), for the Early Universe our results imply a limitation on black hole masses and a contribution of stars born with $M \gtrsim 30\,$M$_\odot$ to its surprisingly strong nitrogen enrichment. 
We propose a scenario in which single stars at low metallicity emit sufficiently hard ionizing radiation to produce He\,{\sc ii} and C\,{\sc iv} lines. In this scenario, late-phase metallicity-independent mass loss produces hot, helium-rich stars. Due to the well-understood metallicity dependence on the radiation-driven winds of hot stars, a window of opportunity would open below 0.2\,Z$_\odot$, where self-stripped helium-rich stars can exist without dense Wolf-Rayet winds that absorb hard ionizing radiation.
}

\maketitle

\section{Introduction} \label{sec:intro}

At the end of their lives, massive stars in the Early Universe might or might not retain their hydrogen-rich outer layers. If they are destined to lose these layers, they are expected to evolve into hot, helium-rich stars. 
These stars emit copious amounts of high-energy photons \citep{Todt15, Kubatova19}, which would contribute to the ionization of their surroundings. This might help explain observed signatures of ionizing radiation -- for example, He\,{\sc ii} and C\,{\sc iv} emission observed in galaxies from low to high redshifts \citep{Stark15, Senchyna19, Saxena20, Mingozzi22, Mingozzi24, Hayes25}. 
Current stellar population synthesis models tend to produce too little ionizing radiation, prompting recent studies to use blackbody models for explaining observations \citep{Olivier22, Cameron24}.
If winds of hot, helium-rich stars are dense, most of this ionizing radiation cannot escape \citep{Sander20}. Such dense winds can give rise to strong emission lines in stellar spectra. Stars showing this emission are classified as Wolf-Rayet (WR) stars, which can affect their host galaxy's spectral appearance \citep{Conti94} and evolution \citep{Crowther07}.

It is sometimes assumed that the hydrogen-rich layers of stars born more massive than $\sim$35\,M$_\odot$ fall back into a black hole (BH) after their supernova explosion \citep{Fryer12}, though this remains uncertain \citep{Lovegrove13, Costa21}.
If the fallback scenario is true, loss of hydrogen-rich layers before the supernova reduces the masses of BHs that massive stars can produce. 
Since BH masses can be measured from gravitational merger events \citep{Abbott16}, a redshift-dependence of those masses could be expected if this mass loss is metallicity dependent. So far no evidence of such a dependence has been found up to a redshift of 1 \citep{Lalleman25}, but future high-sensitivity measurements could provide further insight.

Unfortunately, for galaxies in the Early Universe, detailed studies of the individual stars and their mass loss remain impossible. Fortunately, clues can be found by investigating, for example, red supergiants (RSGs; stars with thick hydrogen envelopes) in nearby metal-poor galaxies.
In the sample studied for this purpose, which includes the Large Magellanic Cloud (LMC), the Small Magellanic Cloud (SMC), M\,31, and the Milky Way (MW), RSGs exist only up to a luminosity limit of $\log (L / L_\odot) \approx 5.6$. This corresponds to an initial mass of $\sim$30\,M$_\odot$ \citep[almost independently of the metallicity;][]{Brott11, Georgy13, Choi16}.
This so-called Humphreys-Davidson limit seems independent of metallicity -- at least down to the metallicity of the SMC where $Z = 1/5$\,Z$_\odot$ \citep[][]{Humphreys79, Humphreys94, Humphreys25, Davies18, McDonald22, deWit24, Bonanos25}.

In the same galaxies, a multitude of hot WR stars (which no longer have thick hydrogen envelopes) have been found to exist at luminosities of $\log (L / L_\odot) > 5.6$. See for example \cite{Rate20} for the MW; \cite{Hainich14, Hainich15, Shenar16, Shenar19} for the Magellanic Clouds; and \cite{Sander14, Neugent23} for M\,31.
Together, the luminosity properties of WR stars and RSGs paint a picture in which stars with metallicities down to $Z = 1/5$\,Z$_\odot$ and initial masses of $M_\mathrm{ini} \gtrsim 30$\,M$_\odot$ 
lose their hydrogen-rich layers and evolve into WR stars.

Given that wind mass-loss rates of hot stars depend on metallicity \citep{Mokiem07, Hainich15, Backs24}, this lack of metallicity dependence in the RSG luminosity limit has been a long-standing puzzle. It could imply that mass loss of cooler evolved stars is strong at low metallicity \citep[e.g.,][]{Lamers88, Yang23, Schootemeijer24}. 
It could also be that internal mixing \citep{Schootemeijer19, Higgins20, Gilkis21, Sabhahit21} and/or binary stripping \citep[][]{Paczynski67, Vanbeveren80, Pauli22} play an important role.

Here, we aim to further investigate the upper luminosity limit of stars that retain their hydrogen-rich envelopes. We expand the sample of galaxies studied for this purpose using archival data from the Hubble Space Telescope (HST) and, importantly, extend the explored metallicity range down to 1/40\,Z$_\odot$ with new observations of I\,Zw\,18 \citep[][]{Hirschauer24, Bortolini24} taken with the James Webb Space Telescope (JWST).
We structure our paper as follows. We discuss observational data of the dwarf galaxies investigated in Sect.\,\ref{sec:data}. In Sect.\,\ref{sec:meth} we discuss our methods, and in Sect.\,\ref{sec:results} we show our results. Finally, we discuss possible caveats and the implications of our results in Sect.\,\ref{sec:discussion} and present our conclusions in Sect.\,\ref{sec:conclusions}.

\section{Data of target galaxies \label{sec:data}}
In this study, we used a variety of archival photometry data. Infrared observations of I Zw 18 were taken with JWST \citep[see][]{Hirschauer24}. From these data, \cite{Bortolini24} performed point source extraction and aperture photometry using the software package DOLPHOT \citep{Dolphin00, Weisz24}.
Here we used the \cite{Bortolini24} catalog, adopting the same source selection criteria (e.g., for signal-to-noise) as described there. From this dataset, we used the data obtained with the $F115W$ and $F200W$ filters. To test our method, we also used optical HST data of I\,Zw\,18 from \cite{Aloisi07}, who transformed them to $V$ and $I$ magnitudes. 

To expand our sample of dwarf galaxies, we searched the data archives of the ACS Nearby Galaxy Survey Treasury \citep[ANGST;][]{Dalcanton09} and the Legacy ExtraGalactic UV Survey \citep[LEGUS;][]{Sabbi18}. 
These HST observations were reduced with DOLPHOT \citep{Dolphin02}, and the source selection criteria are described in the ANGST and LEGUS papers.
From these data archives, we visually selected galaxies that appear to have well-populated RSG branches: NGC\,300, NGC\,3109, NGC\,55, and Sextans\,A from ANGST, and NGC\,5253, NGC\,4395, and Ho\,II from LEGUS. Further inspection of the data showed that NGC\,3109, NGC\,55, Sextans\,A, and Ho\,II have about an order of magnitude fewer bright red sources than the other three dwarf galaxies. Also, in terms of absolute $F814W$ magnitudes, the brightest RSGs in these galaxies are dimmer than in NGC\,300, NGC\,5253, and NGC\,4395 (in case of Sextans\,A, by 2 magnitudes).
This is in line with their relatively small amount of bright main-sequence stars. Therefore, we discarded these galaxies, as they are less useful for drawing statistically significant conclusions. The sample of dwarf galaxies from ANGST and LEGUS in the end thus includes NGC\,300, NGC\,5253, and NGC\,4395. 

As such, at sub-SMC metallicities, our sample of galaxies includes only I\,Zw\,18. The apparent lack of other metal-poor galaxies with sufficiently large massive star populations likely has to do with the positive mass-metallicity correlation in nearby dwarf galaxies, highlighting the unique nature of I\,Zw\,18 \citep[see, e.g.,][their fig.\,8]{McQuinn20}. 

We did not include the MW, because its individual stars can suffer from uncertain distances, which would propagate into uncertain luminosities.
However, we included previous studies that calculated RSGs luminosities in the SMC and LMC \citep{Davies18} and M\,31 \citep{McDonald22}. We compile the properties of the galaxies in our final sample in Table\,\ref{tab:gal_data}. These include the distance modulus (\textit{DM}) and distance, the average visual extinction $A_V$, and the metallicity. These galaxies cover a metallicity range of $1/40 \lesssim Z/Z_\odot \lesssim 1.5 $.

\begin{table}[t]
\caption{\label{tab:gal_data}
Properties of our target galaxies.}
\small
\centering
\begin{tabular}{llllll}
\hline
\hline
Galaxy & $DM$ & Distance & $A_V$ & $Z$ & Lref$^1$\\
 & [mag] & [Mpc] & [mag] & [Z$_\odot$] & \\ 
\hline
I\,Zw\,18 & 31.3\,$^{a)}$ & 18.2 & 0.1\,$^{b)}$ & 1/40\,$^{c)}$ & TW\\
SMC & 18.977\,$^{d)}$ & 0.062 & 0.35\,$^{e)}$ & 1/5\,$^{f)}$ & $^{g)}$, TW\\
NGC\,5253 & 27.48\,$^{h)}$ & 3.1 & 0.3\,$^{i)}$ & 1/3\,$^{j)}$ & TW\\
NGC\,300 & 27.36\,$^{k)}$ & 3.0 & 0.15\,$^{l)}$ & 1/2.5\,$^{m)}$ & TW\\
LMC & 18.476\,$^{n)}$ & 0.050 & (not used) & 1/2\,$^{o)}$ & $^{g)}$\\
NGC\,4395 & 28.15\,$^{p)}$ & 4.3 & 0.3\,$^{q)}$ & 1\,$^{r)}$ & TW\\
M\,31 & 24.407$^{s)}$ & 0.761  & (not used) & 1.5$^{t)}$ & $^{u)}$ \\

\hline
\end{tabular}
\tablefoot{
$^1$: Lref refers to the work (i.e., TW) in which cool SG luminosities were measured.
References: $^{a)}$: \cite{Aloisi07} -- $^{b)}$: \cite{Izotov16}, \cite{Hirschauer24} -- $^{c)}$: from \cite{Lebouteiller13} and from oxygen abundances of \cite{Skillman93} compared to $\log (O/H)_\odot = 8.75$ \citep{Bergemann21} -- $^{d)}$: \cite{Graczyk20} -- $^{e)}$: \cite{Schootemeijer21} -- $^{f)}$: \cite{Venn99} -- $^{g)}$: \cite{Davies18} -- $^{h)}$: \cite{Davidge07} -- $^{i)}$: \cite{Abril-Melgarejo24} -- $^{j)}$: \cite{Monreal-Ibero12} -- $^{k)}$: \cite{Rizzi06} --  $^{l)}$  \cite{Micheva22} -- $^{m)}$: \cite{Gazak15} -- $^{n)}$: \cite{Pietrzynski19} -- $^{o)}$: \cite{Davies15}, \cite{Patrick16}, \cite{Choudhury21} -- $^{p)}$: \cite{Thim04} -- $^{q)}$: based on $A_{263\,\mathrm{nm}}\approx
0.75$ \citep{Nandi23} and $A_\lambda/A_V$ from \cite{Gordon03} -- $^{r)}$: \cite{Cedres02} -- $^{s)}$: \cite{Li21} -- ${^t)}$: based on \cite{Zaritsky94} and \cite{Bergemann21} -- $^{u)}$: \cite{McDonald22}. 
}
\end{table}

\section{Methods \label{sec:meth}}

\subsection{Identifying cool supergiants}
\begin{figure}[t]
\begin{center}
\includegraphics[width=\linewidth]{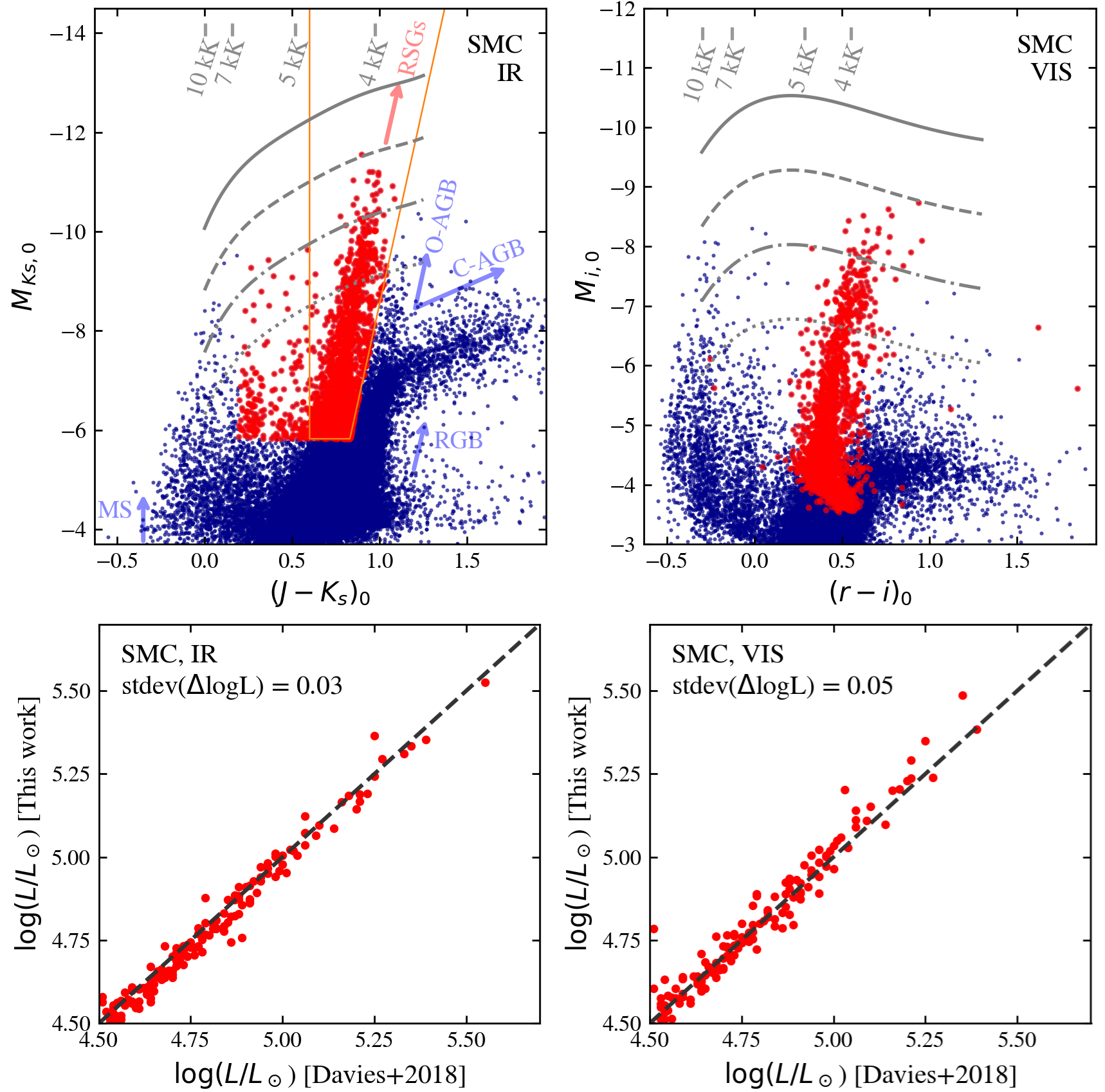} 
 \caption{
 \textit{Top panels:} Extinction-corrected CMDs of SMC sources, based on IR data (left) and optical VIS data (right). The orange lines indicate the RSG cuts in the IR. The red markers indicate sources classified as cool SGs, and the blue markers indicate other sources. The four gray lines represent iso-luminosity lines (Eq.\,\ref{eq:iso_l}). These iso-luminosity lines represent $\log ( L / L_\odot ) = 6.0$, 5.5, 5.0, and 4.5, where the higher values are closer toward the top of the plots. At the top of each panel, gray ticks indicate the temperatures to which they correspond, according to the adopted bolometric corrections. In the top left panel, following \cite{Nally24}, the arrows parallel to various features indicate the main sequence (MS), red giant branch, oxygen (O) and carbon (C) AGB, and the RSGs.
 \textit{Bottom panels:} Luminosities calculated in this work (both in the visual and infrared) as a function of luminosities calculated in earlier work by \cite{Davies18}. The dashed black line indicates where both luminosities would be identical.
 }
 \label{fig:meth_test}
\end{center}
\end{figure}

Often, RSGs are selected with cuts in color-magnitude diagrams (CMDs).
The use of CMD cuts to classify sources in different target galaxies is hindered by three issues that cause systematic shifts:
i) different extinction values, ii) different filters (e.g., $F606W - F814W$ color instead of $F555W - F814W$), and iii) different metallicities, which lead to different effective temperatures.
Therefore, we trained a neural network (NN) on SMC data to identify RSGs, as described below. We began with the source catalog of \cite{Yang19}, who combined a plethora of photometric data on evolved SMC stars. From this catalog, we used \textit{2MASS} \citep{Skrutskie06} and \textit{Skymapper} \citep[data release 1.1;][]{Wolf18} data (Fig.\,\ref{fig:meth_test}). Specifically, we used cuts in an IR CMD constructed with 2MASS $J$ and $K_S$ data to identify RSGs. We adopted the cut separating RSGs and asymptotic giant branch (AGB) stars from \cite{Cioni06}; we also adopted $M_{Ks} < -5.8$ and $J-K_S > 0.6$ for RSGs (see the top left panel of Fig.\,\ref{fig:meth_test}).

\begin{figure*}[t]
\begin{center}
\includegraphics[width=\linewidth]{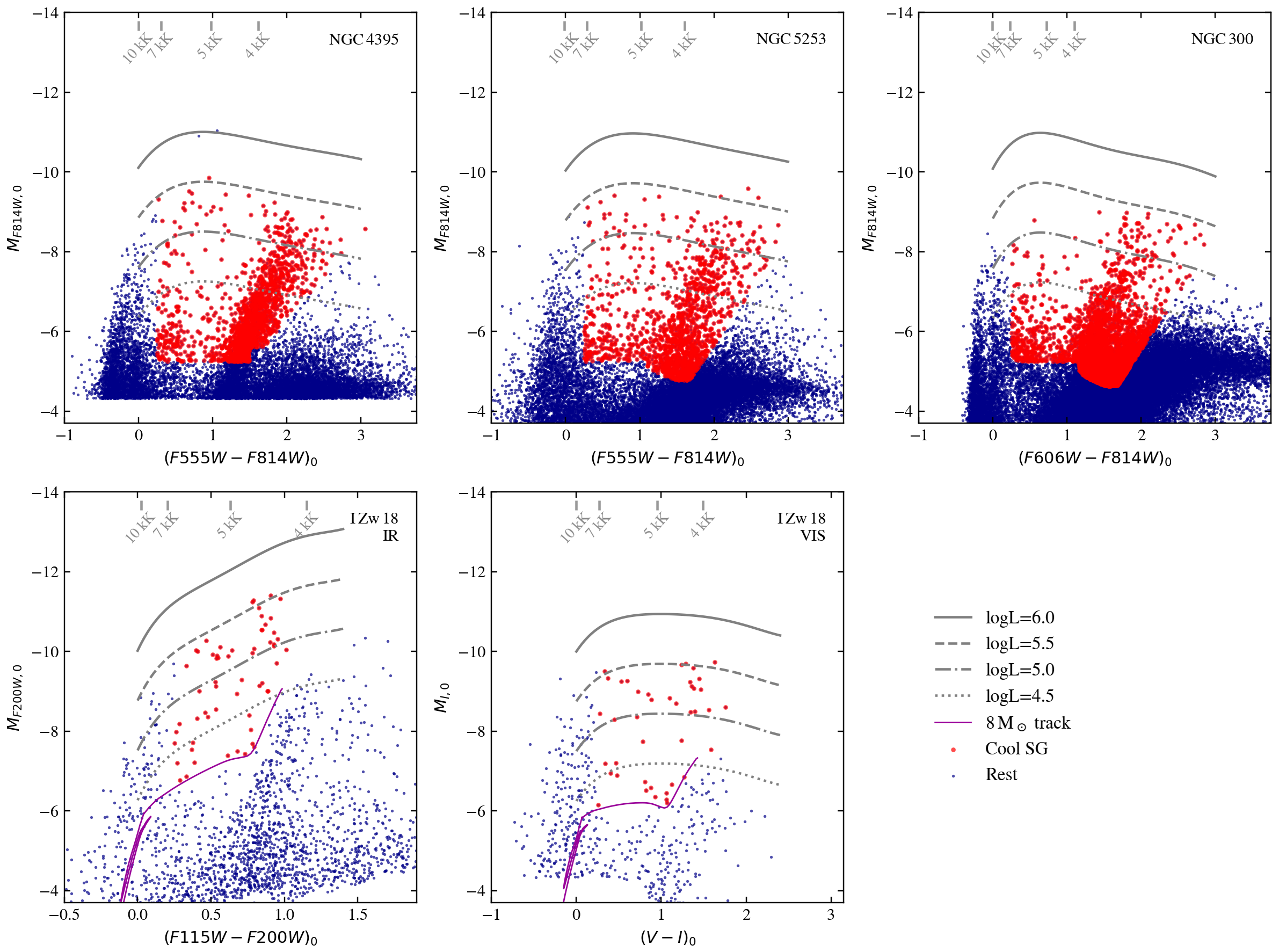} 
 \caption{
 Same as the top panels of Fig.\,\ref{fig:meth_test}, but for VIS data of NGC\,4395, NGC\,5253, and NGC\,300 (\textit{top panels}) and IR and VIS data of I\,Zw\,18 (\textit{bottom panels}). Also, the purple line in the bottom panels represents an 8\,M$_\odot$ MIST track with [Fe/H] = -1.75.)
 }
 \label{fig:cmds_all}
\end{center}
\end{figure*}

We applied a 70-30 train-test split to the SMC sample.
Next, we plotted the RSG and non-RSG sources from the training set in an $r - i$ CMD using the \textit{Skymapper} data. Then we used the {\sc Keras TensorFlow} library \citep{Tensorflow22} to build a NN with two hidden layers of 16 units each. The upper hidden layer used a "relu" activation function, while the lower hidden layer and the output layer used a "sigmoid" activation function. This NN was trained to identify RSGs and non-RSGs based on their position in the $r-i$ CMD. 
Here we chose the $r$ and $i$ filters because they transmit light at similar wavelengths as the $F555W$ and $F814W$ HST filters, which we consider later on.
We applied the NN to the test data to confirm that it properly identified RSGs. For the 100 brightest sources in the test sample, its success rate was 96\%.

The trained and tested NN was then used to identify RSGs in the optical CMDs of NGC\,5253, NGC\,300, and NGC\,4395, compiled from optical HST data. 
We later compare these with models that have temperatures up to 7\,kK, so we extended the color range in the CMDs down to 0.25. Because this includes not only RSGs but also yellow supergiants (YSGs), we hereafter refer to the stars selected by the NN and extended color range as "cool SGs." For the purpose of this study, the crucial aspect of the cool SG selection method is that it includes the highest-luminosity sources. 
This is the case, as demonstrated by the CMDs with highlighted cool SGs (top panels of Fig.\,\ref{fig:meth_test} and Fig.\,\ref{fig:cmds_all}). 
The cool SG selection method might perform suboptimally at $M_\mathrm{abs} \gtrsim -6$, where it appears to include some AGB stars. However, this is not a concern for our analysis later on, where we include only sources at $\log (L / L_\odot) \geq 4.6$.

Two sources near the $\log (L / L_\odot) = 6$ line in NGC\,4395 were also initially classified as cool SGs. These sources are bright enough for astrometric measurements in the Gaia Data Release 3 catalog\footnote{\url{https://gea.esac.esa.int/archive/}} 
\citep{GAIA23}, which indicate that they are foreground sources (e.g., proper motions in the declination direction of $pm_\mathrm{dec}=0.96 \pm 0.11$ and $pm_\mathrm{dec}=6.67 \pm 0.12$, while we expect proper motions consistent with zero at 4.3\,Mpc). Thus, we do not consider these two sources as bright cool SGs in further analyses.

We note that in the optical SMC CMD, we found a few very red sources (with $(r-i)_0> 1$) classified as cool SGs. Inspection of the data showed that they are not particularly bright in the $K_S$ band, arguing against intrinsically luminous sources suffering from high extinction.

In I\,Zw\,18, which has far lower metallicity ($~\sim$1/40\,Z$_\odot$) than the other galaxies in our sample, no well-defined RSG branch appears.  
This is likely a stellar evolution effect, as discussed in detail later. Because of this, we could not apply the NN to identify cool SGs in I\,Zw\,18. Instead, as an alternative method, we used evolutionary tracks of 8\,M$_\odot$ with [Fe/H] = -1.7 from MIST\footnote{\url{https://waps.cfa.harvard.edu/MIST/}} \citep{Dotter16, Choi16} to select evolved sources more massive than 8\,M$_\odot$ (purple lines in the bottom panels of Fig.\,\ref{fig:cmds_all}; these have evolved to near carbon exhaustion). For redder sources, we extrapolated the tracks. Again, we excluded sources bluer than 0.25\,mag. As a test, we also applied this MIST track method to identify cool SGs in NGC\,4395, NGC\,5253, and NGC\,300. Compared to the NN method, we identified marginally more sources as cool SGs at $4.5 < \log (L / L_\odot) < 5.0$, and, most importantly, we identified the same sources as cool SGs above $\log (L / L_\odot) = 5.0$ (our region of interest; Sect\,\ref{sec:ldists}).

\subsection{Calculating luminosities}
To convert the colors and magnitudes from photometric observations into effective temperatures and luminosities, we used bolometric correction (\textit{BC}) tables from MIST \citep{Dotter16, Choi16}. To begin, we selected effective temperatures in the range $3 \leq T_\mathrm{eff}  / \mathrm{kK} \leq 10$, a surface gravity of $\log (g \, \mathrm{cm}^{-1}\,\mathrm{s}^{2}) = 0$, and metallicities that lie closest to the values shown in Table\,\ref{tab:gal_data} (e.g., $[\mathrm{Fe}/\mathrm{H}] = -0.75$ for the SMC).
We then fit fifth-order polynomials to obtain $T_\mathrm{eff}$-color relations at the relevant metallicities. We obtained color-\textit{BC} relations (e.g., a relation between intrinsic $F115W - F200W$ color and $BC_\mathrm{F200W}$) in a similar fashion. We note that for $\log (g \, \mathrm{cm}^{-1}\,\mathrm{s}^{2}) = 1$ we find BCs that are similar within 0.05\,mag.

This allowed us to calculate the luminosity as follows:
\begin{equation}
\label{eq:logl}
    \log (L / L_\odot) = -0.4 \bigl( M_\mathrm{abs, 0} - M_\mathrm{bol, \, \odot} + BC(\mathrm{color}) \bigr) .
\end{equation}
Here, $M_\mathrm{abs, 0} = m_\lambda - DM - A_\lambda$ is the intrinsic absolute magnitude of a source, obtained from the apparent magnitude $m_\lambda$ in the reddest filter used to construct the CMD, the $DM$, and the extinction $A_\lambda = A_V \cdot A_\lambda / A_V$. The values used for $DM$ and $A_V$ are shown in Table\,\ref{tab:gal_data}; the value for $A_\lambda / A_V$ is from \cite{Wang19}. We adopted $M_\mathrm{bol, \, \odot} = 4.74$ for the solar absolute bolometric magnitude. 

Similarly, we drew iso-luminosity lines in our CMDs (Figs.\,\ref{fig:meth_test}, \ref{fig:cmds_all}). We calculated these using
\begin{equation}
 \label{eq:iso_l}
    M_\mathrm{abs, 0}(\mathrm{color}) = -2.5\log (L / L_\odot) + M_\mathrm{bol, \, \odot} - BC(\mathrm{color}) .
\end{equation}

For IR and visible (VIS) data, we tested our method with SMC data from \cite{Davies18}, who used photometry to calculate luminosities of cool SGs in the SMC by integrating over their spectral energy distributions. The \cite{Davies18} catalog provides \textit{2MASS} IR photometry. To obtain VIS photometry, we used CDS-xmatch\footnote{\url{http://cdsxmatch.u-strasbg.fr/}} to cross-correlate the \cite{Davies18} catalog with \cite{Yang19} catalog, which includes \textit{SkyMapper} photometry.
Figure\,\ref{fig:meth_test} shows the correlation between the luminosity from \cite{Davies18} and our method. The black line shows where the values would be equal (i.e., it is not a linear fit). The standard deviation of the difference in $\log (L / L_\odot)$ is 0.05 in the optical and 0.03 in the IR. We conclude that our method accurately calculates cool SG luminosities, especially with IR data.

\section{Results \label{sec:results}}

\subsection{Color-magnitude diagrams}
We present optical CMDs of NGC\,4395, NGC\,5253, and NGC\,300 in the top row of Fig.\,\ref{fig:cmds_all}. Using the iso-luminosity lines for navigation, we show that the cool SGs in these galaxies are abundant up to $\log (L / L_\odot) \approx 5.6$, but absent at higher luminosities. In the bottom left panel of Fig.\,\ref{fig:cmds_all}, we show the IR CMD of I\,Zw\,18 based on JWST data \citep{Hirschauer24, Bortolini24}. At the bright end, the IR CMD shows the same behavior as the higher-metallicity galaxies, with cool SGs present up to $\log (L / L_\odot) \approx 5.6$. We discuss these luminosity distributions in further detail in Sect\,\ref{sec:ldists}.

I\,Zw\,18 is the most challenging of our target galaxies to study because it is the most distant. Therefore, we used optical data from \cite{Aloisi07} to perform an independent measurement of the luminosities of cool SGs in I\,Zw\,18 (lower central panel of Fig.\,\ref{fig:cmds_all}). With the optical data, we find that the brightest RSG has a luminosity of $\log (L / L_\odot) = 5.59$. With the IR data, we obtained almost the same luminosity for the brightest RSG ($\log (L / L_\odot) = 5.57$) -- a difference of only 0.02\,dex. We find this result encouraging.

\begin{figure}[t]
\begin{center}
\includegraphics[width=\linewidth]{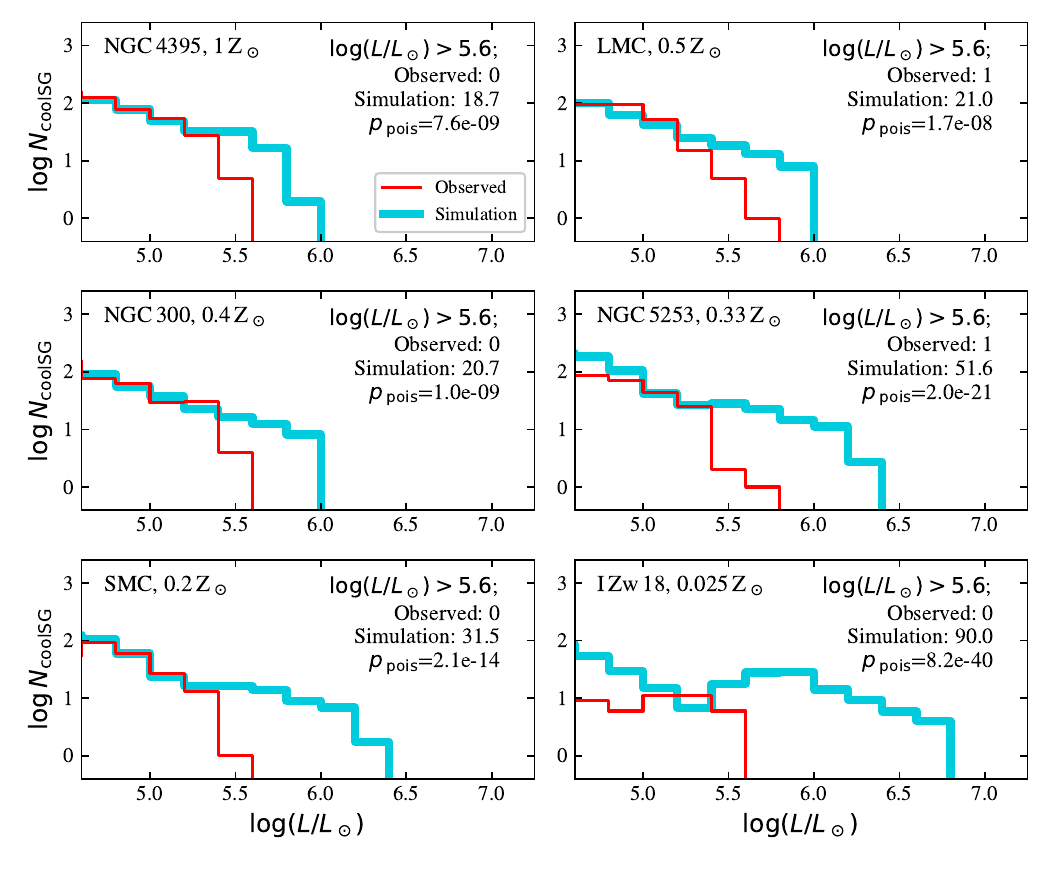} 
 \caption{
 Luminosity distributions of observed sources (red lines) and sources in the theoretical population (light blue lines) in six different galaxies. The observed SMC and LMC sources are taken from \cite{Davies18}; for the remaining sources the luminosities are derived in this work. At the right side of each panel, we list the number of observed and model population sources with $\log (L / L_\odot) > 5.6$, along with the associated Poisson probability of having this many or fewer observed sources (given the theoretically predicted number).
 }
 \label{fig:ldist}
\end{center}
\end{figure}

\subsection{Luminosity distributions \label{sec:ldists}}
In Fig.\,\ref{fig:ldist}, we show the observed luminosity distributions (red lines) of cool SGs in the galaxies discussed in the previous section. For I\,Zw\,18, we consider the luminosities based on the IR observations from JWST. Additionally, we include the LMC and the SMC, for which we obtain the cool SG luminosities directly from \cite{Davies18}.

For each galaxy, we compare this observed luminosity distribution with a model population from BoOST \citep[Bonn Optimized Stellar Tracks;][]{Szecsi22}. These BoOST models include moderate rotation and metallicity-dependent mass loss (rates from \cite{Nieuwenhuijzen90}, scaling with the iron abundance as $(X_\mathrm{Fe}/X_{\mathrm{Fe},\odot})^{0.85}$ following \cite{Vink01}). 
They are an extension of the massive star model grid of \cite{Brott11}.
We picked the closest-metallicity models for galaxies that have no models tailored for them: SMC models for NGC\,5253, LMC models for NGC\,300, and MW models for NGC\,4395.
We then drew random ages and initial masses, where the probability for the initial mass is dictated by the Salpeter initial mass function (IMF) \citep[$p \propto M_\mathrm{ini}^{-2.35}$;][]{Salpeter55}. 
We discarded models with $T_\mathrm{eff}>7$\,kK as well as models with a stellar lifetime shorter than the drawn age. To reduce statistical noise, we assigned each drawn stellar model a weight of 0.05. We continued to draw random models until we had the same number as observed in the luminosity range $5.0 \leq \log (L / L_\odot) \leq 5.4$ (e.g., if 20 such sources were observed, we kept drawing until we obtained 400 theoretical models, each with a statistical weight of 0.05). 
We tabulate numbers of observed and predicted cool SGs in Table\,\ref{tab:gal_sfrs_etc}. Here, we also provide a minimum star formation rate (SFR) per galaxy based on its number of observed cool SGs ("minimum" because stars can also burn helium as hotter objects). See Appendix\,\ref{app:numbers} for details.

\begin{figure}[t]
\begin{center}
\includegraphics[width=\linewidth]{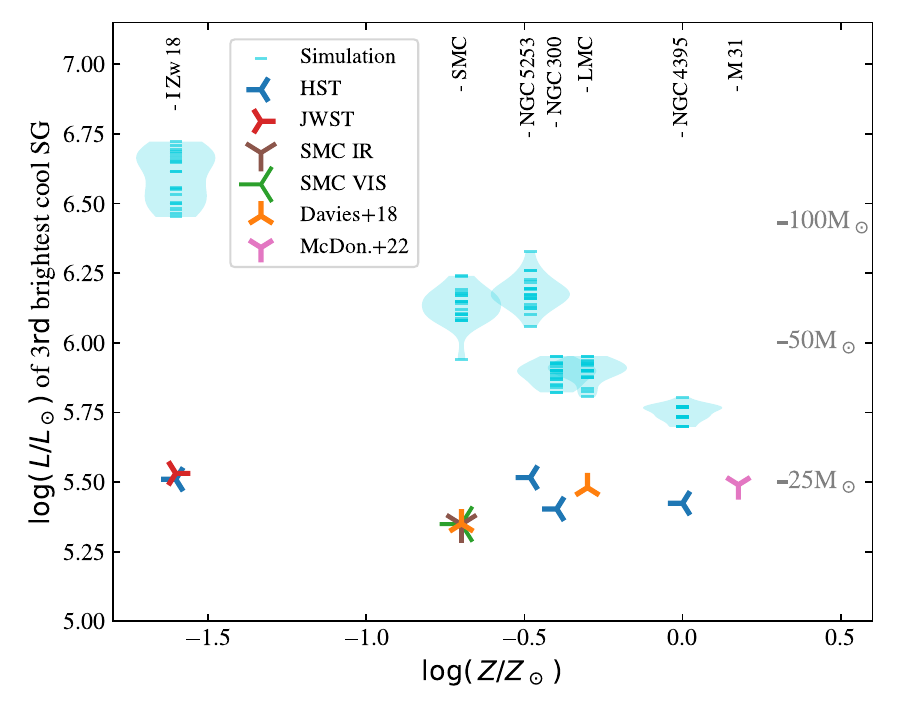} 
 \caption{
 Third brightest cool SG in various galaxies as a function of galaxy metallicity. The data points with the labels HST, JWST, SMC IR, and SMC VIS are from this work. The data points labeled Davies+18 and McDon+22 come from \cite{Davies18}'s work on the SMC and LMC, and \cite{McDonald22}'s work on M\,31, respectively. We note that I\,Zw\,18 and the SMC have multiple data points that are similar within 0.02\,dex. The light blue horizontal lines show the third brightest cool SG in individual simulations. We performed 20 simulations per galaxy, and visualize the number density of simulation outcomes with violin plots. The gray ticks indicate the time-averaged cool SG luminosities of BoOST SMC models with three different initial masses, which are similar across the shown metallicity range.
 }
 \label{fig:ztrends}
\end{center}
\end{figure}

Comparing the observed and theoretical luminosity distributions, we find very good agreement below $\log (L / L_\odot) = 5.4$ (except for I\,Zw\,18). Above $\log (L / L_\odot) = 5.6$, the theoretical distributions contain far more models than observed. The theoretical upper luminosity limits are, in practice, set by the adopted mass-loss rates, which are lower at low metallicity. Therefore, the tension with the (nearly constant) observed upper luminosity limit significantly increases with decreasing metallicity. At the right side of each panel of Fig.\,\ref{fig:ldist}, we list the Poisson probability $p_\mathrm{pois}$ to draw as many or fewer cool SGs as observed above $\log (L / L_\odot) = 5.6$, given the number expected from the theoretical population. For all galaxies, these Poisson probabilities lie below $10^{-7}$, meaning that the absence of observed cool SGs with $\log (L / L_\odot) > 5.6$ cannot plausibly be explained by chance.

We discuss the case of I\,Zw\,18 in more detail. This is the only galaxy where the observed and theoretical luminosity distributions do not match at the low-luminosity end. 
The number of observed cool SGs in the range $4.6 < \log (L / L_\odot) < 5.0$ (corresponding to $M_\mathrm{ini} \approx 15$\,M$_\odot$) is about five times lower than the models predict. We suspect this to be a stellar evolution effect. In this scenario, stars born with $\sim$15\,M$_\odot$ spend a lower fraction of their helium-burning lifetime as cool SGs than do stars with initial masses of $\sim$25\,M$_\odot$ (which have $\log (L / L_\odot) \approx 5.4$ during helium burning).
In fact, such a stellar evolution effect is present in the BOoST models, but manifesting at higher luminosity (causing the bump around $\log (L / L_\odot) \approx 5.8$ in the bottom right panel of Fig.\,\ref{fig:ldist}). We discuss uncertainties in stellar evolution models in Sect.\,\ref{sec:models}. 
Due to these uncertainties, we used BoOST models to test a scenario where all I\,Zw\,18 stars spend a constant 7\% of their helium-burning lifetime as cool SGs. This resulted in a model population harboring approximately ten cool SGs with $\log (L / L_\odot) > 5.6$ and an associated $p_\mathrm{pois} \approx 10^{-4}$ for drawing none. As such, in this test there is less spectacular difference between the luminosity distributions of observed and model populations of bright cool SGs in I\,Zw\,18. Still, the absence of observed cool SGs with $\log (L / L_\odot) > 5.6$ remains highly unlikely by chance.

\subsection{Trend with metallicity}
After calculating luminosity distributions of cool SGs in different galaxies, we now investigate possible trends with metallicity. To do so, we consider the third brightest observed cool SG in each galaxy and plot it against the metallicity of its galaxy. We chose to take the third brightest object because one uncertain object can significantly scatter the upper luminosity limit (see, e.g., the cases of \cite{McDonald22} in M31 and the rapidly-changing object WOH\,G64 \citep{Levesque09,Davies18,Ohnaka24, MunozSanchez24} in the LMC).
Our conclusions would be the same when considering the brightest cool SG instead of the third brightest. 
We note that we included only RSGs from literature studies of the SMC, LMC, and M\,31. Including YSGs would move the observed data point slightly upward in the LMC, where \cite{Humphreys23} found three YSGs with $\log (L/ L_\odot) = 5.7$. No such bright YSGs were found in the SMC \citep{Neugent10}, M31 \citep{Drout09}, and our Fig.\,\ref{fig:cmds_all}, where they would lie directly above the dashed gray line and be easily observable.

The observed metallicity dependence -- or lack thereof -- in the cool SG upper luminosity limit is shown in Fig.\,\ref{fig:ztrends}. The average luminosity of the third brightest cool SG in our sample galaxies is $\log (L / L_\odot) = 5.46$ with a standard deviation of only 0.06.
 
For comparison, we also consider theoretical cool SG populations based on BoOST models. Similar to that described in Sect.\,\ref{sec:ldists}, for each galaxy we drew cool SGs until we had as many in the range $5.0 \leq \log (L / L_\odot) \leq 5.4$ as observed (the only difference here is that each non-rejected model has a statistical weight of 1). Then, we identified the third brightest cool SG (orange lines in Fig.\,\ref{fig:ztrends}). We repeated this 20 times for each galaxy and indicate the probability density of the blue lines with violin plots.
The NGC\,5253 simulations tend to have brighter cool SGs than the SMC simulations because NGC\,5253 contains more dim cool SGs. The same models are used for these galaxies.
Other than that, the model populations show a clear trend, where brighter RSGs are expected to be present in more metal-poor galaxies. This evidently conflicts with our observational results, which lack bright cool SGs ($\log (L / L_\odot) \gtrsim 5.6$). We discuss the astrophysical implications of this at the end of Sect.\,\ref{sec:implications}.

\section{Discussion \label{sec:discussion}}

\subsection{Sensitivity to warm supergiants in I Zw 18 \label{sec:warmsgs}}
Above, we found no cool SGs with $\log (L / L_\odot) > 5.6$ or $T_\mathrm{eff} < 7$\,kK in I\,Zw\,18. For hotter stars at such luminosities, we are not able to calculate the luminosities as accurately, but they might still appear in the CMD -- if they exist. We investigate this further in this section.

\begin{figure}[]
\begin{center}
\includegraphics[width=0.93\linewidth]{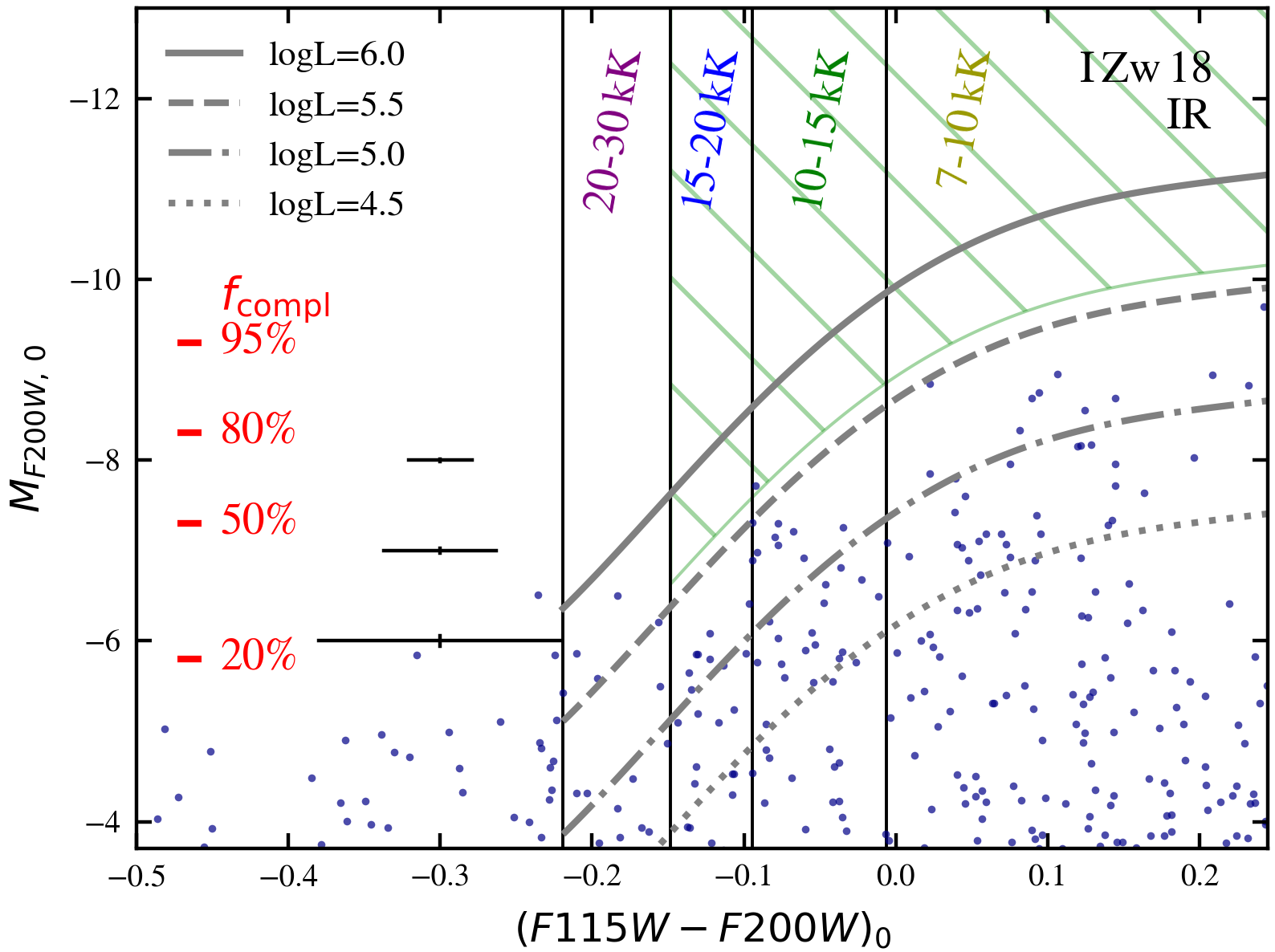} 
 \caption{
 Similar to the bottom left panel of Fig.\,\ref{fig:cmds_all}, but showing only the infrared data of I\,Zw\,18 and zoomed in on the bluer side of the CMD. Here, we show iso-luminosity lines extending to bluer colors. The temperature ranges associated with the color intervals bordered by the vertical black lines are listed in the plot. The green hatching highlights the empty region associated with $T_\mathrm{eff} < 20$\,kK and $\log ( L / L_\odot ) > 5.6$. The red ticks indicate the absolute $F200W$ magnitudes at which \cite{Bortolini24} find a completeness ($f_\mathrm{compl}$) of 20\%, 50\%, 80\%, and 95\% in the densest region of I\,Zw\,18, R1. With black crosses we show typical observational errors from \cite{Bortolini24} at absolute $F115W$ magnitudes of -8, -7, and -6, also estimated for R1.
   }
 \label{fig:izw18_hotstars}
\end{center}
\end{figure}

In Fig.\,\ref{fig:izw18_hotstars}, we show the blue side of the IR CMD of I\,Zw\,18. We also indicate regimes associated with different temperatures and luminosities based on the MIST BCs. This figure shows that also at bluer colors ($F115W-F200W \gtrsim -0.15$; temperatures of $\sim$20\,kK and below), the CMD locus associated with $\log (L / L_\odot) \gtrsim 5.6$ remains more or less empty (only a few border cases appear). The CMD implies the presence of about 25 less-luminous warm SGs with $5.0 < \log (L / L_\odot) < 5.6$.
In this part of the CMD, the completeness is still expected to be on the order of 50\% (Fig.\,\ref{fig:izw18_hotstars}; see also Sect.\,\ref{sec:completeness} for further discussions on completeness).
We refrain from making quantitative statements for bluer sources (hotter than $\sim$20\,kK) as their inferred luminosity becomes increasingly sensitive to color. At the same time, errors and completeness progressively worsen with decreasing brightness in the $F200W$ filter. 
This analysis shows that the CMD of I\,Zw\,18 lacks a sizable population of bright warm SGs ($\log (L / L_\odot) > 5.6$; $7 \leq T_\mathrm{eff} / \mathrm{kK} \leq 20$). 

\begin{figure*}[]
\begin{center}
\includegraphics[width=0.85\linewidth]{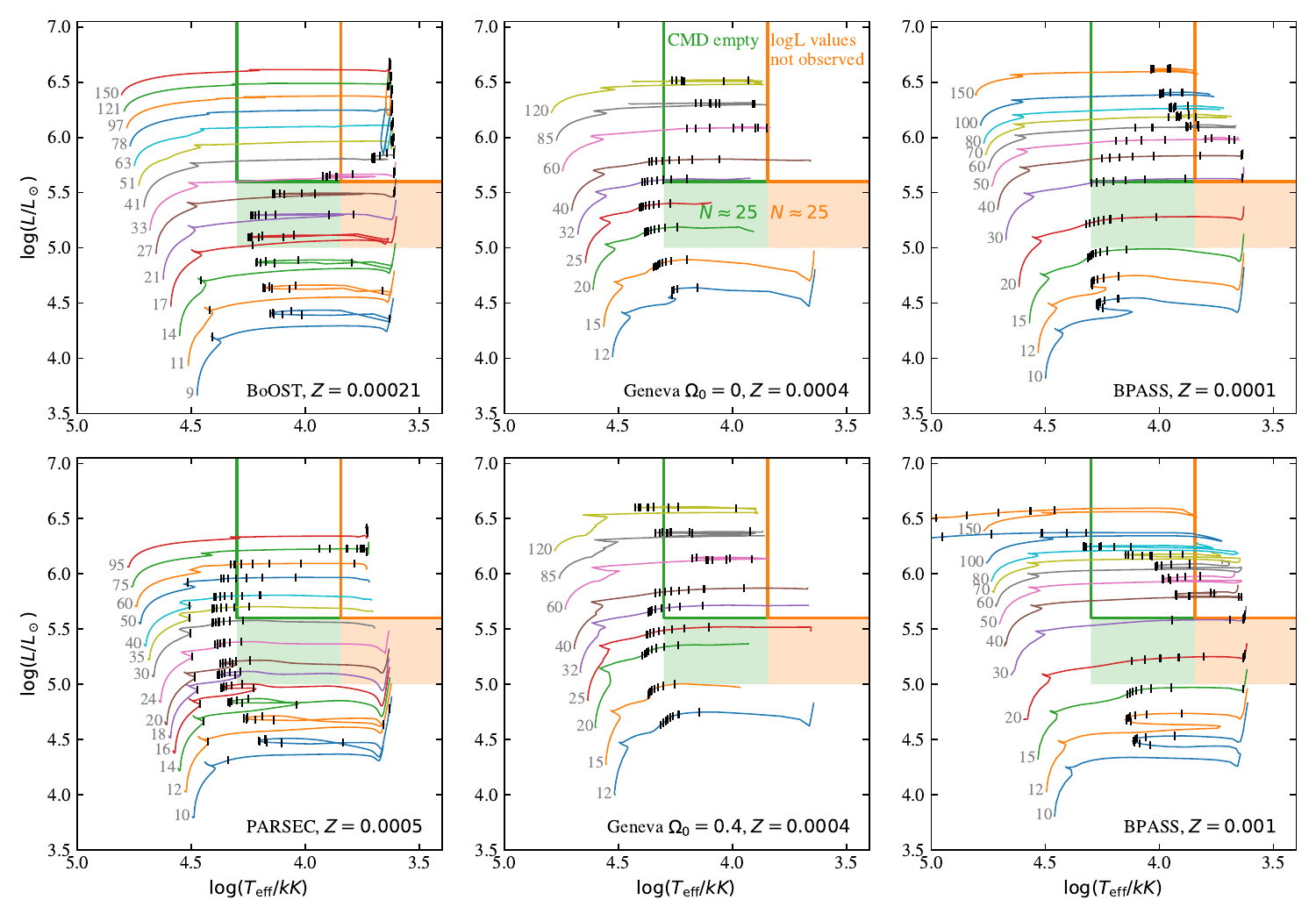} 
 \caption{
 Hertzsprung-Russell diagrams showing evolutionary tracks with a metallicity similar to I\,Zw\,18. The gray numbers indicate the initial masses of the evolutionary models in solar units. The black ticks indicate the model positions at 0.925, 0.935, \ldots, 0.995 times the age of the final model, indicating where helium-burning models spend most of their time. The empty region in the CMD of I\,Zw\,18 associated with bright warm sources ($\log ( L / L_\odot ) > 5.6$ and $ 7 \leq T_\mathrm{eff}/\mathrm{kK} < 20$; Fig.\,\ref{fig:izw18_hotstars}) is highlighted with an empty green box. The area where we found no bright cool SGs is highlighted with an empty orange box. The areas where we inferred approximately 25 warm and cool SGs are highlighted with filled green and orange boxes, respectively.
  }
 \label{fig:model_hrds}
\end{center}
\end{figure*}

\subsection{Comparison with evolutionary models of I Zw 18 \label{sec:models}}

In Fig.\,\ref{fig:model_hrds} we show Hertzsprung-Russell diagrams (HRDs) of various evolutionary models with a metallicity similar to I\,Zw\,18's. These include the BoOST models used to synthesize theoretical populations of cool SGs. Additionally, we display PARSEC models \citep[][$Z=0.0005$]{Bressan12,Tang14} and Geneva models \citep[][$Z=0.0004$, with and without rotation]{Groh19}. Finally, we show BPASS single-star models \citep{Eldridge17} with both $Z=0.0001$ and $Z=0.001$, since neither is particularly close to I\,Zw\,18's metallicity of $Z\approx$ 0.0004 and no models with an in-between value have been computed.
We refrain from showing MIST models \citep{Dotter16, Choi16}, because most downloaded in the mass range 10-150\,M$_\odot$ did not evolve through helium burning. For reference, we show the regions of the HRD that appeared unpopulated according to our CMD analysis of warm SGs (Fig.\,\ref{fig:izw18_hotstars}) and our cool SG luminosity measurements (Sect.\,\ref{sec:ldists}). We also highlight the HRD regions in which we inferred the presence of $\sim$25 SGs (both warm and cool) in the luminosity range $5.0 < \log (L / L_\odot) < 5.6$.

The evolutionary models differ in helium-burning temperatures, where BoOST models are particularly cooler than others during this phase. We attribute this to differences in internal mixing assumptions. Overshooting and (semi-)convection are known to drastically affect post-MS evolution \citep[see, e.g.,][]{Schootemeijer19, Higgins20,Farrell22}. Also, rotational mixing \citep{Maeder00, Heger00} could affect the evolutionary paths.
In the range $5.0 < \log (L / L_\odot) < 5.4$, PARSEC and Geneva models only predict cool SGs below 7\,kK to exist after central helium exhaustion, which lasts less than 0.1\% of their stellar lifetime. Given our inference of at least 20 cool SGs in that luminosity range, PARSEC and Geneva models predict tens of thousands of hydrogen- and helium-burning progenitor stars to exist. This seems unrealistic, given that their combined luminosity would exceed I\,Zw\,18's total luminosity of around 10$^8$\,L$_\odot$\footnote{Obtained using Eq.\,\ref{eq:logl}, using $B=15.98$ from \cite{deVaucouleurs91, deVaucouleurs95} and $BC=0$, which should be appropriate for its color of $B-V\approx 0$ \citep{Papaderos02} in an order-of-magnitude estimate.}.
Therefore, Geneva and PARSEC models at I\,Zw\,18 metallicity probably underestimate cool SG lifetime.
In contrast, BoOST models -- seemingly correctly -- predict longer cool SG lifetimes; however, they overpredict the number of bright cool SGs (Sect.\,\ref{sec:ldists}).
In the BPASS models with $Z=0.001$, the blueward evolution is caused by strong RSG mass loss, after which the mass-loss rates become lower, preventing most models from evolving into stars hotter than $\sim$20\,kK. 
At the metallicity of I\,Zw\,18 or lower ($Z  \lesssim 0.0004$), none of the sets of evolutionary models shown in Fig.\,\ref{fig:model_hrds} become hot enough to be WR stars. However, observational evidence for WR stars in I\,Zw\,18 (Sect\,\ref{sec:sfh}) exists.

In theory, the absence of cool SGs with $\log (L / L_\odot) > 5.6$ in I\,Zw\,18 could be a stellar evolution effect unrelated to mass loss. This scenario requires the following initial mass trend: stars with $M_\mathrm{ini} \lesssim 30$\,M$_\odot$ produce observed cool SGs, while higher-mass stars retain their hydrogen-rich envelopes but evolve into hotter stars with temperatures above 20\,kK.
However, no set of evolutionary models at I\,Zw\,18 metallicity exhibits this trend. If anything, they show the opposite: brighter helium-burning models are cooler (see the black ticks in Fig.\,\ref{fig:model_hrds}). These models, however, remain uncertain, as discussed in the previous paragraph. If they wrongly predict the temperature trend with mass during helium burning, we cannot exclude that $M_\mathrm{ini} \gtrsim 30$\,M$_\odot$ stars in I\,Zw\,18 retain their hydrogen-rich layers yet burn helium at $T_\mathrm{eff} > 20$\,kK.

\subsection{Possible caveats}

\subsubsection{Variability:}

Our results could be affected by intrinsic variability of cool SGs. Betelgeuse, for example, is known to display $V$ band magnitudes ranging from $0.2 \lesssim V \lesssim 1.0$ over the last 15 years, even reaching $V\approx 1.7$ during the Great Dimming in 2020 \citep{Jadlovski24}.
Here, we briefly discuss literature results on variability of the brightest cool SG in the MW, M\,31, and the Magellanic Clouds.

In the MW, we investigate the five brightest sources from \cite{Messineo19} with epoch photometry\footnote{\url{https://gea.esac.esa.int/archive/}} in Gaia DR3 \citep{GAIA23}. We find that the standard deviations of the $G_{\rm{RP}}$ magnitudes of these five sources are between 0.05 and 0.25 (with 0.13 as both average and median). 
The $G_{\rm{RP}}$ band has a similar central wavelength as the $I$ band and the $F814W$ band, which are used to construct Fig.\,\ref{fig:cmds_all}. In the $G_{\rm{BP}}$ band (central wavelength similar to the $V$ and $F555W$ bands), the variations were about twice as large as in the $G_{\rm{RP}}$ band.

Beyond the MW, \cite{Yang18} found that the brightest RSGs in the LMC have a median absolute deviation of about 0.1\,mag in the infrared $WISE1$ band. 
\cite{Soraisam18} calculated the root mean square deviation (RMSD) from the mean of the $R$-band magnitude of RSGs in M\,31. Analyzing the five brightest RSGs in their table\,1, we find RMSD values in the range of 0.01\,mag to 0.37\,mag (median: 0.13, average: 0.16). 
Looking into the five brightest RSGs in the SMC, we find that they have a median dispersion (or variance) in the $G_{\rm{RP}}$ band of 0.08\,mag in the \cite{Maiz23} catalog.
For both the M\,31 and the SMC case, we get similar results when including more RSGs.

We conclude that the typical variability of the brightest RSGs in the Magellanic Clouds, M\,31, and MW seems to be on the order of 0.1-0.15\,mag, with no clear metallicity trend. This translates into 0.03 - 0.05\,dex in $\log ( L / L_\odot )$, which is too small to significantly affect our results.
Furthermore, in the near-IR, RSG variability is significantly smaller than at optical wavelengths
\citep[see][for further discussion]{Wang25}, which adds strength to the luminosities determined with JWST data for the RSG population of I\,Zw\,18.

\subsubsection{Crowding and completeness in I Zw 18 \label{sec:completeness}}

At $\sim$18\,Mpc, I\,Zw\,18 is the most distant galaxy in our sample, making it prone to crowding issues. This could result in an overestimated brightness for individual sources, or it could prevent sources from being detected. Thus, \cite{Bortolini24} investigated errors and completeness in their JWST observations, which are also used in this paper. To do so, they performed artificial star tests (ASTs). In short, these ASTs randomly placed one individual artificial star at a time into the image, with the probability of its spatial positioning following the observed surface brightness. Different magnitudes were explored. For each artificial star, the image was reprocessed. The study investigated how often artificial stars of different magnitudes were recovered. This provided magnitude-dependent completeness fractions and errors for a few different regions, from the outskirts of I\,Zw\,18 to its dense central area. 

Here, we consider the completeness fraction in the densest region, called R1 \citep[see Fig.\,1 of ][]{Bortolini24}. It contains the vast majority of the cool SGs for which we derived $\log (L / L_\odot) > 5$. R1 is located in the central part of the galaxy's main body and has an angular size of about 3.5\arcsec\ (corresponding to $\sim$300\,pc at the distance of I\,Zw\,18). For comparison, the angular resolution of JWST is about 0.1\arcsec\ ; moreover, the cluster NGC\,330 in the SMC would have a projected diameter of $\sim$0.5\arcsec\ if it were at the distance of I\,Zw\,18.

\cite{Bortolini24} estimated via ASTs that R1 has a completeness fraction of 95\% or higher (as also indicated in Fig. \,\ref{fig:izw18_hotstars}) for absolute magnitudes of $M_{F200W} < -9$ (corresponding to $\log (L / L_\odot) \gtrsim 5$ for cool SGs). 
Therefore, despite the high concentration of sources in R1, the ASTs suggest that the brightest cool SGs can still be recovered. 
This result is likely a consequence of how the DOLPHOT software works: it first identifies the brightest sources, rarely missing them. We note that for the data of \cite{Bortolini24}, no sources -- which we would otherwise classify as cool or warm SGs with $\log (L / L_\odot) \gtrsim 5.5$ -- in I\,Zw\,18 are rejected by quality cuts (see Sect.\,\ref{sec:meth}). We also note that in R1, \cite{Annibali13} find similar completeness as a function of absolute $I$ magnitude for the data of \cite{Aloisi07} shown in the lower middle panel of Fig.\,\ref{fig:cmds_all}.
These high estimates for the completeness fraction would imply that our main result -- the absence of bright cool SGs -- is not strongly affected by completeness issues.
As an additional test, we inspected the JWST image to search for potential unresolved clusters. We find one region in R1 with diffuse emission, but in the $F200W$ image its integrated flux is lower than that of the brightest cool SGs in R1. Therefore, it is not possible that this diffuse emission region hosts a cool SG brighter than the brightest ones detected. We find no other candidates for unresolved clusters hosting bright cool SGs.

The brightest cool SGs may cluster together much more than assumed in the ASTs of \cite{Bortolini24}. This could lead us to overestimate their individual luminosities. While it is difficult to rule this out, it is worth mentioning that RSGs in the LMC and M\,31 are more isolated than O stars and WR stars \citep{Smith15, Aadland18, Martin25}. Also, inspecting data from \cite{Davies18} shows that RSGs in the SMC and LMC are rather evenly spread out.

In Sect.\,\ref{sec:ldists}, we noticed that, in the range $ 4.6 < \log (L / L_\odot) < 5$, I\,Zw\,18 contains about five times fewer cool SGs than the theoretical population. This could be explained in part by the declining completeness fraction of the observed sources. However, since the sources in question have absolute magnitudes of $M_{F200W} \approx -8$ or brighter, where the completeness is still expected to be around 80\%, a lower completeness fraction can only partially explain this relatively low number.

\subsubsection{Reliability of luminosity measurements} 
\cite{Hirschauer24} found that some of their brightest sources are dusty star candidates, for which they adopted a threshold of $F115W - F444W > 0.5$. Dust can cause self-extinction, which, if severe enough, could cause our method to underestimate stellar luminosities. To investigate if this poses a potential problem, we looked at SMC and LMC data from \cite{Davies18}. In these data, we found no obvious correlation between $J-WISE2$ color (which uses filters with central wavelengths similar to the $F115W$ and $F444W$ filters\footnote{\url{http://svo2.cab.inta-csic.es/theory/fps/}}) and visual extinction $A_V$. This implies $F444W$ excess does not significantly affect $A_V$ values. Moreover, to derive cool SG luminosities in I\,Zw\,18 we used $F115W$ and $F200W$ magnitudes, which are affected much less by extinction than the $V$ band \citep[$A_{F200W} / A_V \approx0.08$ according to][]{Wang19}. 
In addition, I\,Zw\,18 contains only a small amount of dust \citep{Cannon02}, as expected given its low metallicity.
We conclude that self-extinction is unlikely to significantly affect the luminosity measurements based on IR data from JWST. 

Apart from self-extinction, variable extinction within a galaxy could also affect luminosity measurements. The $A_V$ values found by \cite{Davies18} for SMC RSGs have a standard deviation of 0.15\,mag. This corresponds to an uncertainty of 0.06\,dex in luminosity (less if measurement errors caused extra scatter). For the galaxies shown in Fig.\,\ref{fig:cmds_all}, this uncertainty is smaller because longer-wavelength filters were used \citep[where extinction is smaller;][]{Wang19}, particularly for the IR observations of I\,Zw\,18. If the variable extinction in the galaxies shown in Fig.\,\ref{fig:cmds_all} is comparable to the SMC (the total $A_V$ values are comparable or smaller; see Table\,\ref{tab:gal_data}), we do not expect it to affect our results.

Furthermore, using JWST IR data, we found that the luminosity of the brightest (and third brightest) cool SG was within 0.02\,dex to that obtained from the optical data of \cite{Aloisi07}. This result would be unexpected if extinction strongly affected our measurements, or if unresolved nearby main sequence stars were to significantly contribute to the flux.
Also, our tests in the SMC (Fig.\,\ref{fig:meth_test}) suggest reasonably high accuracy in our luminosity measurements. Moreover, in the SMC three different methods (IR CMD, optical CMD, and the results from \cite{Davies18}) yielded a luminosity of the third brightest cool SG within 0.01\,dex. 

The ASTs by \cite{Bortolini24} discussed in the previous section suggest that the errors for sources with  $M_{F200W} \leq -8$ are on the order of 0.001\,mag. This is too low to significantly affect the cool SG luminosities we derived.

\subsubsection{Star-formation history of I Zw 18 \label{sec:sfh}}

The SFR in I\,Zw\,18 is thought to have varied over time, with a relatively large number of stars forming recently, but some also around 1\,Gyr ago or earlier \citep{Aloisi07, Annibali13, Bortolini24}. The recent starburst in I\,Zw\,18 may have been triggered by the interaction of its main body with its component C \citep{Kim17, McQuinn20, Bortolini24}.
It is, in principle, possible that a recent halt in star formation caused the absence of the brightest and most short-lived cool SGs, although this explanation would require fine-tuning to reproduce the upper luminosity limit of $\log (L / L_\odot) \approx 5.6$, which is also observed in the other galaxies.

In this context, it is relevant that the presence of WR stars -- which are also thought to be luminous helium-burning stars -- has been inferred from the spectra of I\,Zw\,18. Features of WC (WR stars with strong carbon lines) and WN (WR stars with strong nitrogen lines) stars have been reported, and it was deduced that I\,Zw\,18 hosts some tens of WR stars in total \citep{Izotov97, Legrand97, deMello98, Brown02}. The exact number of WR stars in I\,Zw\,18 is quite uncertain. First, these numbers have been calculated using emission line fluxes per WR star, which are highly uncertain at I\,Zw\,18 metallicity \citep[see also][]{GonzalezTora25}. Second, these numbers are based on distances of around 11\,Mpc, which would result in roughly three times fewer WR stars than if a more recent distance determination of 18\,Mpc were adopted \citep{Aloisi07}.

We can speculate that the WR stars in I\,Zw\,18 are more luminous than $\log (L / L_\odot) \approx 5.6$, based on the following arguments. First, in the MW and its direct surroundings, the luminosity of the dimmest WR star increases with decreasing metallicity \citep{Shenar20}, where $\log (L_\mathrm{WR, \, min} / L_\odot)$ has a value of 4.9 in the MW, 5.2 in the LMC, and 5.6 in the SMC. Second, a hot stripped star with $\log ( L / L_\odot) = 5.1$, which does not manifest itself as a WR star, has been detected in the SMC \citep{Gotberg23}. This further supports the idea that, at low metallicity, WR stars  must be more luminous to drive winds dense enough to produce the characteristic WR emission lines. In that scenario, the WR stars in I\,Zw\,18, with $\log (L / L_\odot) > 5.6$, can be expected to have higher initial masses and thus shorter lifetimes than the brightest observed cool SGs ($\log (L / L_\odot) \approx 5.6$).
From this it would follow that stars younger than the brightest observed cool SGs do exist in I\,Zw\,18. In that case, the absence of cool SGs with $\log (L / L_\odot) \gtrsim 5.6$ in I\,Zw\,18 is not due to star formation history; rather, it results from such luminous helium-burning stars manifesting as different types of objects such as WR stars. We note that at the metallicity of I\,Zw\,18, a luminosity gap may well exist between the brightest cool SG and the dimmest WR star. In that gap, we would expect evolved stars to be hot, helium-rich stars or intermediate-temperature objects (see Sect.\,\ref{sec:stellar_populations}). For convenience, we provide definitions of different types of stars we discussed in Table\,\ref{tab:types_of_star}. 

\begin{table}[t]
\caption{\label{tab:types_of_star}
Overview of the types of stars discussed in this paper. }
\small
\centering
\begin{tabular}{lll}
\hline
\hline
Name & $T_\mathrm{eff}$ [kK] & Comment\\
\hline
Cool SG & $<7$ & Includes red and yellow SGs \\
Warm SG & 7 to 20  & \\
Intermediate $T_\mathrm{eff}$ & 20 to $\sim$50 & too hot to detect in I\,Zw\,18;\\
 & & too cold for He$^+$ ionization\\ 
Hot He-rich star & $\gtrsim 50$  & has lost its H-rich envelope \\
WR star & $\gtrsim 50$ & a hot He-rich star with winds dense\\ 
& & enough to form emission lines \\

\hline
\end{tabular}
\end{table}

\subsection{Astrophysical implications of the absence of bright cool and warm SGs \label{sec:implications}}

Above, we found no cool and warm SGs in low-metallicity galaxies with $\log (L / L_\odot) \gtrsim 5.6$, even though we would expect to see them based on theoretical arguments. Also, the stellar evolution models analyzed here predict that this absence of bright cool and warm SGs is not a stellar evolution effect.

Apart from I\,Zw\,18 (Sect. \ref{sec:sfh}), WR features have also been reported in the other galaxies we analyzed, where no cool SGs above $\log (L / L_\odot) = 5.6$ were found. NGC\,5253 is said to host on the order of 30 to 40 WR stars \citep{Schaerer97, Westmoquette13}. \cite{Schild91} and \cite{Breysacher97} derived a similar number of WR stars for NGC\,300. Finally, \cite{Brinchmann08} reported WR features in the spectrum of NGC\,4395. In the SMC, LMC, and M\,31, many WR stars have been observed as individual sources \citep{Crowther07}.

To summarize, in the galaxies considered in this study, there is i) an absence of bright cool SGs, ii) a presence of WR stars, and iii) evidence that low-metallicity WR stars are expected to be luminous. Together, this implies that low-metallicity massive stars born with $M_\mathrm{ini} \gtrsim 30$\,M$_\odot$ evolve into hot stars ($T_\mathrm{eff} > 7$\,kK), in many cases WR stars. In Sect.\,\ref{sec:stellar_evolution} we discuss, from a stellar-evolution point of view, the most likely cause of this behavior. The implications for stellar populations are discussed in Sect.\,\ref{sec:stellar_populations}.

\subsubsection{Implications for stellar evolution \label{sec:stellar_evolution}}

\paragraph{Late-phase metallicity-independent mass loss.}
When \cite{Humphreys79} noticed that cool and warm SGs in the LMC and MW had a similar upper luminosity limit, they attributed this to mass loss. Specifically, they proposed that late-phase metallicity-independent mass loss from luminous blue variable (LBV) stars is at work. This has been supported by \cite{Davies18}, who investigated cool SGs in the SMC and LMC. Red supergiant mass loss could play a similar role \citep[][]{Yang23, Zapartas25, Antoniadis25}. 
Metallicity-independent mass loss appears a straightforward explanation for our results in I\,Zw\,18, where we find that the upper luminosity limit for cool SGs is no different than anywhere else.
Unfortunately, mass-loss rates of evolved stars are difficult to constrain \citep{Smith14}, making it challenging to securely confirm or reject this hypothesis. 

Several studies advocate a (near-)independence of these mass-loss rates on metallicity by mechanisms related to, for example, helium opacity peaks, pulsations, or turbulence in extended stellar atmospheres \citep[e.g.,][]{Goldman17, Jiang21, Davidson20, Kee21, Cheng24, vanLoon25, Pauli26}.
In this context, it is interesting to mention that the presence of LBV stars has been inferred in very low-metallicity galaxies, even at metallicities comparable to that of I\,Zw\,18 \citep{Izotov09, Pustilnik08, Pustilnik25}.

\paragraph{Internal mixing.}
It has also been proposed that a combination of metallicity-dependent mass loss and internal mixing processes sets the upper luminosity limit of cool SGs in the Magellanic Clouds and the MW \citep{Higgins20, Sabhahit21, Gilkis21, Schootemeijer19}. However, this could lead either to metallicity-dependent upper luminosity limits \citep{Higgins20} or a need to fine-tune the mass-loss rates to maintain metallicity independence \citep{Sabhahit21}. For the latter, explaining a constant upper luminosity limit of cool SGs over the broader metallicity range shown in Fig.\,\ref{fig:ztrends} (spanning almost 2\,dex, compared to the 0.7\,dex range they considered from the SMC to the MW) may be difficult.
\cite{Gilkis21} found that models with high overshooting values can explain the observed upper cool SG luminosity limits in the MW, LMC, and SMC. However, they found that, especially at SMC metallicity, they predicted too many SGs around $T_\mathrm{eff} \approx 10$\,kK. This led them to conclude that something was missing in the physics assumptions in their models. 

Rotational mixing could also play a role, in particular if it is strong enough to induce chemically homogeneous evolution \citep[CHE; e.g.,][]{Brott11}. Chemically homogeneous evolution models remain hot and evolve into hot, helium-rich stars (potentially with WR-type spectra) rather than cool SGs. Some of the most hydrogen-poor SMC WR stars could be consistent with CHE \citep[see, e.g.,][]{Martins09, Schootemeijer18, Sharpe24, Boco25}. Therefore, CHE has been invoked to (help) explain the cool SG upper luminosity limit in the SMC \citep{Ramachandran19, Boco25}. We emphasize that to fully explain the observed cool SG luminosity limit, all stars above $\sim$30\,M$_\odot$ must undergo CHE. 
However, in the SMC, the properties of these stars in HRDs \citep{Blaha89, Massey95, Schootemeijer21, Bestenlehner25} are the opposite of what a population CHE models would predict. The area near the zero-age main sequence in those HRDs, where CHE models spend around 90\% of their lifetime, is almost completely void. And at cooler temperatures at which CHE models spend little to no time, several tens of stars are observed above the 30\,M$_\odot$ track. Another challenge to this CHE scenario is that 90\% of massive SMC stars appear to rotate too slowly to undergo CHE \citep{Boco25}. Furthermore, CHE models struggle to simultaneously explain the observed projected rotation velocity, temperature, and surface abundances of most SMC WR stars \citep{Hainich15, Vink17, Schootemeijer18, Martins23}.

\paragraph{Binary interaction.} 
Most massive stars are observed in binaries \citep{Sana12, Sana25}.
The luminosity distribution of cool SGs would be affected by binary interaction if it prevents stars from evolving into cool SGs.
However, a high binary fraction among all massive stars would then result in a systematic shift in the cool SG luminosity distribution rather than the observed drop at $\log (L/L_\odot) \approx 5.6$. To explain this drop, we see two possibilities. First, the binary fraction could change with initial mass, where above 30\,M$_\odot$ it approaches unity, whereas at lower masses it does not. The second possibility is that the binary fraction is near unity for all massive stars, but that the outcome of binary interaction would be different above 30\,M$_\odot$ (never cool SGs) and below 30\,M$_\odot$ (cool SG production possible).

Whether either of the two explanations works in practice remains to be seen. Regarding the first explanation, in the BLOeM sample of massive SMC stars \citep{Shenar24}, the binary fraction appears constant and does not increase above 30\,M$_\odot$ \citep[][Fig. E1 of the latter]{Britavskiy25, Sana25, Villasenor25, Bestenlehner25}. These results are based on nine out of 25 epochs of the survey; we expect a more definitive result upon its completion. 
For the second explanation, at first glance, models of accretor stars \citep{Schneider24} and mergers \citep{Menon24} do not appear to show a sharp transition in the lifetimes of cool SGs around 30\,M$_\odot$.
It is also worth mentioning that, if binary stripping were the only method capable of removing the outer layers of evolved massive stars, it would be difficult to explain why 50-60\% of WR stars appear to be genuinely single in both the MW \citep{Deshmukh24} and the SMC \citep[][see sect.\,7.1 of the latter for a detailed discussion]{Foellmi03, Schootemeijer24}. 
Despite these potential challenges, major uncertainties exist in the prevalence of distant and low-mass binary companions, as well as in the physics of mass accretion and merger processes. This argues against drawing strong conclusions on the impact of binaries on the observed upper luminosity limit of cool SGs.

\paragraph{Interpretation and possible implications.} All explanations discussed in this section have their own drawbacks, but, in our view, metallicity-independent late-phase mass loss would require the smallest amount of fine-tuning. Therefore, we argue that metallicity-independent mass loss is the most straightforward explanation for a metallicity-independent cool SG luminosity limit (Fig.\,\ref{fig:ztrends}). To solidify this assessment or to disprove it, we encourage future studies on, for example, binary population synthesis, mass-loss rates of evolved massive stars at low metallicity, or more bias-corrected measurements of binary fractions of metal-poor WR stars and their hydrogen-burning progenitor stars.

FRANEC \citep{Limongi18} and MIST \citep{Dotter16, Choi16} models adopt late-phase mass loss of massive stars that is independent of metallicity. If our interpretation is correct, this assumption is favored over the metallicity-dependence of late-phase mass loss in the PARSEC \citep{Bressan12, Tang14}, Geneva \citep{Georgy13, Groh19}, and Bonn models \citep{Brott11, Schootemeijer19, Szecsi22}.
Some studies assume that hydrogen-rich layers of stars with initial masses above 30\,M$_\odot$ would end up in their BH remnants \citep{Fryer12}.
If instead their hydrogen-rich layers are shed even at low metallicity, this would also set a limitation on their final stellar masses before core collapse and, therefore, on BH remnant masses \citep[see, e.g.,][their fig.\,10]{Schootemeijer24}.

\subsubsection{Implications for He$^+$-ionizing emission from stellar populations \label{sec:stellar_populations}}

For real stellar populations, the important question is whether massive stars evolve into hot, helium-rich stars, rather than whether, for example, binary or wind stripping is at work. 
Below we discuss potential consequences for ionizing radiation and chemical evolution in galaxies for a scenario in which low-metallicity stars born more massive than $\sim$30\,M$_\odot$ evolve into hot, helium-rich stars.

Previous studies \citep{Szecsi15, Szecsi25, Kubatova19} have already proposed that massive stars with I\,Zw\,18 metallicity could become hot enough to produce hard ionizing radiation, while their low metallicity prevents them from launching dense winds.
In their scenario, massive stars undergoing extremely rapid rotation experience CHE induced by efficient rotational mixing. 
Here, we propose a similar scenario, but one in which wind-stripping, rather than CHE, creates the hot stars with weak winds that allow hard ionizing radiation to escape the stellar wind, and thereby lead to emission in, for example, He\,{\sc ii} and C\,{\sc iv}.
This wind stripping would be achieved by late-phase metallicity-independent mass loss 
\citep[as in models from][and see Sect.\,\ref{sec:stellar_evolution}]{Limongi18, Pauli26}.

\begin{figure}[]
\begin{center}
\includegraphics[width=\linewidth]{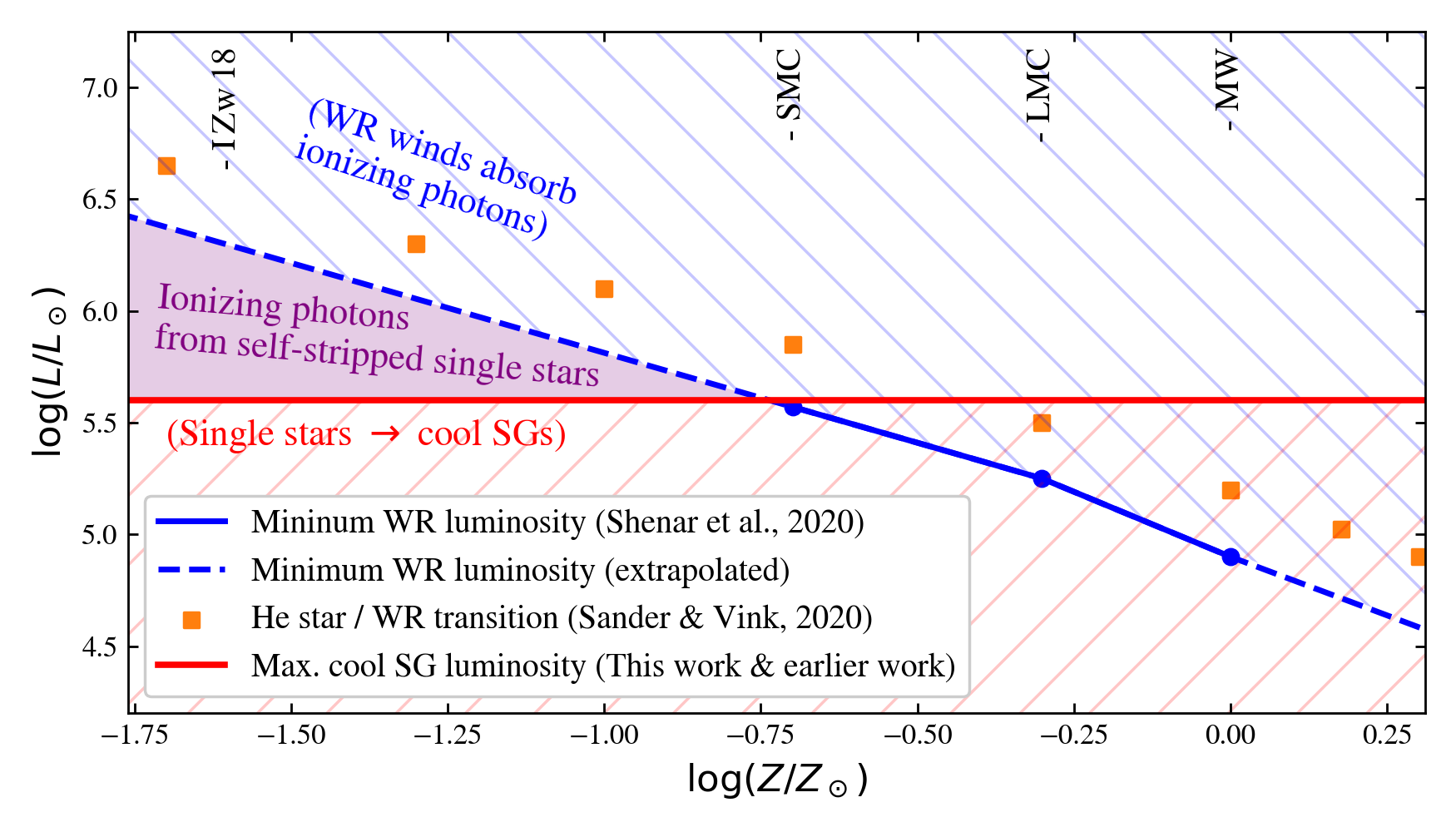} 
 \caption{
 Diagram illustrating the window of opportunity for the production of hard ionizing radiation in our single-star wind-stripping scenario (highlighted with purple shading). The blue- and red-hatched regions indicate where dense WR winds and evolution into cool SGs would prevent the production of hard ionizing radiation, respectively. The values for minimum WR luminosity (blue line) are taken from \cite{Shenar20} and the He star and WR transition in orange are the points where \cite{Sander20} predict the onset of multiple scattering ($\eta = 1$), below which winds are expected to be weak.
   }
 \label{fig:iotons}
\end{center}
\end{figure}

We illustrate this scenario in Fig.\,\ref{fig:iotons}. The figure shows where we expect single stars to produce hard ionizing radiation as a function of metallicity and luminosity.
Since our findings and those in the literature imply that cool SGs (which are too cool to produce hard ionizing photons) exist up to a metallicity-independent upper luminosity limit, we cross out the area with $\log (L / L_\odot) < 5.6$. 
Furthermore, \cite{Sander20} predict that the onset of optically thick WR winds can dramatically reduce the He$^+$-ionizing flux (by a factor of 10$^{10}$; their fig.\,22).
We thus also cross out the part in Fig.\,\ref{fig:iotons} where WR winds are expected for hot, helium-rich stars. Observations show that the onset of the WR phenomenon is metallicity dependent \citep{Hainich14, Hainich15, Hamann19, Shenar20}. 
We adopt the minimum WR luminosity from \cite{Shenar20} and extrapolate outside of their considered metallicity range. 
It is reassuring that, apart from a slight offset, the result of this extrapolation is similar to boundary between helium stars and WR stars in wind models from \cite{Sander20}. 
This metallicity dependence of the onset of WR winds would introduce a window of opportunity for the production of hard ionizing radiation at $\log(Z / Z_\odot) < 0.75$. This is marked with purple shading in Fig.\,\ref{fig:iotons}.
We provide a rough estimate for the rate at which He$^+$ ionizing photons could be emitted in this window of opportunity: 

\begin{equation} \label{eq:woo} 
    Q(\mathrm{He}^+) = \max \! \big[ - 2.1\cdot10^{51} \big( \log(Z / Z_\odot) + 0.75 \big), \ 0 \big] \ \mathrm{s}^{-1} \ \bigg(\frac{\mathrm{M}_\odot}{\mathrm{yr}} \bigg)^{-1}.
\end{equation}
Equation\,\ref{eq:woo} is scaled to an SFR of 1\,M$_\odot$/yr, and it is based on the IMF of \cite{Kroupa01} and the models of \cite{Sander20}. See Appendix\,\ref{app:numbers} for details.

Our scenario has a few caveats. 
First, in some cases, WR stars might emit more He$^+$-ionizing photons than models by \cite{Sander20} suggest \citep{Sander25, GonzalezTora25, Sixtos23}.
Second, our scenario relies on helium-burning massive stars in I\,Zw\,18 with $\log (L / L_\odot) > 5.6$ evolving into hot, helium-rich stars, whereas we have only shown their absence up to $\sim$20\,kK. 
It could be that some helium-burning stars with $M_\mathrm{ini} \gtrsim 30$\,M$_\odot$ evolve into intermediate-temperature objects instead (such cases have been reported in binary systems: one by \cite{Pauli22b}, and four dimmer counterparts at $\log (L / L_\odot) < 5.0$ by \cite{Ramachandran23, Ramachandran24}).
Also, since about 70\% of massive stars are thought to be in binaries \citep[][]{Sana12, Sana25}, binary stripping could produce the majority of hot and luminous helium-rich stars above $\log (L / L_\odot) = 5.6$ while also contributing at lower luminosities \citep[][]{Gotberg18, HovisAfflerbach25}.

On the other hand, the number of binary-stripped stars at low metallicities might be relatively low if low-$Z$ stars tend to stay compact until the late helium-burning phase \citep{HovisAfflerbach25}. Also, \cite{Sana12} estimated that only 30\% of massive stars undergo binary stripping, whereas \cite{Schuermann24} found that binaries might be even more likely to merge upon interaction than previously thought. Therefore, it is plausible that wind stripping is more important than binary stripping for producing luminous helium-rich stars at low metallicity. This is corroborated by the case of the SMC, which hosts five binary WR stars and seven single ones \citep{Foellmi03, Schootemeijer24}. 
Moreover, an advantage of the wind-stripping scenario is that it does not rely on CHE, which has not been proven to occur in nature (Sect.\,\ref{sec:stellar_evolution}).

Popular spectral synthesis tools for stellar populations rely on the evolutionary models 
discussed in Sect.\,\ref{sec:models} \citep[BPASS \citep{Eldridge17}, and Geneva models in case of Starburst99;][]{Leitherer99, Leitherer14}. As such,
these codes are built on single-star tracks that do not well represent massive stellar evolution at I\,Zw\,18 metallicity, and, in fact, entirely overestimate the number of cool and/or warm SGs. 
While BPASS can account for binary interaction, it does not include the sharp decrease in wind mass loss below the single-scattering limit $\eta = 1$ that \cite{Sander20} predict. BPASS can even predict lower ionizing photon emission rates at low metallicities than Starburst99 (see below).
If all low-metallicity stars with $M_\mathrm{ini} \gtrsim 30$\,M$_\odot$ can instead evolve into hot stars with weak winds, then their predicted emission rates of hard ionizing photons from spectral synthesis codes could be significantly underestimated. We investigate this below for the case of I\,Zw\,18.

\subsubsection{Implications for He$^+$-ionizing emission in I Zw 18 \label{sec:izw18_iotons}}

By measuring the integrated narrow-line He\,{\sc ii} emission of I\,Zw\,18, \cite{Kehrig15} estimated that its ensemble of ionizing sources emits He$^+$-ionizing photons at a rate of $Q(\mathrm{He}^+)\,=\,1.3\cdot10^{50}\,\mathrm{s}^{-1}$. They concluded that an unrealistically large population of WR stars (more than 400) is required to explain this number.
For a weak-wind helium star around $\log (L / L_\odot ) = 6$, a value close to $Q(\mathrm{He}^+)\,=\,10^{49}\,\mathrm{s}^{-1}$ is expected \citep{Sander20}.
Explaining the inferred He$^+$-ionization emission rate in I\,Zw\,18 would require only about 20 of such stars, which appears consistent with our findings on the expected number of helium-burning stars in Sects.\,\ref{sec:ldists} and \ref{sec:warmsgs}. 

We now compare this to the predicted He$^+$-ionizing emission rate of I\,Zw\,18 from spectral population synthesis codes. 
Adopting the recent SFR of 0.6\,$M_\odot$/yr from \cite{Bortolini24}, Starburst99 predicts $Q(\mathrm{He}^+)\,\approx\,2.5\cdot10^{50}\,\mathrm{s}^{-1}$ \citep[][their fig.\,82e]{Leitherer99}, while BPASS predicts $Q(\mathrm{He}^+)\,\approx\,1.5\cdot10^{50}\,\mathrm{s}^{-1}$ \citep[][their fig\,11]{Stanway16}. For both predictions, we used the lowest-metallicity underlying models ($Z=0.001$) and note that this metallicity is higher by about a factor of 3 than that of I\,Zw\,18. In Starburst99, the He$^+$-ionizing photons are produced by wind-stripped WR stars born with 85\,M$_\odot$ or more \citep[models from][]{Meynet94}, while in BPASS, they mostly come from binary-stripped stars and accretor stars undergoing CHE following spin-up triggered by binary mass transfer \citep{Eldridge11, Stanway16}.

This good agreement between the He$^+$-ionizing emission rates from observations in I\,Zw\,18 and the theoretical predictions of BPASS and Starburst99 does not signal a need for extra emission by weak-wind self-stripped stars (purple shading in Fig.\,\ref{fig:iotons}). However, those theoretical values are proportional to the SFR, which, in the case of \cite{Bortolini24}, is based on PARSEC models. As seen in Sect.\,\ref{sec:models}, these models likely strongly underestimate cool SG lifetimes at I\,Zw\,18 metallicity, which would lead to an overestimated SFR. 
Similarly, the value of $\sim$1\,M$_\odot$/yr from \cite{Annibali13} is based on evolutionary models from \cite{Fagotto94} that live as cool SGs only briefly (i.e., during carbon burning).
In Appendix\,\ref{app:numbers}, we find a minimum SFR of 0.019\,M$_\odot$/yr for I\,Zw\,18, assuming that the 28 observed cool SGs at $5.0 < \log ( L / L_\odot ) < 5.6$ are in a helium-burning phase lasting 7\% of their lifetime. If the true SFR of I\,Zw\,18 were closer to this value, BPASS and Starburst99 could underpredict $Q$(He$^+$) by about a factor 10, and extra sources of this emission would indeed be required.
According to Eq.\,\ref{eq:woo}, the SFR matching the $Q(\mathrm{He}^+)$ value from \cite{Kehrig15} would be $\sim$0.06\,M$_\odot$/yr, implying that a lower SFR can also account for the extreme ionizing photon conditions in I\,Zw\,18.
Future short-wavelength observations of massive MS stars in I\,Zw\,18 may provide a less model-dependent value of its SFR. This would give vital clues for our self-stripping scenario, as well as for Starburst99 and BPASS models.

\subsubsection{Implications for nitrogen abundances in high-redshift galaxies}

Puzzlingly, high nitrogen abundances and [N/O] ratios have been observed in high-redshift low-metallicity galaxies \citep[e.g.,][]{Bunker23, Cameron23, Arellano-Cordova25}. \cite{Vink23} argues that CNO-processed winds of very massive stars ($M > 100$\,M$_\odot$) are a compelling explanation for these high nitrogen abundances. According to \cite{Vink23}, alternative explanations have different drawbacks: AGB \citep[][]{Timmes95} stars cannot explain the observed Na-N anticorrelation; supermassive stars \citep[$M>1000$\,M$_\odot$][]{Nandal24a} and stars undergoing rotationally induced CHE \citep[][see also Sect.\,\ref{sec:stellar_evolution}]{Roy21, Nandal24b} have never been directly observed; and stars with $M<100$\,M$_\odot$ have too weak winds.
If the absence of bright cool SGs above $\sim$30\,M$_\odot$ at low metallicities is indeed caused by the loss of their hydrogen-rich layers (either via winds or inefficient binary mass transfer, including those that are CNO-processed), then stars above this mass limit could also contribute to the observed nitrogen enrichment in high-redshift galaxies.

We estimated the potential relative contribution to nitrogen enrichment from stars in the initial mass range $30 \ldots 100$\,M$\odot$ compared to very massive stars. For this purpose we investigated a model with $M_\mathrm{ini} = 32$\,M$_\odot$ from \cite{Schootemeijer19}, with semi-convection and overshooting parameters of 1 and 0.33, respectively, that had just completed central hydrogen burning. We find that it has about 7\,M$_\odot$ of hydrogen-rich layers in CNO equilibrium (i.e., the difference between the masses of the inner mixing regions at the start and end of core hydrogen-burning phase). We compared this to a similar model of $M_\mathrm{ini} = 100$\,M$_\odot$ and find $\sim$28\,M$_\odot$ of hydrogen-rich layers in CNO equilibrium, a proportionality of $\propto M_\mathrm{ini}^{1.18}$. 
Combining this with the IMF of \cite{Kroupa01} and integrating, we find that stars born with $30 \ldots 100$\,M$_\odot$ could contribute as much as very massive stars above $M_\mathrm{ini}=100$\,M$_\odot$ if the upper mass limit in nature were 280\,M$_\odot$ (in reality, it might be higher or lower and depend on metallicity). 
From this rough estimate we conclude that, if very massive stars can drive nitrogen enrichment in high-redshift galaxies, stars with $ M_\mathrm{ini} = 30 \ldots 100$\,M$_\odot$ may also significantly contribute.

\section{Conclusions \label{sec:conclusions}}

In this study, we used archival HST and JWST photometry data of four dwarf galaxies to derive the luminosity distributions of their cool SG ($T_\mathrm{eff} / \mathrm{kK} < 7$) populations. 
We combined this sample with cool SGs in the SMC, LMC, and M\,31 with known luminosities.
The resulting ensemble of the studied galaxies covers a metallicity range of $1/40 \lesssim Z / Z_\odot \lesssim 1.5$. The same picture emerges at all metallicities, both in this work and in literature studies: cool SGs with $\log (L / L_\odot) > 5.6$ are absent, while many cool SGs are observed at lower luminosities. By including I\,Zw\,18, our work extends the boundary from $Z_\odot/5$ to $Z_\odot/40$.

We found that theoretical stellar populations predict that helium-burning stars with $\log (L / L_\odot) > 5.6$ should be present. 
In I\,Zw\,18 ($Z/Z_\odot \approx 1/40$), we investigated whether the absence of cool SGs at these luminosities could be explained by the most massive stars evolving into warm SGs ($7 < T_\mathrm{eff} / \mathrm{kK} < 20$).
This does not seem to be the case, because the locus in the CMD where such stars would be expected is empty.
This implies that in I\,Zw\,18, stars born more massive than $\sim$30\,M$_\odot$ evolve into neither cool nor warm SGs.

In the literature, the presence of a multitude of WR stars has been spectroscopically inferred for all the galaxies investigated, including I\,Zw\,18.
Combining this with our findings on the absence of bright cool and warm SGs and previous studies in M\,31, the MW, and the Magellanic Clouds, a convergent picture emerges in which, even at extremely low metallicities, stars above $\sim$30\,M$_\odot$ evolve into hot stars instead. 
How hot and helium-rich these stars tend to be, and how often they display WR features, remains to be determined.
But it seems likely that they lose most of their hydrogen-rich outer layers, since all the considered evolutionary models above 30\,M$_\odot$ retaining those layers manifest as cool or warm SGs.

From a stellar-evolution point of view, we argued that the most straightforward explanation for the metallicity-independent cool SG upper luminosity limit is metallicity-independent mass loss during the LBV or cool SG phase. 
This would limit the masses of BH descendants from (extremely) low-metallicity massive stars.
Also, our results indicate that low-metallicity massive stars born with $30 \lesssim M_\mathrm{ini}/M_\odot \lesssim 100$ may contribute to explaining the observed high nitrogen abundances in high-redshift galaxies.

As a possible pathway to produce hard ionizing radiation at low metallicities, we propose a scenario in which single stars undergo self-stripping. There, independent of metallicity, evolved stars above the observed cool SG luminosity limit of $\log (L / L_\odot) \approx 5.6$ shed their hydrogen-rich layers and become hot objects. Since the minimum luminosity for the WR phenomenon to arise increases with decreasing metallicity, below SMC metallicity a window of opportunity opens for hot, helium-rich stars whose hard ionizing photons avoid absorption in a dense WR wind.
This could help explain the signatures of hard ionizing photons observed in low-metallicity galaxies, such as He\,{\sc ii} and C\,{\sc iv} emission.

\begin{acknowledgements}
We thank our anonymous referee for carefully reading the manuscript and providing a constructive report with helpful feedback. 
This work is based in part on observations made with the NASA/ESA/CSA James Webb Space Telescope. The data were obtained from the Mikulski Archive for Space Telescopes at the Space Telescope Science Institute, which is operated by the Association of Universities for Research in Astronomy, Inc., under NASA contract NAS 5-03127 for JWST. These observations are associated with program \#1233. The specific observations analyzed can be accessed via DOI: \href{https://archive.stsci.edu/doi/resolve/resolve.html?doi=10.17909/3c1d-6182}{10.17909/3c1d-6182}. Moreover, this research is based in part on observations made with the NASA/ESA Hubble Space Telescope obtained from the Space Telescope Science Institute, which is operated by the Association of Universities for Research in Astronomy, Inc., under NASA contract NAS 5–26555. These observations are associated with programs \#13664, GO-10915, and DD-11307. This research was supported in part by grant NSF PHY-2309135 to the Kavli Institute for Theoretical Physics (KITP).
LRP acknowledges support by grants PID2019-105552RB-C41 and PID2022-137779OB-C41 funded by MCIN/AEI/10.13039/501100011033 by "ERDF A way of making Europe". LRP acknowledges support from grant PID2022-140483NB-C22 funded by MCIN/AEI/10.13039/501100011033.
\end{acknowledgements}

\bibliography{bib}

@ARTICLE{Aadland18,
       author = {{Aadland}, Erin and {Massey}, Philip and {Neugent}, Kathryn F. and
         {Drout}, Maria R.},
        title = "{Shedding Light on the Isolation of Luminous Blue Variables}",
      journal = {\aj},
         year = "2018",
        month = "Dec",
       volume = {156},
       number = {6},
          eid = {294},
        pages = {294},
          doi = {10.3847/1538-3881/aaeb96},
archivePrefix = {arXiv},
       eprint = {1810.11169},
 primaryClass = {astro-ph.SR},
       adsurl = {https://ui.adsabs.harvard.edu/abs/2018AJ....156..294A}
}

@ARTICLE{Abbott16,
   author = {{Abbott}, B.~P. and {Abbott}, R. and {Abbott}, T.~D. and {Abernathy}, M.~R. and 
	{Acernese}, F. and {Ackley}, K. and {Adams}, C. and {Adams}, T. and 
	{Addesso}, P. and {Adhikari}, R.~X. and et al.},
    title = "{{\GG{201602}}Observation of Gravitational Waves from a Binary Black Hole Merger}",
  journal = {Physical Review Letters},
archivePrefix = "arXiv",
   eprint = {1602.03837},
 primaryClass = "gr-qc",
     year = 2016,
    month = feb,
   volume = 116,
   number = 6,
      eid = {061102},
    pages = {061102},
      doi = {10.1103/PhysRevLett.116.061102},
   adsurl = {http://adsabs.harvard.edu/abs/2016PhRvL.116f1102A}
}

@ARTICLE{Abril-Melgarejo24,
       author = {{Abril-Melgarejo}, Valentina and {James}, Bethan L. and {Aloisi}, Alessandra and {Mingozzi}, Matilde and {Lebouteiller}, Vianney and {Hernandez}, Svea and {Kumari}, Nimisha and {AAS Journals Data Editors}},
        title = "{Mapping Multiphase Metals in Star-forming Galaxies: A Spatially Resolved UV+Optical Study of NGC 5253}",
      journal = {\apj},
         year = 2024,
        month = oct,
       volume = {973},
       number = {2},
          eid = {173},
        pages = {173},
          doi = {10.3847/1538-4357/ad5e79},
archivePrefix = {arXiv},
       eprint = {2406.16553},
 primaryClass = {astro-ph.GA},
       adsurl = {https://ui.adsabs.harvard.edu/abs/2024ApJ...973..173A}
}

@ARTICLE{Aloisi07,
       author = {{Aloisi}, A. and {Clementini}, G. and {Tosi}, M. and {Annibali}, F. and {Contreras}, R. and {Fiorentino}, G. and {Mack}, J. and {Marconi}, M. and {Musella}, I. and {Saha}, A. and {Sirianni}, M. and {van der Marel}, R.~P.},
        title = "{I Zw 18 Revisited with HST ACS and Cepheids: New Distance and Age}",
      journal = {\apjl},
         year = 2007,
        month = oct,
       volume = {667},
       number = {2},
        pages = {L151-L154},
          doi = {10.1086/522368},
archivePrefix = {arXiv},
       eprint = {0707.2371},
 primaryClass = {astro-ph},
       adsurl = {https://ui.adsabs.harvard.edu/abs/2007ApJ...667L.151A}
}

@ARTICLE{Annibali13,
       author = {{Annibali}, F. and {Cignoni}, M. and {Tosi}, M. and {van der Marel}, R.~P. and {Aloisi}, A. and {Clementini}, G. and {Contreras Ramos}, R. and {Fiorentino}, G. and {Marconi}, M. and {Musella}, I.},
        title = "{The Star Formation History of the Very Metal-poor Blue Compact Dwarf I Zw 18 from HST/ACS Data}",
      journal = {\aj},
         year = 2013,
        month = dec,
       volume = {146},
       number = {6},
          eid = {144},
        pages = {144},
          doi = {10.1088/0004-6256/146/6/144},
archivePrefix = {arXiv},
       eprint = {1303.3909},
 primaryClass = {astro-ph.CO},
       adsurl = {https://ui.adsabs.harvard.edu/abs/2013AJ....146..144A}
}

@ARTICLE{Antoniadis25,
       author = {{Antoniadis}, K. and {Zapartas}, E. and {Bonanos}, A.~Z. and {Maravelias}, G. and {Vlassis}, S. and {Mu{\~n}oz-Sanchez}, G. and {Nally}, C. and {Meixner}, M. and {Jones}, O.~C. and {Lenki{\'c}}, L. and {Kavanagh}, P.~J.},
        title = "{Investigating the metallicity dependence of the mass-loss rate relation of red supergiants}",
      journal = {\aap},
         year = 2025,
        month = oct,
       volume = {702},
          eid = {A178},
        pages = {A178},
          doi = {10.1051/0004-6361/202554416},
archivePrefix = {arXiv},
       eprint = {2503.05876},
 primaryClass = {astro-ph.SR},
       adsurl = {https://ui.adsabs.harvard.edu/abs/2025A&A...702A.178A}
}

@ARTICLE{Arellano-Cordova25,
       author = {{Arellano-C{\'o}rdova}, K.~Z. and {Berg}, D.~A. and {Mingozzi}, M. and {James}, B.~L. and {Vincenzo}, F. and {Rogers}, N.~S.~J. and {Skillman}, E.~D. and {Amor{\'\i}n}, R.~O. and {Cullen}, F. and {Flury}, S.~R. and {Abril-Melgarejo}, V. and {Chisholm}, J. and {Heckman}, T. and {Hayes}, M.~J. and {Hernandez}, S. and {Kumari}, N. and {Kobayashi}, C. and {Leitherer}, C. and {Martin}, C.~L. and {Martinez}, Z. and {Nanayakkara}, T. and {Parker}, K.~S. and {Senchyna}, P. and {Scarlata}, C. and {Stephenson}, M.~G. and {Wofford}, A. and {Xu}, X. and {Zhu}, P.},
        title = "{CLASSY XII: nitrogen enrichment shaped by gas density and feedback}",
      journal = {\mnras},
         year = 2025,
        month = dec,
       volume = {544},
       number = {2},
        pages = {1588-1607},
          doi = {10.1093/mnras/staf1723},
archivePrefix = {arXiv},
       eprint = {2507.11658},
 primaryClass = {astro-ph.GA},
       adsurl = {https://ui.adsabs.harvard.edu/abs/2025MNRAS.544.1588A}
}

@ARTICLE{Backs24,
       author = {{Backs}, F. and {Brands}, S.~A. and {de Koter}, A. and {Kaper}, L. and {Vink}, J.~S. and {Puls}, J. and {Sundqvist}, J. and {Tramper}, F. and {Sana}, H. and {Bernini-Peron}, M. and {Bestenlehner}, J.~M. and {Crowther}, P.~A. and {Hawcroft}, C. and {Ignace}, R. and {Kuiper}, R. and {van Loon}, J. Th. and {Mahy}, L. and {Marcolino}, W. and {Najarro}, F. and {Oskinova}, L.~M. and {Pauli}, D. and {Ramachandran}, V. and {Sander}, A.~A.~C. and {Verhamme}, O.},
        title = "{X-Shooting ULLYSES: Massive stars at low metallicity: VI. Atmosphere and mass-loss properties of O-type giants in the Small Magellanic Cloud}",
      journal = {\aap},
         year = 2024,
        month = dec,
       volume = {692},
          eid = {A88},
        pages = {A88},
          doi = {10.1051/0004-6361/202451893},
archivePrefix = {arXiv},
       eprint = {2411.06884},
 primaryClass = {astro-ph.SR},
       adsurl = {https://ui.adsabs.harvard.edu/abs/2024A&A...692A..88B}
}

@ARTICLE{Bergemann21,
       author = {{Bergemann}, Maria and {Hoppe}, Richard and {Semenova}, Ekaterina and {Carlsson}, Mats and {Yakovleva}, Svetlana A. and {Voronov}, Yaroslav V. and {Bautista}, Manuel and {Nemer}, Ahmad and {Belyaev}, Andrey K. and {Leenaarts}, Jorrit and {Mashonkina}, Lyudmila and {Reiners}, Ansgar and {Ellwarth}, Monika},
        title = "{Solar oxygen abundance}",
      journal = {\mnras},
         year = 2021,
        month = dec,
       volume = {508},
       number = {2},
        pages = {2236-2253},
          doi = {10.1093/mnras/stab2160},
archivePrefix = {arXiv},
       eprint = {2109.01143},
 primaryClass = {astro-ph.SR},
       adsurl = {https://ui.adsabs.harvard.edu/abs/2021MNRAS.508.2236B}
}

@ARTICLE{Bestenlehner25,
       author = {{Bestenlehner}, J.~M. and {Crowther}, Paul A. and {Bronner}, V.~A. and {Sim{\'o}n-D{\'\i}az}, S. and {Lennon}, D.~J. and {Bodensteiner}, J. and {Langer}, N. and {Marchant}, P. and {Sana}, H. and {Schneider}, F.~R.~N. and {Shenar}, T.},
        title = "{Binarity at LOw Metallicity (BLOeM): pipeline-determined physical properties of OB stars}",
      journal = {\mnras},
         year = 2025,
        month = jul,
       volume = {540},
       number = {4},
        pages = {3523-3548},
          doi = {10.1093/mnras/staf900},
archivePrefix = {arXiv},
       eprint = {2506.00117},
 primaryClass = {astro-ph.SR},
       adsurl = {https://ui.adsabs.harvard.edu/abs/2025MNRAS.540.3523B}
}

@ARTICLE{Blaha89,
   author = {{Blaha}, C. and {Humphreys}, R.~M.},
    title = "{A comparison of the luminosity functions in U, B, and V and their relationship to the initial mass function for the Galaxy and the Magellanic Clouds}",
  journal = {\aj},
     year = 1989,
    month = nov,
   volume = 98,
    pages = {1598-1608},
      doi = {10.1086/115244},
   adsurl = {http://adsabs.harvard.edu/abs/1989AJ.....98.1598B}
}

@ARTICLE{Boco25,
       author = {{Boco}, Lumen and {Mapelli}, Michela and {Sander}, Andreas A.~C. and {Mesini}, Sofia and {Ramachandran}, Varsha and {Torniamenti}, Stefano and {Korb}, Erika and {Liu}, Boyuan and {Sabhahit}, Gautham N. and {Vink}, Jorick S.},
        title = "{Metal-poor single Wolf-Rayet stars: The interplay of optically thick winds and rotation}",
      journal = {\aap},
         year = 2025,
        month = nov,
       volume = {703},
          eid = {A243},
        pages = {A243},
          doi = {10.1051/0004-6361/202556187},
archivePrefix = {arXiv},
       eprint = {2507.00137},
 primaryClass = {astro-ph.SR},
       adsurl = {https://ui.adsabs.harvard.edu/abs/2025A&A...703A.243B}
}

@ARTICLE{Bonanos25,
       author = {{Bonanos}, Alceste Z.},
        title = "{Red Supergiants in the Milky Way and Nearby Galaxies}",
      journal = {Galaxies},
         year = 2025,
        month = jun,
       volume = {13},
       number = {3},
          eid = {66},
        pages = {66},
          doi = {10.3390/galaxies13030066},
archivePrefix = {arXiv},
       eprint = {2507.15964},
 primaryClass = {astro-ph.GA},
       adsurl = {https://ui.adsabs.harvard.edu/abs/2025Galax..13...66B}
}

@ARTICLE{Bortolini24,
       author = {{Bortolini}, Giacomo and {{\"O}stlin}, G{\"o}ran and {Habel}, Nolan and {Hirschauer}, Alec S. and {Jones}, Olivia C. and {Justtanont}, Kay and {Meixner}, Margaret and {Boyer}, Martha L. and {Blommaert}, Joris A.~D.~L. and {Crouzet}, Nicolas and {Lenki{\'c}}, Laura and {Nally}, Conor and {Sargent}, Beth A. and {van der Werf}, Paul and {G{\"u}del}, Manuel and {Henning}, Thomas and {Lagage}, Pierre O.},
        title = "{Imaging of I Zw 18 by JWST: II. Spatially resolved star formation history}",
      journal = {\aap},
         year = 2024,
        month = sep,
       volume = {689},
          eid = {A146},
        pages = {A146},
          doi = {10.1051/0004-6361/202450632},
archivePrefix = {arXiv},
       eprint = {2406.17429},
 primaryClass = {astro-ph.GA},
       adsurl = {https://ui.adsabs.harvard.edu/abs/2024A&A...689A.146B}
}

@ARTICLE{Bressan12,
       author = {{Bressan}, Alessandro and {Marigo}, Paola and {Girardi}, L{\'e}o. and {Salasnich}, Bernardo and {Dal Cero}, Claudia and {Rubele}, Stefano and {Nanni}, Ambra},
        title = "{PARSEC: stellar tracks and isochrones with the PAdova and TRieste Stellar Evolution Code}",
      journal = {\mnras},
         year = 2012,
        month = nov,
       volume = {427},
       number = {1},
        pages = {127-145},
          doi = {10.1111/j.1365-2966.2012.21948.x},
archivePrefix = {arXiv},
       eprint = {1208.4498},
 primaryClass = {astro-ph.SR},
       adsurl = {https://ui.adsabs.harvard.edu/abs/2012MNRAS.427..127B}
}

@ARTICLE{Breysacher97,
       author = {{Breysacher}, J. and {Azzopardi}, M. and {Testor}, G. and {Muratorio}, G.},
        title = "{Wolf-Rayet stars detected in five associations of NGC 300.}",
      journal = {\aap},
         year = 1997,
        month = oct,
       volume = {326},
        pages = {976-981},
       adsurl = {https://ui.adsabs.harvard.edu/abs/1997A&A...326..976B}
}

@ARTICLE{Brinchmann08,
       author = {{Brinchmann}, J. and {Kunth}, D. and {Durret}, F.},
        title = "{Galaxies with Wolf-Rayet signatures in the low-redshift Universe. A survey using the Sloan Digital Sky Survey}",
      journal = {\aap},
         year = 2008,
        month = jul,
       volume = {485},
       number = {3},
        pages = {657-677},
          doi = {10.1051/0004-6361:200809783},
archivePrefix = {arXiv},
       eprint = {0805.1073},
 primaryClass = {astro-ph},
       adsurl = {https://ui.adsabs.harvard.edu/abs/2008A&A...485..657B}
}

@ARTICLE{Britavskiy25,
       author = {{Britavskiy}, N. and {Mahy}, L. and {Lennon}, D.~J. and {Patrick}, L.~R. and {Sana}, H. and {Villase{\~n}or}, J.~I. and {Shenar}, T. and {Bodensteiner}, J. and {Bernini-Peron}, M. and {Berlanas}, S.~R. and {Bowman}, D.~M. and {Crowther}, P.~A. and {de Mink}, S.~E. and {Evans}, C.~J. and {G{\"o}tberg}, Y. and {Holgado}, G. and {Johnston}, C. and {Keszthelyi}, Z. and {Klencki}, J. and {Langer}, N. and {Mandel}, I. and {Menon}, A. and {Moe}, M. and {Oskinova}, L.~M. and {Pauli}, D. and {Pawlak}, M. and {Ramachandran}, V. and {Renzo}, M. and {Sander}, A.~A.~C. and {Schneider}, F.~R.~N. and {Schootemeijer}, A. and {Sen}, K. and {Sim{\'o}n-D{\'\i}az}, S. and {van Loon}, J. Th. and {Vink}, J.~S.},
        title = "{Binarity at LOw Metallicity (BLOeM): Multiplicity of early B-type supergiants in the Small Magellanic Cloud}",
      journal = {\aap},
         year = 2025,
        month = jun,
       volume = {698},
          eid = {A40},
        pages = {A40},
          doi = {10.1051/0004-6361/202452963},
archivePrefix = {arXiv},
       eprint = {2502.12239},
 primaryClass = {astro-ph.SR},
       adsurl = {https://ui.adsabs.harvard.edu/abs/2025A&A...698A..40B}
}

@ARTICLE{Brott11,
   author = {{Brott}, I. and {de Mink}, S.~E. and {Cantiello}, M. and {Langer}, N. and 
	{de Koter}, A. and {Evans}, C.~J. and {Hunter}, I. and {Trundle}, C. and 
	{Vink}, J.~S.},
    title = "{Rotating massive main-sequence stars. I. Grids of evolutionary models and isochrones}",
  journal = {\aap},
archivePrefix = "arXiv",
   eprint = {1102.0530},
 primaryClass = "astro-ph.SR",
     year = 2011,
    month = jun,
   volume = 530,
      eid = {A115},
    pages = {A115},
      doi = {10.1051/0004-6361/201016113},
   adsurl = {http://adsabs.harvard.edu/abs/2011A%26A...530A.115B}
}

@ARTICLE{Brown02,
       author = {{Brown}, Thomas M. and {Heap}, Sara R. and {Hubeny}, Ivan and {Lanz}, Thierry and {Lindler}, Don},
        title = "{Isolating Clusters with Wolf-Rayet Stars in I Zw 18}",
      journal = {\apjl},
         year = 2002,
        month = nov,
       volume = {579},
       number = {2},
        pages = {L75-L78},
          doi = {10.1086/345336},
archivePrefix = {arXiv},
       eprint = {astro-ph/0210089},
 primaryClass = {astro-ph},
       adsurl = {https://ui.adsabs.harvard.edu/abs/2002ApJ...579L..75B}
}

@ARTICLE{Bunker23,
       author = {{Bunker}, Andrew J. and {Saxena}, Aayush and {Cameron}, Alex J. and {Willott}, Chris J. and {Curtis-Lake}, Emma and {Jakobsen}, Peter and {Carniani}, Stefano and {Smit}, Renske and {Maiolino}, Roberto and {Witstok}, Joris and {Curti}, Mirko and {D'Eugenio}, Francesco and {Jones}, Gareth C. and {Ferruit}, Pierre and {Arribas}, Santiago and {Charlot}, Stephane and {Chevallard}, Jacopo and {Giardino}, Giovanna and {de Graaff}, Anna and {Looser}, Tobias J. and {L{\"u}tzgendorf}, Nora and {Maseda}, Michael V. and {Rawle}, Tim and {Rix}, Hans-Walter and {Del Pino}, Bruno Rodr{\'\i}guez and {Alberts}, Stacey and {Egami}, Eiichi and {Eisenstein}, Daniel J. and {Endsley}, Ryan and {Hainline}, Kevin and {Hausen}, Ryan and {Johnson}, Benjamin D. and {Rieke}, George and {Rieke}, Marcia and {Robertson}, Brant E. and {Shivaei}, Irene and {Stark}, Daniel P. and {Sun}, Fengwu and {Tacchella}, Sandro and {Tang}, Mengtao and {Williams}, Christina C. and {Willmer}, Christopher N.~A. and {Baker}, William M. and {Baum}, Stefi and {Bhatawdekar}, Rachana and {Bowler}, Rebecca and {Boyett}, Kristan and {Chen}, Zuyi and {Circosta}, Chiara and {Helton}, Jakob M. and {Ji}, Zhiyuan and {Kumari}, Nimisha and {Lyu}, Jianwei and {Nelson}, Erica and {Parlanti}, Eleonora and {Perna}, Michele and {Sandles}, Lester and {Scholtz}, Jan and {Suess}, Katherine A. and {Topping}, Michael W. and {{\"U}bler}, Hannah and {Wallace}, Imaan E.~B. and {Whitler}, Lily},
        title = "{JADES NIRSpec Spectroscopy of GN-z11: Lyman-{\ensuremath{\alpha}} emission and possible enhanced nitrogen abundance in a z = 10.60 luminous galaxy}",
      journal = {\aap},
         year = 2023,
        month = sep,
       volume = {677},
          eid = {A88},
        pages = {A88},
          doi = {10.1051/0004-6361/202346159},
archivePrefix = {arXiv},
       eprint = {2302.07256},
 primaryClass = {astro-ph.GA},
       adsurl = {https://ui.adsabs.harvard.edu/abs/2023A&A...677A..88B}
}

@ARTICLE{Cameron23,
       author = {{Cameron}, Alex J. and {Katz}, Harley and {Rey}, Martin P. and {Saxena}, Aayush},
        title = "{Nitrogen enhancements 440 Myr after the big bang: supersolar N/O, a tidal disruption event, or a dense stellar cluster in GN-z11?}",
      journal = {\mnras},
         year = 2023,
        month = aug,
       volume = {523},
       number = {3},
        pages = {3516-3525},
          doi = {10.1093/mnras/stad1579},
archivePrefix = {arXiv},
       eprint = {2302.10142},
 primaryClass = {astro-ph.GA},
       adsurl = {https://ui.adsabs.harvard.edu/abs/2023MNRAS.523.3516C}
}

@ARTICLE{Cameron24,
       author = {{Cameron}, Alex J. and {Katz}, Harley and {Witten}, Callum and {Saxena}, Aayush and {Laporte}, Nicolas and {Bunker}, Andrew J.},
        title = "{Nebular dominated galaxies: insights into the stellar initial mass function at high redshift}",
      journal = {\mnras},
         year = 2024,
        month = oct,
       volume = {534},
       number = {1},
        pages = {523-543},
          doi = {10.1093/mnras/stae1547},
archivePrefix = {arXiv},
       eprint = {2311.02051},
 primaryClass = {astro-ph.GA},
       adsurl = {https://ui.adsabs.harvard.edu/abs/2024MNRAS.534..523C}
}

@ARTICLE{Cannon02,
       author = {{Cannon}, John M. and {Skillman}, Evan D. and {Garnett}, Donald R. and {Dufour}, Reginald J.},
        title = "{Dust in I Zw 18 from Hubble Space Telescope Narrowband Imaging}",
      journal = {\apj},
         year = 2002,
        month = feb,
       volume = {565},
       number = {2},
        pages = {931-940},
          doi = {10.1086/324691},
archivePrefix = {arXiv},
       eprint = {astro-ph/0109558},
 primaryClass = {astro-ph},
       adsurl = {https://ui.adsabs.harvard.edu/abs/2002ApJ...565..931C}
}

@ARTICLE{Cedres02,
       author = {{Cedr{\'e}s}, B. and {Cepa}, J.},
        title = "{Distributions, equivalent widths and metallicities of the H II regions in the spiral galaxies NGC 5457 and NGC 4395}",
      journal = {\aap},
         year = 2002,
        month = sep,
       volume = {391},
        pages = {809-821},
          doi = {10.1051/0004-6361:20020588},
       adsurl = {https://ui.adsabs.harvard.edu/abs/2002A&A...391..809C}
}

@ARTICLE{Cheng24,
       author = {{Cheng}, Shelley J. and {Goldberg}, Jared A. and {Cantiello}, Matteo and {Bauer}, Evan B. and {Renzo}, Mathieu and {Conroy}, Charlie},
        title = "{A Model for Eruptive Mass Loss in Massive Stars}",
      journal = {\apj},
         year = 2024,
        month = oct,
       volume = {974},
       number = {2},
          eid = {270},
        pages = {270},
          doi = {10.3847/1538-4357/ad701e},
archivePrefix = {arXiv},
       eprint = {2405.12274},
 primaryClass = {astro-ph.SR},
       adsurl = {https://ui.adsabs.harvard.edu/abs/2024ApJ...974..270C}
}

@ARTICLE{Choi16,
   author = {{Choi}, J. and {Dotter}, A. and {Conroy}, C. and {Cantiello}, M. and 
	{Paxton}, B. and {Johnson}, B.~D.},
    title = "{Mesa Isochrones and Stellar Tracks (MIST). I. Solar-scaled Models}",
  journal = {\apj},
archivePrefix = "arXiv",
   eprint = {1604.08592},
 primaryClass = "astro-ph.SR",
     year = 2016,
    month = jun,
   volume = 823,
      eid = {102},
    pages = {102},
      doi = {10.3847/0004-637X/823/2/102},
   adsurl = {http://adsabs.harvard.edu/abs/2016ApJ...823..102C}
}

@ARTICLE{Choudhury21,
       author = {{Choudhury}, Samyaday and {de Grijs}, Richard and {Bekki}, Kenji and {Cioni}, Maria-Rosa L. and {Ivanov}, Valentin D. and {van Loon}, Jacco Th and {Miller}, Amy E. and {Niederhofer}, Florian and {Oliveira}, Joana M. and {Ripepi}, Vincenzo and {Sun}, Ning-Chen and {Subramanian}, Smitha},
        title = "{The VMC survey - XLIV: mapping metallicity trends in the large magellanic cloud using near-infrared passbands}",
      journal = {\mnras},
         year = 2021,
        month = nov,
       volume = {507},
       number = {4},
        pages = {4752-4763},
          doi = {10.1093/mnras/stab2446},
archivePrefix = {arXiv},
       eprint = {2108.10529},
 primaryClass = {astro-ph.GA},
       adsurl = {https://ui.adsabs.harvard.edu/abs/2021MNRAS.507.4752C}
}

@ARTICLE{Cioni06,
       author = {{Cioni}, M. -R.~L. and {Girardi}, L. and {Marigo}, P. and {Habing}, H.~J.},
        title = "{AGB stars in the Magellanic Clouds. III. The rate of star formation across the Small Magellanic Cloud}",
      journal = {\aap},
         year = 2006,
        month = jun,
       volume = {452},
       number = {1},
        pages = {195-201},
          doi = {10.1051/0004-6361:20054699},
archivePrefix = {arXiv},
       eprint = {astro-ph/0602483},
 primaryClass = {astro-ph},
       adsurl = {https://ui.adsabs.harvard.edu/abs/2006A&A...452..195C}
}

@ARTICLE{Conti94,
       author = {{Conti}, Peter S. and {Vacca}, William D.},
        title = "{HST UV Imaging of the Starburst Regions in the Wolf-Rayet Galaxy He 2-10: Newly Formed Globular Clusters?}",
      journal = {\apjl},
         year = 1994,
        month = mar,
       volume = {423},
        pages = {L97},
          doi = {10.1086/187245},
       adsurl = {https://ui.adsabs.harvard.edu/abs/1994ApJ...423L..97C}
}

@ARTICLE{Costa21,
       author = {{Costa}, Guglielmo and {Bressan}, Alessandro and {Mapelli}, Michela and {Marigo}, Paola and {Iorio}, Giuliano and {Spera}, Mario},
        title = "{Formation of GW190521 from stellar evolution: the impact of the hydrogen-rich envelope, dredge-up, and $^{12}$C({\ensuremath{\alpha}}, {\ensuremath{\gamma}})$^{16}$O rate on the pair-instability black hole mass gap}",
      journal = {\mnras},
         year = 2021,
        month = mar,
       volume = {501},
       number = {3},
        pages = {4514-4533},
          doi = {10.1093/mnras/staa3916},
archivePrefix = {arXiv},
       eprint = {2010.02242},
 primaryClass = {astro-ph.SR},
       adsurl = {https://ui.adsabs.harvard.edu/abs/2021MNRAS.501.4514C}
}

@ARTICLE{Crowther07,
   author = {{Crowther}, P.~A.},
    title = "{Physical Properties of Wolf-Rayet Stars}",
  journal = {\araa},
   eprint = {astro-ph/0610356},
     year = 2007,
    month = sep,
   volume = 45,
    pages = {177-219},
      doi = {10.1146/annurev.astro.45.051806.110615},
   adsurl = {http://adsabs.harvard.edu/abs/2007ARA%26A..45..177C}
}

@ARTICLE{Dalcanton09,
       author = {{Dalcanton}, Julianne J. and {Williams}, Benjamin F. and {Seth}, Anil C. and {Dolphin}, Andrew and {Holtzman}, Jon and {Rosema}, Keith and {Skillman}, Evan D. and {Cole}, Andrew and {Girardi}, L{\'e}o and {Gogarten}, Stephanie M. and {Karachentsev}, Igor D. and {Olsen}, Knut and {Weisz}, Daniel and {Christensen}, Charlotte and {Freeman}, Ken and {Gilbert}, Karoline and {Gallart}, Carme and {Harris}, Jason and {Hodge}, Paul and {de Jong}, Roelof S. and {Karachentseva}, Valentina and {Mateo}, Mario and {Stetson}, Peter B. and {Tavarez}, Maritza and {Zaritsky}, Dennis and {Governato}, Fabio and {Quinn}, Thomas},
        title = "{The ACS Nearby Galaxy Survey Treasury}",
      journal = {\apjs},
         year = 2009,
        month = jul,
       volume = {183},
       number = {1},
        pages = {67-108},
          doi = {10.1088/0067-0049/183/1/67},
archivePrefix = {arXiv},
       eprint = {0905.3737},
 primaryClass = {astro-ph.GA},
       adsurl = {https://ui.adsabs.harvard.edu/abs/2009ApJS..183...67D}
}

@ARTICLE{Davidge07,
       author = {{Davidge}, T.~J.},
        title = "{NGC 5253 and ESO 269-G058: Dwarf Galaxies with a Past}",
      journal = {\aj},
         year = 2007,
        month = nov,
       volume = {134},
       number = {5},
        pages = {1799-1812},
          doi = {10.1086/521984},
archivePrefix = {arXiv},
       eprint = {0709.3761},
 primaryClass = {astro-ph},
       adsurl = {https://ui.adsabs.harvard.edu/abs/2007AJ....134.1799D}
}

@ARTICLE{Davidson20,
       author = {{Davidson}, Kris},
        title = "{Radiation-Driven Stellar Eruptions}",
      journal = {Galaxies},
         year = 2020,
        month = feb,
       volume = {8},
       number = {1},
          eid = {10},
        pages = {10},
          doi = {10.3390/galaxies8010010},
archivePrefix = {arXiv},
       eprint = {2009.02340},
 primaryClass = {astro-ph.SR},
       adsurl = {https://ui.adsabs.harvard.edu/abs/2020Galax...8...10D}
}

@ARTICLE{Davies15,
       author = {{Davies}, Ben and {Kudritzki}, Rolf-Peter and {Gazak}, Zach and
         {Plez}, Bertrand and {Bergemann}, Maria and {Evans}, Chris and
         {Patrick}, Lee},
        title = "{Red Supergiants as Cosmic Abundance Probes: The Magellanic Clouds}",
      journal = {\apj},
         year = 2015,
        month = jun,
       volume = {806},
       number = {1},
          eid = {21},
        pages = {21},
          doi = {10.1088/0004-637X/806/1/21},
archivePrefix = {arXiv},
       eprint = {1504.03694},
 primaryClass = {astro-ph.GA},
       adsurl = {https://ui.adsabs.harvard.edu/abs/2015ApJ...806...21D}
}

@ARTICLE{Davies18,
   author = {{Davies}, B. and {Crowther}, P.~A. and {Beasor}, E.~R.},
    title = "{The luminosities of cool supergiants in the Magellanic Clouds, and the Humphreys-Davidson limit revisited}",
  journal = {\mnras},
archivePrefix = "arXiv",
   eprint = {1804.06417},
 primaryClass = "astro-ph.SR",
     year = 2018,
    month = aug,
   volume = 478,
    pages = {3138-3148},
      doi = {10.1093/mnras/sty1302},
   adsurl = {http://adsabs.harvard.edu/abs/2018MNRAS.478.3138D}
}

@ARTICLE{deMello98,
       author = {{de Mello}, Du{\'\i}lia F. and {Schaerer}, Daniel and {Heldmann}, Jennifer and {Leitherer}, Claus},
        title = "{Searching for Wolf-Rayet Stars in I ZW 18: the Origin of He II Emission}",
      journal = {\apj},
         year = 1998,
        month = nov,
       volume = {507},
       number = {1},
        pages = {199-209},
          doi = {10.1086/306317},
archivePrefix = {arXiv},
       eprint = {astro-ph/9805342},
 primaryClass = {astro-ph},
       adsurl = {https://ui.adsabs.harvard.edu/abs/1998ApJ...507..199D}
}

@ARTICLE{Deshmukh24,
       author = {{Deshmukh}, K. and {Sana}, H. and {M{\'e}rand}, A. and {Bordier}, E. and {Langer}, N. and {Bodensteiner}, J. and {Dsilva}, K. and {Frost}, A.~J. and {Gosset}, E. and {Le Bouquin}, J. -B. and {Lefever}, R.~R. and {Mahy}, L. and {Patrick}, L.~R. and {Reggiani}, M. and {Sander}, A.~A.~C. and {Shenar}, T. and {Tramper}, F. and {Villase{\~n}or}, J.~I. and {Waisberg}, I.},
        title = "{Investigating 39 Galactic Wolf-Rayet stars with VLTI/GRAVITY: Uncovering a long-period binary desert}",
      journal = {\aap},
         year = 2024,
        month = dec,
       volume = {692},
          eid = {A109},
        pages = {A109},
          doi = {10.1051/0004-6361/202452352},
archivePrefix = {arXiv},
       eprint = {2409.15212},
 primaryClass = {astro-ph.SR},
       adsurl = {https://ui.adsabs.harvard.edu/abs/2024A&A...692A.109D}
}

@BOOK{deVaucouleurs91,
       author = {{de Vaucouleurs}, Gerard and {de Vaucouleurs}, Antoinette and {Corwin}, Jr., Herold G. and {Buta}, Ronald J. and {Paturel}, Georges and {Fouque}, Pascal},
        title = "{Third Reference Catalogue of Bright Galaxies}",
         year = 1991,
       adsurl = {https://ui.adsabs.harvard.edu/abs/1991rc3..book.....D}
}

@dataset{deVaucouleurs95,
       author = {{de Vaucouleurs}, G. and {de Vaucouleurs}, A. and {Corwin}, H.~G. and {Buta}, R.~J. and {Paturel}, G. and {Fouque}, P.},
        title = "{VizieR Online Data Catalog: Third Reference Cat. of Bright Galaxies (RC3) (de Vaucouleurs+ 1991)}",
 howpublished = {VizieR On-line Data Catalog: VII/155.  Originally published in: Springer-Verlag: New York, (1991)},
         year = 1995,
        month = feb,
          eid = {VII/155},
       adsurl = {https://ui.adsabs.harvard.edu/abs/1995yCat.7155....0D}
}

@ARTICLE{deWit24,
       author = {{de Wit}, S. and {Bonanos}, A.~Z. and {Antoniadis}, K. and {Zapartas}, E. and {Ruiz}, A. and {Britavskiy}, N. and {Christodoulou}, E. and {De}, K. and {Maravelias}, G. and {Munoz-Sanchez}, G. and {Tsopela}, A.},
        title = "{Investigating episodic mass loss in evolved massive stars: II. Physical properties of red supergiants at subsolar metallicity}",
      journal = {\aap},
         year = 2024,
        month = sep,
       volume = {689},
          eid = {A46},
        pages = {A46},
          doi = {10.1051/0004-6361/202449607},
archivePrefix = {arXiv},
       eprint = {2402.12442},
 primaryClass = {astro-ph.SR},
       adsurl = {https://ui.adsabs.harvard.edu/abs/2024A&A...689A..46D}
}

@ARTICLE{Dolphin00,
       author = {{Dolphin}, Andrew E.},
        title = "{WFPC2 Stellar Photometry with HSTPHOT}",
      journal = {\pasp},
         year = 2000,
        month = oct,
       volume = {112},
       number = {776},
        pages = {1383-1396},
          doi = {10.1086/316630},
archivePrefix = {arXiv},
       eprint = {astro-ph/0006217},
 primaryClass = {astro-ph},
       adsurl = {https://ui.adsabs.harvard.edu/abs/2000PASP..112.1383D}
}

@ARTICLE{Dolphin02,
       author = {{Dolphin}, A.~E.},
        title = "{Numerical methods of star formation history measurement and applications to seven dwarf spheroidals}",
      journal = {\mnras},
         year = 2002,
        month = may,
       volume = {332},
       number = {1},
        pages = {91-108},
          doi = {10.1046/j.1365-8711.2002.05271.x},
archivePrefix = {arXiv},
       eprint = {astro-ph/0112331},
 primaryClass = {astro-ph},
       adsurl = {https://ui.adsabs.harvard.edu/abs/2002MNRAS.332...91D}
}

@ARTICLE{Dotter16,
       author = {{Dotter}, Aaron},
        title = "{MESA Isochrones and Stellar Tracks (MIST) 0: Methods for the Construction of Stellar Isochrones}",
      journal = {\apjs},
         year = "2016",
        month = "Jan",
       volume = {222},
       number = {1},
          eid = {8},
        pages = {8},
          doi = {10.3847/0067-0049/222/1/8},
archivePrefix = {arXiv},
       eprint = {1601.05144},
 primaryClass = {astro-ph.SR},
       adsurl = {https://ui.adsabs.harvard.edu/abs/2016ApJS..222....8D}
}

@ARTICLE{Drout09,
   author = {{Drout}, M.~R. and {Massey}, P. and {Meynet}, G. and {Tokarz}, S. and 
	{Caldwell}, N.},
    title = "{Yellow Supergiants in the Andromeda Galaxy (M31)}",
  journal = {\apj},
archivePrefix = "arXiv",
   eprint = {0907.5471},
 primaryClass = "astro-ph.SR",
     year = 2009,
    month = sep,
   volume = 703,
    pages = {441-460},
      doi = {10.1088/0004-637X/703/1/441},
   adsurl = {http://adsabs.harvard.edu/abs/2009ApJ...703..441D}
}

@ARTICLE{Eldridge11,
       author = {{Eldridge}, John J. and {Langer}, Norbert and {Tout}, Christopher A.},
        title = "{Runaway stars as progenitors of supernovae and gamma-ray bursts}",
      journal = {\mnras},
         year = 2011,
        month = jul,
       volume = {414},
       number = {4},
        pages = {3501-3520},
          doi = {10.1111/j.1365-2966.2011.18650.x},
archivePrefix = {arXiv},
       eprint = {1103.1877},
 primaryClass = {astro-ph.SR},
       adsurl = {https://ui.adsabs.harvard.edu/abs/2011MNRAS.414.3501E}
}

@ARTICLE{Eldridge17,
       author = {{Eldridge}, J.~J. and {Stanway}, E.~R. and {Xiao}, L. and {McClelland}, L.~A.~S. and {Taylor}, G. and {Ng}, M. and {Greis}, S.~M.~L. and {Bray}, J.~C.},
        title = "{Binary Population and Spectral Synthesis Version 2.1: Construction, Observational Verification, and New Results}",
      journal = {\pasa},
         year = 2017,
        month = nov,
       volume = {34},
          eid = {e058},
        pages = {e058},
          doi = {10.1017/pasa.2017.51},
archivePrefix = {arXiv},
       eprint = {1710.02154},
 primaryClass = {astro-ph.SR},
       adsurl = {https://ui.adsabs.harvard.edu/abs/2017PASA...34...58E}
}

@ARTICLE{Fagotto94,
       author = {{Fagotto}, F. and {Bressan}, A. and {Bertelli}, G. and {Chiosi}, C.},
        title = "{Evolutionary sequences of stellar models with new radiative opacities. III. Z=0.0004 and Z=0.05}",
      journal = {\aaps},
         year = 1994,
        month = apr,
       volume = {104},
        pages = {365-376},
       adsurl = {https://ui.adsabs.harvard.edu/abs/1994A&AS..104..365F}
}

@ARTICLE{Farrell22,
       author = {{Farrell}, Eoin and {Groh}, Jose H. and {Meynet}, Georges and {Eldridge}, J.~J.},
        title = "{Numerical experiments to help understand cause and effect in massive star evolution}",
      journal = {\mnras},
         year = 2022,
        month = may,
       volume = {512},
       number = {3},
        pages = {4116-4135},
          doi = {10.1093/mnras/stac538},
archivePrefix = {arXiv},
       eprint = {2109.02488},
 primaryClass = {astro-ph.SR},
       adsurl = {https://ui.adsabs.harvard.edu/abs/2022MNRAS.512.4116F}
}

@ARTICLE{Foellmi03,
   author = {{Foellmi}, C. and {Moffat}, A.~F.~J. and {Guerrero}, M.~A.},
    title = "{Wolf-Rayet binaries in the Magellanic Clouds and implications for massive-star evolution - I. Small Magellanic Cloud}",
  journal = {\mnras},
     year = 2003,
    month = jan,
   volume = 338,
    pages = {360-388},
      doi = {10.1046/j.1365-8711.2003.06052.x},
   adsurl = {http://adsabs.harvard.edu/abs/2003MNRAS.338..360F}
}

@ARTICLE{Fryer12,
       author = {{Fryer}, Chris L. and {Belczynski}, Krzysztof and {Wiktorowicz}, Grzegorz and {Dominik}, Michal and {Kalogera}, Vicky and {Holz}, Daniel E.},
        title = "{Compact Remnant Mass Function: Dependence on the Explosion Mechanism and Metallicity}",
      journal = {\apj},
         year = 2012,
        month = apr,
       volume = {749},
       number = {1},
          eid = {91},
        pages = {91},
          doi = {10.1088/0004-637X/749/1/91},
archivePrefix = {arXiv},
       eprint = {1110.1726},
 primaryClass = {astro-ph.SR},
       adsurl = {https://ui.adsabs.harvard.edu/abs/2012ApJ...749...91F}
}

@ARTICLE{GAIA23,
       author = {{Gaia Collaboration} and {Vallenari}, A. and {Brown}, A.~G.~A. and {Prusti}, T. and {de Bruijne}, J.~H.~J. and {Arenou}, F. and {Babusiaux}, C. and {Biermann}, M. and {Creevey}, O.~L. and {Ducourant}, C. and {Evans}, D.~W. and {Eyer}, L. and {Guerra}, R. and {Hutton}, A. and {Jordi}, C. and {Klioner}, S.~A. and {Lammers}, U.~L. and {Lindegren}, L. and {Luri}, X. and {Mignard}, F. and {Panem}, C. and {Pourbaix}, D. and {Randich}, S. and {Sartoretti}, P. and {Soubiran}, C. and {Tanga}, P. and {Walton}, N.~A. and {Bailer-Jones}, C.~A.~L. and {Bastian}, U. and {Drimmel}, R. and {Jansen}, F. and {Katz}, D. and {Lattanzi}, M.~G. and {van Leeuwen}, F. and {Bakker}, J. and {Cacciari}, C. and {Casta{\~n}eda}, J. and {De Angeli}, F. and {Fabricius}, C. and {Fouesneau}, M. and {Fr{\'e}mat}, Y. and {Galluccio}, L. and {Guerrier}, A. and {Heiter}, U. and {Masana}, E. and {Messineo}, R. and {Mowlavi}, N. and {Nicolas}, C. and {Nienartowicz}, K. and {Pailler}, F. and {Panuzzo}, P. and {Riclet}, F. and {Roux}, W. and {Seabroke}, G.~M. and {Sordo}, R. and {Th{\'e}venin}, F. and {Gracia-Abril}, G. and {Portell}, J. and {Teyssier}, D. and {Altmann}, M. and {Andrae}, R. and {Audard}, M. and {Bellas-Velidis}, I. and {Benson}, K. and {Berthier}, J. and {Blomme}, R. and {Burgess}, P.~W. and {Busonero}, D. and {Busso}, G. and {C{\'a}novas}, H. and {Carry}, B. and {Cellino}, A. and {Cheek}, N. and {Clementini}, G. and {Damerdji}, Y. and {Davidson}, M. and {de Teodoro}, P. and {Nu{\~n}ez Campos}, M. and {Delchambre}, L. and {Dell'Oro}, A. and {Esquej}, P. and {Fern{\'a}ndez-Hern{\'a}ndez}, J. and {Fraile}, E. and {Garabato}, D. and {Garc{\'\i}a-Lario}, P. and {Gosset}, E. and {Haigron}, R. and {Halbwachs}, J. -L. and {Hambly}, N.~C. and {Harrison}, D.~L. and {Hern{\'a}ndez}, J. and {Hestroffer}, D. and {Hodgkin}, S.~T. and {Holl}, B. and {Jan{\ss}en}, K. and {Jevardat de Fombelle}, G. and {Jordan}, S. and {Krone-Martins}, A. and {Lanzafame}, A.~C. and {L{\"o}ffler}, W. and {Marchal}, O. and {Marrese}, P.~M. and {Moitinho}, A. and {Muinonen}, K. and {Osborne}, P. and {Pancino}, E. and {Pauwels}, T. and {Recio-Blanco}, A. and {Reyl{\'e}}, C. and {Riello}, M. and {Rimoldini}, L. and {Roegiers}, T. and {Rybizki}, J. and {Sarro}, L.~M. and {Siopis}, C. and {Smith}, M. and {Sozzetti}, A. and {Utrilla}, E. and {van Leeuwen}, M. and {Abbas}, U. and {{\'A}brah{\'a}m}, P. and {Abreu Aramburu}, A. and {Aerts}, C. and {Aguado}, J.~J. and {Ajaj}, M. and {Aldea-Montero}, F. and {Altavilla}, G. and {{\'A}lvarez}, M.~A. and {Alves}, J. and {Anders}, F. and {Anderson}, R.~I. and {Anglada Varela}, E. and {Antoja}, T. and {Baines}, D. and {Baker}, S.~G. and {Balaguer-N{\'u}{\~n}ez}, L. and {Balbinot}, E. and {Balog}, Z. and {Barache}, C. and {Barbato}, D. and {Barros}, M. and {Barstow}, M.~A. and {Bartolom{\'e}}, S. and {Bassilana}, J. -L. and {Bauchet}, N. and {Becciani}, U. and {Bellazzini}, M. and {Berihuete}, A. and {Bernet}, M. and {Bertone}, S. and {Bianchi}, L. and {Binnenfeld}, A. and {Blanco-Cuaresma}, S. and {Blazere}, A. and {Boch}, T. and {Bombrun}, A. and {Bossini}, D. and {Bouquillon}, S. and {Bragaglia}, A. and {Bramante}, L. and {Breedt}, E. and {Bressan}, A. and {Brouillet}, N. and {Brugaletta}, E. and {Bucciarelli}, B. and {Burlacu}, A. and {Butkevich}, A.~G. and {Buzzi}, R. and {Caffau}, E. and {Cancelliere}, R. and {Cantat-Gaudin}, T. and {Carballo}, R. and {Carlucci}, T. and {Carnerero}, M.~I. and {Carrasco}, J.~M. and {Casamiquela}, L. and {Castellani}, M. and {Castro-Ginard}, A. and {Chaoul}, L. and {Charlot}, P. and {Chemin}, L. and {Chiaramida}, V. and {Chiavassa}, A. and {Chornay}, N. and {Comoretto}, G. and {Contursi}, G. and {Cooper}, W.~J. and {Cornez}, T. and {Cowell}, S. and {Crifo}, F. and {Cropper}, M. and {Crosta}, M. and {Crowley}, C. and {Dafonte}, C. and {Dapergolas}, A. and {David}, M. and {David}, P. and {de Laverny}, P. and {De Luise}, F. and {De March}, R. and {De Ridder}, J. and {de Souza}, R. and {de Torres}, A. and {del Peloso}, E.~F. and {del Pozo}, E. and {Delbo}, M. and {Delgado}, A. and {Delisle}, J. -B. and {Demouchy}, C. and {Dharmawardena}, T.~E. and {Di Matteo}, P. and {Diakite}, S. and {Diener}, C. and {Distefano}, E. and {Dolding}, C. and {Edvardsson}, B. and {Enke}, H. and {Fabre}, C. and {Fabrizio}, M. and {Faigler}, S. and {Fedorets}, G. and {Fernique}, P. and {Fienga}, A. and {Figueras}, F. and {Fournier}, Y. and {Fouron}, C. and {Fragkoudi}, F. and {Gai}, M. and {Garcia-Gutierrez}, A. and {Garcia-Reinaldos}, M. and {Garc{\'\i}a-Torres}, M. and {Garofalo}, A. and {Gavel}, A. and {Gavras}, P. and {Gerlach}, E. and {Geyer}, R. and {Giacobbe}, P. and {Gilmore}, G. and {Girona}, S. and {Giuffrida}, G. and {Gomel}, R. and {Gomez}, A. and {Gonz{\'a}lez-N{\'u}{\~n}ez}, J. and {Gonz{\'a}lez-Santamar{\'\i}a}, I. and {Gonz{\'a}lez-Vidal}, J.~J. and {Granvik}, M. and {Guillout}, P. and {Guiraud}, J. and {Guti{\'e}rrez-S{\'a}nchez}, R. and {Guy}, L.~P. and {Hatzidimitriou}, D. and {Hauser}, M. and {Haywood}, M. and {Helmer}, A. and {Helmi}, A. and {Sarmiento}, M.~H. and {Hidalgo}, S.~L. and {Hilger}, T. and {H{\l}adczuk}, N. and {Hobbs}, D. and {Holland}, G. and {Huckle}, H.~E. and {Jardine}, K. and {Jasniewicz}, G. and {Jean-Antoine Piccolo}, A. and {Jim{\'e}nez-Arranz}, {\'O}. and {Jorissen}, A. and {Juaristi Campillo}, J. and {Julbe}, F. and {Karbevska}, L. and {Kervella}, P. and {Khanna}, S. and {Kontizas}, M. and {Kordopatis}, G. and {Korn}, A.~J. and {K{\'o}sp{\'a}l}, {\'A}. and {Kostrzewa-Rutkowska}, Z. and {Kruszy{\'n}ska}, K. and {Kun}, M. and {Laizeau}, P. and {Lambert}, S. and {Lanza}, A.~F. and {Lasne}, Y. and {Le Campion}, J. -F. and {Lebreton}, Y. and {Lebzelter}, T. and {Leccia}, S. and {Leclerc}, N. and {Lecoeur-Taibi}, I. and {Liao}, S. and {Licata}, E.~L. and {Lindstr{\o}m}, H.~E.~P. and {Lister}, T.~A. and {Livanou}, E. and {Lobel}, A. and {Lorca}, A. and {Loup}, C. and {Madrero Pardo}, P. and {Magdaleno Romeo}, A. and {Managau}, S. and {Mann}, R.~G. and {Manteiga}, M. and {Marchant}, J.~M. and {Marconi}, M. and {Marcos}, J. and {Marcos Santos}, M.~M.~S. and {Mar{\'\i}n Pina}, D. and {Marinoni}, S. and {Marocco}, F. and {Marshall}, D.~J. and {Martin Polo}, L. and {Mart{\'\i}n-Fleitas}, J.~M. and {Marton}, G. and {Mary}, N. and {Masip}, A. and {Massari}, D. and {Mastrobuono-Battisti}, A. and {Mazeh}, T. and {McMillan}, P.~J. and {Messina}, S. and {Michalik}, D. and {Millar}, N.~R. and {Mints}, A. and {Molina}, D. and {Molinaro}, R. and {Moln{\'a}r}, L. and {Monari}, G. and {Mongui{\'o}}, M. and {Montegriffo}, P. and {Montero}, A. and {Mor}, R. and {Mora}, A. and {Morbidelli}, R. and {Morel}, T. and {Morris}, D. and {Muraveva}, T. and {Murphy}, C.~P. and {Musella}, I. and {Nagy}, Z. and {Noval}, L. and {Oca{\~n}a}, F. and {Ogden}, A. and {Ordenovic}, C. and {Osinde}, J.~O. and {Pagani}, C. and {Pagano}, I. and {Palaversa}, L. and {Palicio}, P.~A. and {Pallas-Quintela}, L. and {Panahi}, A. and {Payne-Wardenaar}, S. and {Pe{\~n}alosa Esteller}, X. and {Penttil{\"a}}, A. and {Pichon}, B. and {Piersimoni}, A.~M. and {Pineau}, F. -X. and {Plachy}, E. and {Plum}, G. and {Poggio}, E. and {Pr{\v{s}}a}, A. and {Pulone}, L. and {Racero}, E. and {Ragaini}, S. and {Rainer}, M. and {Raiteri}, C.~M. and {Rambaux}, N. and {Ramos}, P. and {Ramos-Lerate}, M. and {Re Fiorentin}, P. and {Regibo}, S. and {Richards}, P.~J. and {Rios Diaz}, C. and {Ripepi}, V. and {Riva}, A. and {Rix}, H. -W. and {Rixon}, G. and {Robichon}, N. and {Robin}, A.~C. and {Robin}, C. and {Roelens}, M. and {Rogues}, H.~R.~O. and {Rohrbasser}, L. and {Romero-G{\'o}mez}, M. and {Rowell}, N. and {Royer}, F. and {Ruz Mieres}, D. and {Rybicki}, K.~A. and {Sadowski}, G. and {S{\'a}ez N{\'u}{\~n}ez}, A. and {Sagrist{\`a} Sell{\'e}s}, A. and {Sahlmann}, J. and {Salguero}, E. and {Samaras}, N. and {Sanchez Gimenez}, V. and {Sanna}, N. and {Santove{\~n}a}, R. and {Sarasso}, M. and {Schultheis}, M. and {Sciacca}, E. and {Segol}, M. and {Segovia}, J.~C. and {S{\'e}gransan}, D. and {Semeux}, D. and {Shahaf}, S. and {Siddiqui}, H.~I. and {Siebert}, A. and {Siltala}, L. and {Silvelo}, A. and {Slezak}, E. and {Slezak}, I. and {Smart}, R.~L. and {Snaith}, O.~N. and {Solano}, E. and {Solitro}, F. and {Souami}, D. and {Souchay}, J. and {Spagna}, A. and {Spina}, L. and {Spoto}, F. and {Steele}, I.~A. and {Steidelm{\"u}ller}, H. and {Stephenson}, C.~A. and {S{\"u}veges}, M. and {Surdej}, J. and {Szabados}, L. and {Szegedi-Elek}, E. and {Taris}, F. and {Taylor}, M.~B. and {Teixeira}, R. and {Tolomei}, L. and {Tonello}, N. and {Torra}, F. and {Torra}, J. and {Torralba Elipe}, G. and {Trabucchi}, M. and {Tsounis}, A.~T. and {Turon}, C. and {Ulla}, A. and {Unger}, N. and {Vaillant}, M.~V. and {van Dillen}, E. and {van Reeven}, W. and {Vanel}, O. and {Vecchiato}, A. and {Viala}, Y. and {Vicente}, D. and {Voutsinas}, S. and {Weiler}, M. and {Wevers}, T. and {Wyrzykowski}, {\L}. and {Yoldas}, A. and {Yvard}, P. and {Zhao}, H. and {Zorec}, J. and {Zucker}, S. and {Zwitter}, T.},
        title = "{Gaia Data Release 3. Summary of the content and survey properties}",
      journal = {\aap},
         year = 2023,
        month = jun,
       volume = {674},
          eid = {A1},
        pages = {A1},
          doi = {10.1051/0004-6361/202243940},
archivePrefix = {arXiv},
       eprint = {2208.00211},
 primaryClass = {astro-ph.GA},
       adsurl = {https://ui.adsabs.harvard.edu/abs/2023A&A...674A...1G}
}

@ARTICLE{Gazak15,
       author = {{Gazak}, J. Zachary and {Kudritzki}, Rolf and {Evans}, Chris and {Patrick}, Lee and {Davies}, Ben and {Bergemann}, Maria and {Plez}, Bertrand and {Bresolin}, Fabio and {Bender}, Ralf and {Wegner}, Michael and {Bonanos}, Alceste Z. and {Williams}, Stephen J.},
        title = "{Red Supergiants as Cosmic Abundance Probes: The Sculptor Galaxy NGC 300}",
      journal = {\apj},
         year = 2015,
        month = jun,
       volume = {805},
       number = {2},
          eid = {182},
        pages = {182},
          doi = {10.1088/0004-637X/805/2/182},
archivePrefix = {arXiv},
       eprint = {1505.00871},
 primaryClass = {astro-ph.GA},
       adsurl = {https://ui.adsabs.harvard.edu/abs/2015ApJ...805..182G}
}

@ARTICLE{Georgy13,
   author = {{Georgy}, C. and {Ekstr{\"o}m}, S. and {Eggenberger}, P. and 
	{Meynet}, G. and {Haemmerl{\'e}}, L. and {Maeder}, A. and {Granada}, A. and 
	{Groh}, J.H. and {Hirschi}, R. and {Mowlavi}, N. and {Yusof}, N. and 
	{Charbonnel}, C. and {Decressin}, T. and {Barblan}, F.},
    title = "{Grids of stellar models with rotation. III. Models from 0.8 to 120 M$_{⊙}$ at a metallicity Z = 0.002}",
  journal = {\aap},
archivePrefix = "arXiv",
   eprint = {1308.2914},
 primaryClass = "astro-ph.SR",
     year = 2013,
    month = oct,
   volume = 558,
      eid = {A103},
    pages = {A103},
      doi = {10.1051/0004-6361/201322178},
   adsurl = {http://adsabs.harvard.edu/abs/2013A%26A...558A.103G}
}

@ARTICLE{Gilkis21,
       author = {{Gilkis}, Avishai and {Shenar}, Tomer and {Ramachandran}, Varsha and {Jermyn}, Adam S. and {Mahy}, Laurent and {Oskinova}, Lidia M. and {Arcavi}, Iair and {Sana}, Hugues},
        title = "{The excess of cool supergiants from contemporary stellar evolution models defies the metallicity-independent Humphreys-Davidson limit}",
      journal = {\mnras},
         year = 2021,
        month = may,
       volume = {503},
       number = {2},
        pages = {1884-1896},
          doi = {10.1093/mnras/stab383},
archivePrefix = {arXiv},
       eprint = {2102.03102},
 primaryClass = {astro-ph.SR},
       adsurl = {https://ui.adsabs.harvard.edu/abs/2021MNRAS.503.1884G}
}

@ARTICLE{Goldman17,
       author = {{Goldman}, Steven R. and {van Loon}, Jacco Th. and {Zijlstra}, Albert A. and {Green}, James A. and {Wood}, Peter R. and {Nanni}, Ambra and {Imai}, Hiroshi and {Whitelock}, Patricia A. and {Matsuura}, Mikako and {Groenewegen}, Martin A.~T. and {G{\'o}mez}, Jos{\'e} F.},
        title = "{The wind speeds, dust content, and mass-loss rates of evolved AGB and RSG stars at varying metallicity}",
      journal = {\mnras},
         year = 2017,
        month = feb,
       volume = {465},
       number = {1},
        pages = {403-433},
          doi = {10.1093/mnras/stw2708},
archivePrefix = {arXiv},
       eprint = {1610.05761},
 primaryClass = {astro-ph.SR},
       adsurl = {https://ui.adsabs.harvard.edu/abs/2017MNRAS.465..403G}
}

@ARTICLE{GonzalezTora25,
       author = {{Gonz{\'a}lez-Tor{\`a}}, G. and {Sander}, A.~A.~C. and {Egorova}, E. and {Lefever}, R.~R. and {Ramachandran}, V. and {Egorov}, O.~V. and {Josiek}, J. and {Sch{\"o}sser}, E.~C. and {Bernini-Peron}, M. and {Kreckel}, K. and {Wofford}, A. and {Telford}, O.~G. and {Senchyna}, P. and {Leitherer}, C. and {Liang}, F.-H. and {Blanc}, G.~A. and {Drory}, N. and {Fern{\'a}ndez-Trincado}, J.~G. and {Johnston}, E.~J. and {Mej{\'\i}a-Narv{\'a}ez}, A.~J. and {Sanchez}, S.~F.},
        title = "{SDSS-V LVM: Detectability of Wolf-Rayet stars and their He II ionizing flux in low-metallicity environments: I. The weak-lined, early-type WN3 stars in the SMC}",
      journal = {\aap},
         year = 2025,
        month = nov,
       volume = {703},
          eid = {L11},
        pages = {L11},
          doi = {10.1051/0004-6361/202557041},
archivePrefix = {arXiv},
       eprint = {2509.04569},
 primaryClass = {astro-ph.GA},
       adsurl = {https://ui.adsabs.harvard.edu/abs/2025A&A...703L..11G}
}

@ARTICLE{Gordon03,
   author = {{Gordon}, K.~D. and {Clayton}, G.~C. and {Misselt}, K.~A. and 
	{Landolt}, A.~U. and {Wolff}, M.~J.},
    title = "{A Quantitative Comparison of the Small Magellanic Cloud, Large Magellanic Cloud, and Milky Way Ultraviolet to Near-Infrared Extinction Curves}",
  journal = {\apj},
   eprint = {astro-ph/0305257},
     year = 2003,
    month = sep,
   volume = 594,
    pages = {279-293},
      doi = {10.1086/376774},
   adsurl = {http://adsabs.harvard.edu/abs/2003ApJ...594..279G}
}

@ARTICLE{Gotberg18,
       author = {{G{\"o}tberg}, Y. and {de Mink}, S.~E. and {Groh}, J.~H. and
         {Kupfer}, T. and {Crowther}, P.~A. and {Zapartas}, E. and {Renzo}, M.},
        title = "{Spectral models for binary products: Unifying subdwarfs and Wolf-Rayet stars as a sequence of stripped-envelope stars}",
      journal = {\aap},
         year = 2018,
        month = jul,
       volume = {615},
          eid = {A78},
        pages = {A78},
          doi = {10.1051/0004-6361/201732274},
archivePrefix = {arXiv},
       eprint = {1802.03018},
 primaryClass = {astro-ph.SR},
       adsurl = {https://ui.adsabs.harvard.edu/abs/2018A&A...615A..78G}
}

@ARTICLE{Gotberg23,
       author = {{G{\"o}tberg}, Y. and {Drout}, M.~R. and {Ji}, A.~P. and {Groh}, J.~H. and {Ludwig}, B.~A. and {Crowther}, P.~A. and {Smith}, N. and {de Koter}, A. and {de Mink}, S.~E.},
        title = "{Stellar Properties of Observed Stars Stripped in Binaries in the Magellanic Clouds}",
      journal = {\apj},
         year = 2023,
        month = dec,
       volume = {959},
       number = {2},
          eid = {125},
        pages = {125},
          doi = {10.3847/1538-4357/ace5a3},
archivePrefix = {arXiv},
       eprint = {2307.00074},
 primaryClass = {astro-ph.SR},
       adsurl = {https://ui.adsabs.harvard.edu/abs/2023ApJ...959..125G}
}

@ARTICLE{Graczyk20,
       author = {{Graczyk}, Dariusz and {Pietrzy{\'n}ski}, Grzegorz and {Thompson}, Ian B. and {Gieren}, Wolfgang and {Zgirski}, Bart{\l}omiej and {Villanova}, Sandro and {G{\'o}rski}, Marek and {Wielg{\'o}rski}, Piotr and {Karczmarek}, Paulina and {Narloch}, Weronika and {Pilecki}, Bogumi{\l} and {Taormina}, Monica and {Smolec}, Rados{\l}aw and {Suchomska}, Ksenia and {Gallenne}, Alexandre and {Nardetto}, Nicolas and {Storm}, Jesper and {Kudritzki}, Rolf-Peter and {Ka{\l}uszy{\'n}ski}, Miko{\l}aj and {Pych}, Wojciech},
        title = "{A Distance Determination to the Small Magellanic Cloud with an Accuracy of Better than Two Percent Based on Late-type Eclipsing Binary Stars}",
      journal = {\apj},
         year = 2020,
        month = nov,
       volume = {904},
       number = {1},
          eid = {13},
        pages = {13},
          doi = {10.3847/1538-4357/abbb2b},
archivePrefix = {arXiv},
       eprint = {2010.08754},
 primaryClass = {astro-ph.GA},
       adsurl = {https://ui.adsabs.harvard.edu/abs/2020ApJ...904...13G}
}

@ARTICLE{Groh19,
       author = {{Groh}, J.~H. and {Ekstr{\"o}m}, S. and {Georgy}, C. and {Meynet}, G. and {Choplin}, A. and {Eggenberger}, P. and {Hirschi}, R. and {Maeder}, A. and {Murphy}, L.~J. and {Boian}, I. and {Farrell}, E.~J.},
        title = "{Grids of stellar models with rotation. IV. Models from 1.7 to 120 M$_{{\ensuremath{\odot}}}$ at a metallicity Z = 0.0004}",
      journal = {\aap},
         year = 2019,
        month = jul,
       volume = {627},
          eid = {A24},
        pages = {A24},
          doi = {10.1051/0004-6361/201833720},
archivePrefix = {arXiv},
       eprint = {1904.04009},
 primaryClass = {astro-ph.SR},
       adsurl = {https://ui.adsabs.harvard.edu/abs/2019A&A...627A..24G}
}

@ARTICLE{Hainich14,
       author = {{Hainich}, R. and {R{\"u}hling}, U. and {Todt}, H. and {Oskinova}, L.~M. and {Liermann}, A. and {Gr{\"a}fener}, G. and {Foellmi}, C. and {Schnurr}, O. and {Hamann}, W. -R.},
        title = "{The Wolf-Rayet stars in the Large Magellanic Cloud. A comprehensive analysis of the WN class}",
      journal = {\aap},
         year = 2014,
        month = may,
       volume = {565},
          eid = {A27},
        pages = {A27},
          doi = {10.1051/0004-6361/201322696},
archivePrefix = {arXiv},
       eprint = {1401.5474},
 primaryClass = {astro-ph.SR},
       adsurl = {https://ui.adsabs.harvard.edu/abs/2014A&A...565A..27H}
}

@ARTICLE{Hainich15,
   author = {{Hainich}, R. and {Pasemann}, D. and {Todt}, H. and {Shenar}, T. and 
	{Sander}, A. and {Hamann}, W.-R.},
    title = "{Wolf-Rayet stars in the Small Magellanic Cloud. I. Analysis of the single WN stars}",
  journal = {\aap},
archivePrefix = "arXiv",
   eprint = {1507.04000},
 primaryClass = "astro-ph.SR",
     year = 2015,
    month = sep,
   volume = 581,
      eid = {A21},
    pages = {A21},
      doi = {10.1051/0004-6361/201526241},
   adsurl = {http://adsabs.harvard.edu/abs/2015A%26A...581A..21H}
}

@ARTICLE{Hamann19,
       author = {{Hamann}, W. -R. and {Gr{\"a}fener}, G. and {Liermann}, A. and {Hainich}, R. and {Sander}, A.~A.~C. and {Shenar}, T. and {Ramachandran}, V. and {Todt}, H. and {Oskinova}, L.~M.},
        title = "{The Galactic WN stars revisited. Impact of Gaia distances on fundamental stellar parameters}",
      journal = {\aap},
         year = 2019,
        month = may,
       volume = {625},
          eid = {A57},
        pages = {A57},
          doi = {10.1051/0004-6361/201834850},
archivePrefix = {arXiv},
       eprint = {1904.04687},
 primaryClass = {astro-ph.SR},
       adsurl = {https://ui.adsabs.harvard.edu/abs/2019A&A...625A..57H}
}

@ARTICLE{Harris04,
       author = {{Harris}, Jason and {Zaritsky}, Dennis},
        title = "{The Star Formation History of the Small Magellanic Cloud}",
      journal = {\aj},
         year = "2004",
        month = "Mar",
       volume = {127},
       number = {3},
        pages = {1531-1544},
          doi = {10.1086/381953},
archivePrefix = {arXiv},
       eprint = {astro-ph/0312100},
 primaryClass = {astro-ph},
       adsurl = {https://ui.adsabs.harvard.edu/abs/2004AJ....127.1531H}
}

@ARTICLE{Harris09,
       author = {{Harris}, Jason and {Zaritsky}, Dennis},
        title = "{The Star Formation History of the Large Magellanic Cloud}",
      journal = {\aj},
         year = 2009,
        month = nov,
       volume = {138},
       number = {5},
        pages = {1243-1260},
          doi = {10.1088/0004-6256/138/5/1243},
archivePrefix = {arXiv},
       eprint = {0908.1422},
 primaryClass = {astro-ph.CO},
       adsurl = {https://ui.adsabs.harvard.edu/abs/2009AJ....138.1243H}
}

@ARTICLE{Hayes25,
       author = {{Hayes}, Matthew J. and {Saldana-Lopez}, Alberto and {Citro}, Annalisa and {James}, Bethan L. and {Mingozzi}, Matilde and {Scarlata}, Claudia and {Martinez}, Zorayda and {Berg}, Danielle A.},
        title = "{On the Average Ultraviolet Emission-line Spectra of High-redshift Galaxies: Hot and Cold, Carbon-poor, Nitrogen Modest, and Oozing Ionizing Photons}",
      journal = {\apj},
         year = 2025,
        month = mar,
       volume = {982},
       number = {1},
          eid = {14},
        pages = {14},
          doi = {10.3847/1538-4357/adaea1},
archivePrefix = {arXiv},
       eprint = {2411.09262},
 primaryClass = {astro-ph.GA},
       adsurl = {https://ui.adsabs.harvard.edu/abs/2025ApJ...982...14H}
}

@ARTICLE{Heger00,
   author = {{Heger}, A. and {Langer}, N. and {Woosley}, S.~E.},
    title = "{Presupernova Evolution of Rotating Massive Stars. I. Numerical Method and Evolution of the Internal Stellar Structure}",
  journal = {\apj},
   eprint = {astro-ph/9904132},
     year = 2000,
    month = jan,
   volume = 528,
    pages = {368-396},
      doi = {10.1086/308158},
   adsurl = {http://adsabs.harvard.edu/abs/2000ApJ...528..368H}
}

@ARTICLE{Helou04,
       author = {{Helou}, G. and {Roussel}, H. and {Appleton}, P. and {Frayer}, D. and {Stolovy}, S. and {Storrie-Lombardi}, L. and {Hurt}, R. and {Lowrance}, P. and {Makovoz}, D. and {Masci}, F. and {Surace}, J. and {Gordon}, K.~D. and {Alonso-Herrero}, A. and {Engelbracht}, C.~W. and {Misselt}, K. and {Rieke}, G. and {Rieke}, M. and {Willner}, S.~P. and {Pahre}, M. and {Ashby}, M.~L.~N. and {Fazio}, G.~G. and {Smith}, H.~A.},
        title = "{The Anatomy of Star Formation in NGC 300}",
      journal = {\apjs},
         year = 2004,
        month = sep,
       volume = {154},
       number = {1},
        pages = {253-258},
          doi = {10.1086/422640},
archivePrefix = {arXiv},
       eprint = {astro-ph/0408248},
 primaryClass = {astro-ph},
       adsurl = {https://ui.adsabs.harvard.edu/abs/2004ApJS..154..253H}
}

@ARTICLE{Higgins20,
       author = {{Higgins}, Erin R. and {Vink}, Jorick S.},
        title = "{Theoretical investigation of the Humphreys-Davidson limit at high and low metallicity}",
      journal = {\aap},
         year = 2020,
        month = mar,
       volume = {635},
          eid = {A175},
        pages = {A175},
          doi = {10.1051/0004-6361/201937374},
archivePrefix = {arXiv},
       eprint = {2002.07204},
 primaryClass = {astro-ph.SR},
       adsurl = {https://ui.adsabs.harvard.edu/abs/2020A&A...635A.175H}
}

@ARTICLE{Hirschauer24,
       author = {{Hirschauer}, Alec S. and {Crouzet}, Nicolas and {Habel}, Nolan and {Lenki{\'c}}, Laura and {Nally}, Conor and {Jones}, Olivia C. and {Bortolini}, Giacomo and {Boyer}, Martha L. and {Justtanont}, Kay and {Meixner}, Margaret and {{\"O}stlin}, G{\"o}ran and {Wright}, Gillian S. and {Azzollini}, Ruyman and {Blommaert}, Joris A.~D.~L. and {Brandl}, Bernhard and {Decin}, Leen and {Nayak}, Omnarayani and {Royer}, Pierre and {Sargent}, B.~A. and {van der Werf}, Paul},
        title = "{Imaging of I Zw 18 by JWST. I. Detecting Dusty Stellar Populations}",
      journal = {\aj},
         year = 2024,
        month = jul,
       volume = {168},
       number = {1},
          eid = {23},
        pages = {23},
          doi = {10.3847/1538-3881/ad4967},
archivePrefix = {arXiv},
       eprint = {2403.06980},
 primaryClass = {astro-ph.GA},
       adsurl = {https://ui.adsabs.harvard.edu/abs/2024AJ....168...23H}
}

@ARTICLE{HovisAfflerbach25,
       author = {{Hovis-Afflerbach}, B. and {G{\"o}tberg}, Y. and {Schootemeijer}, A. and {Klencki}, J. and {Strom}, A.~L. and {Ludwig}, B.~A. and {Drout}, M.~R.},
        title = "{The mass distribution of stars stripped in binaries: The effect of metallicity}",
      journal = {\aap},
         year = 2025,
        month = may,
       volume = {697},
          eid = {A239},
        pages = {A239},
          doi = {10.1051/0004-6361/202453185},
archivePrefix = {arXiv},
       eprint = {2412.05356},
 primaryClass = {astro-ph.SR},
       adsurl = {https://ui.adsabs.harvard.edu/abs/2025A&A...697A.239H}
}

@ARTICLE{Humphreys79,
       author = {{Humphreys}, R.~M. and {Davidson}, K.},
        title = "{Studies of luminous stars in nearby galaxies. III. Comments on the evolution of the most massive stars in the Milky Way and the Large Magellanic Cloud.}",
      journal = {\apj},
         year = 1979,
        month = sep,
       volume = {232},
        pages = {409-420},
          doi = {10.1086/157301},
       adsurl = {https://ui.adsabs.harvard.edu/abs/1979ApJ...232..409H}
}

@ARTICLE{Humphreys94,
       author = {{Humphreys}, Roberta M. and {Davidson}, Kris},
        title = "{The Luminous Blue Variables: Astrophysical Geysers}",
      journal = {\pasp},
         year = 1994,
        month = oct,
       volume = {106},
        pages = {1025},
          doi = {10.1086/133478},
       adsurl = {https://ui.adsabs.harvard.edu/abs/1994PASP..106.1025H}
}

@ARTICLE{Humphreys23,
       author = {{Humphreys}, Roberta M. and {Jones}, Terry J. and {Martin}, John C.},
        title = "{Yellow Supergiants and Post-red Supergiant Evolution in the Large Magellanic Cloud}",
      journal = {\aj},
         year = 2023,
        month = aug,
       volume = {166},
       number = {2},
          eid = {50},
        pages = {50},
          doi = {10.3847/1538-3881/acdd6c},
archivePrefix = {arXiv},
       eprint = {2306.07336},
 primaryClass = {astro-ph.SR},
       adsurl = {https://ui.adsabs.harvard.edu/abs/2023AJ....166...50H}
}

@ARTICLE{Humphreys25,
       author = {{Humphreys}, Roberta M. and {Davidson}, Kris},
        title = "{Reflections on the upper Hertzsprung-Russell Diagram}",
      journal = {\apss},
         year = 2025,
        month = sep,
       volume = {370},
       number = {9},
          eid = {100},
        pages = {100},
          doi = {10.1007/s10509-025-04492-x},
       adsurl = {https://ui.adsabs.harvard.edu/abs/2025Ap&SS.370..100H}
}

@ARTICLE{Izotov97,
       author = {{Izotov}, Yuri I. and {Foltz}, Craig B. and {Green}, Richard F. and {Guseva}, Natalia G. and {Thuan}, Trinh X.},
        title = "{I Zw 18: A New Wolf-Rayet Galaxy}",
      journal = {\apjl},
         year = 1997,
        month = sep,
       volume = {487},
       number = {1},
        pages = {L37-L40},
          doi = {10.1086/310872},
archivePrefix = {arXiv},
       eprint = {astro-ph/9707272},
 primaryClass = {astro-ph},
       adsurl = {https://ui.adsabs.harvard.edu/abs/1997ApJ...487L..37I}
}

@ARTICLE{Izotov09,
       author = {{Izotov}, Yuri I. and {Thuan}, Trinh X.},
        title = "{Luminous Blue Variable Stars in the two Extremely Metal-Deficient Blue Compact Dwarf Galaxies DDO 68 and PHL 293B}",
      journal = {\apj},
         year = 2009,
        month = jan,
       volume = {690},
       number = {2},
        pages = {1797-1806},
          doi = {10.1088/0004-637X/690/2/1797},
archivePrefix = {arXiv},
       eprint = {0809.3077},
 primaryClass = {astro-ph},
       adsurl = {https://ui.adsabs.harvard.edu/abs/2009ApJ...690.1797I}
}

@ARTICLE{Izotov16,
       author = {{Izotov}, Y.~I. and {Thuan}, T.~X.},
        title = "{Near-infrared spectroscopy of a large sample of low-metallicity blue compact dwarf galaxies}",
      journal = {\mnras},
         year = 2016,
        month = mar,
       volume = {457},
       number = {1},
        pages = {64-73},
          doi = {10.1093/mnras/stv2957},
archivePrefix = {arXiv},
       eprint = {1512.06191},
 primaryClass = {astro-ph.GA},
       adsurl = {https://ui.adsabs.harvard.edu/abs/2016MNRAS.457...64I}
}

@ARTICLE{Jadlovski24,
       author = {{Jadlovsk{\'y}}, Daniel and {Granzer}, Thomas and {Weber}, Michael and {Kravchenko}, Kateryna and {Krti{\v{c}}ka}, Ji{\v{r}}{\'\i} and {Dupree}, Andrea K. and {Chiavassa}, Andrea and {Strassmeier}, Klaus G. and {Poppenh{\"a}ger}, Katja},
        title = "{The Great Dimming of Betelgeuse: The photosphere as revealed by tomography over the past 15 yr}",
      journal = {\aap},
         year = 2024,
        month = may,
       volume = {685},
          eid = {A124},
        pages = {A124},
          doi = {10.1051/0004-6361/202348846},
archivePrefix = {arXiv},
       eprint = {2312.02816},
 primaryClass = {astro-ph.SR},
       adsurl = {https://ui.adsabs.harvard.edu/abs/2024A&A...685A.124J}
}

@ARTICLE{Jiang21,
       author = {{Jiang}, Yan-Fei and {Cantiello}, Matteo and {Bildsten}, Lars and {Quataert}, Eliot and {Blaes}, Omer and {Stone}, James},
        title = "{Outbursts of luminous blue variable stars from variations in the helium opacity}",
      journal = {\nat},
         year = 2018,
        month = sep,
       volume = {561},
       number = {7724},
        pages = {498-501},
          doi = {10.1038/s41586-018-0525-0},
archivePrefix = {arXiv},
       eprint = {1809.10187},
 primaryClass = {astro-ph.SR},
       adsurl = {https://ui.adsabs.harvard.edu/abs/2018Natur.561..498J}
}

@ARTICLE{Kang25,
       author = {{Kang}, Xiaoyu and {Kudritzki}, Rolf-Peter and {Gong}, Xiaobo and {Zhang}, Fenghui},
        title = "{Evolution and star formation history of NGC 300 from a chemical evolution model with radial gas inflows}",
      journal = {\aap},
         year = 2025,
        month = sep,
       volume = {701},
          eid = {A2},
        pages = {A2},
          doi = {10.1051/0004-6361/202554108},
archivePrefix = {arXiv},
       eprint = {2507.10245},
 primaryClass = {astro-ph.GA},
       adsurl = {https://ui.adsabs.harvard.edu/abs/2025A&A...701A...2K}
}

@ARTICLE{Kee21,
       author = {{Kee}, N.~D. and {Sundqvist}, J.~O. and {Decin}, L. and {de Koter}, A. and {Sana}, H.},
        title = "{Analytic, dust-independent mass-loss rates for red supergiant winds initiated by turbulent pressure}",
      journal = {\aap},
         year = 2021,
        month = feb,
       volume = {646},
          eid = {A180},
        pages = {A180},
          doi = {10.1051/0004-6361/202039224},
       adsurl = {https://ui.adsabs.harvard.edu/abs/2021A&A...646A.180K}
}

@ARTICLE{Kehrig15,
       author = {{Kehrig}, C. and {V{\'\i}lchez}, J.~M. and {P{\'e}rez-Montero}, E. and {Iglesias-P{\'a}ramo}, J. and {Brinchmann}, J. and {Kunth}, D. and {Durret}, F. and {Bayo}, F.~M.},
        title = "{The Extended He II {\ensuremath{\lambda}}4686-emitting Region in IZw 18 Unveiled: Clues for Peculiar Ionizing Sources}",
      journal = {\apjl},
         year = 2015,
        month = mar,
       volume = {801},
       number = {2},
          eid = {L28},
        pages = {L28},
          doi = {10.1088/2041-8205/801/2/L28},
archivePrefix = {arXiv},
       eprint = {1502.00522},
 primaryClass = {astro-ph.GA},
       adsurl = {https://ui.adsabs.harvard.edu/abs/2015ApJ...801L..28K}
}

@ARTICLE{Kim17,
       author = {{Kim}, Jinhyub and {Chung}, Aeree and {Wong}, O. Ivy and {Lee}, Bumhyun and {Sung}, Eon-Chang and {Staveley-Smith}, Lister},
        title = "{HI properties and star formation history of a fly-by pair of blue compact dwarf galaxies}",
      journal = {\aap},
         year = 2017,
        month = sep,
       volume = {605},
          eid = {A54},
        pages = {A54},
          doi = {10.1051/0004-6361/201730664},
archivePrefix = {arXiv},
       eprint = {1706.01902},
 primaryClass = {astro-ph.GA},
       adsurl = {https://ui.adsabs.harvard.edu/abs/2017A&A...605A..54K}
}

@ARTICLE{Kroupa01,
       author = {{Kroupa}, Pavel},
        title = "{On the variation of the initial mass function}",
      journal = {\mnras},
         year = "2001",
        month = "Apr",
       volume = {322},
       number = {2},
        pages = {231-246},
          doi = {10.1046/j.1365-8711.2001.04022.x},
archivePrefix = {arXiv},
       eprint = {astro-ph/0009005},
 primaryClass = {astro-ph},
       adsurl = {https://ui.adsabs.harvard.edu/abs/2001MNRAS.322..231K}
}

@ARTICLE{Kubatova19,
       author = {{Kub{\'a}tov{\'a}}, B. and {Sz{\'e}csi}, D. and {Sander}, A.~A.~C. and {Kub{\'a}t}, J. and {Tramper}, F. and {Krti{\v{c}}ka}, J. and {Kehrig}, C. and {Hamann}, W. -R. and {Hainich}, R. and {Shenar}, T.},
        title = "{Low-metallicity massive single stars with rotation. II. Predicting spectra and spectral classes of chemically homogeneously evolving stars}",
      journal = {\aap},
         year = 2019,
        month = mar,
       volume = {623},
          eid = {A8},
        pages = {A8},
          doi = {10.1051/0004-6361/201834360},
archivePrefix = {arXiv},
       eprint = {1810.01267},
 primaryClass = {astro-ph.SR},
       adsurl = {https://ui.adsabs.harvard.edu/abs/2019A&A...623A...8K}
}

@ARTICLE{Lalleman25,
       author = {{Lalleman}, M. and {Turbang}, K. and {Callister}, T. and {van Remortel}, N.},
        title = "{No evidence that the binary black hole mass distribution evolves with redshift}",
      journal = {\aap},
         year = 2025,
        month = jun,
       volume = {698},
          eid = {A85},
        pages = {A85},
          doi = {10.1051/0004-6361/202553941},
archivePrefix = {arXiv},
       eprint = {2501.10295},
 primaryClass = {astro-ph.HE},
       adsurl = {https://ui.adsabs.harvard.edu/abs/2025A&A...698A..85L}
}

@ARTICLE{Lamers88,
   author = {{Lamers}, H.~J.~G.~L.~M. and {Fitzpatrick}, E.~L.},
    title = "{The relationship between the Eddington limit, the observed upper luminosity limit for massive stars, and the luminous blue variables}",
  journal = {\apj},
     year = 1988,
    month = jan,
   volume = 324,
    pages = {279-287},
      doi = {10.1086/165894},
   adsurl = {http://adsabs.harvard.edu/abs/1988ApJ...324..279L}
}

@ARTICLE{Lebouteiller13,
       author = {{Lebouteiller}, V. and {Heap}, S. and {Hubeny}, I. and {Kunth}, D.},
        title = "{Chemical enrichment and physical conditions in I Zw 18}",
      journal = {\aap},
         year = 2013,
        month = may,
       volume = {553},
          eid = {A16},
        pages = {A16},
          doi = {10.1051/0004-6361/201220948},
archivePrefix = {arXiv},
       eprint = {1302.4746},
 primaryClass = {astro-ph.CO},
       adsurl = {https://ui.adsabs.harvard.edu/abs/2013A&A...553A..16L}
}

@ARTICLE{Legrand97,
       author = {{Legrand}, F. and {Kunth}, D. and {Roy}, J. -R. and {Mas-Hesse}, J.~M. and {Walsh}, J.~R.},
        title = "{Detection of WR stars in the metal-poor starburst galaxy IZw 18.}",
      journal = {\aap},
         year = 1997,
        month = oct,
       volume = {326},
        pages = {L17-L20},
          doi = {10.48550/arXiv.astro-ph/9707279},
archivePrefix = {arXiv},
       eprint = {astro-ph/9707279},
 primaryClass = {astro-ph},
       adsurl = {https://ui.adsabs.harvard.edu/abs/1997A&A...326L..17L}
}

@ARTICLE{Leitherer99,
       author = {{Leitherer}, Claus and {Schaerer}, Daniel and {Goldader}, Jeffrey D. and {Delgado}, Rosa M. Gonz{\'a}lez and {Robert}, Carmelle and {Kune}, Denis Foo and {de Mello}, Du{\'\i}lia F. and {Devost}, Daniel and {Heckman}, Timothy M.},
        title = "{Starburst99: Synthesis Models for Galaxies with Active Star Formation}",
      journal = {\apjs},
         year = 1999,
        month = jul,
       volume = {123},
       number = {1},
        pages = {3-40},
          doi = {10.1086/313233},
archivePrefix = {arXiv},
       eprint = {astro-ph/9902334},
 primaryClass = {astro-ph},
       adsurl = {https://ui.adsabs.harvard.edu/abs/1999ApJS..123....3L}
}

@ARTICLE{Leitherer14,
       author = {{Leitherer}, Claus and {Ekstr{\"o}m}, Sylvia and {Meynet}, Georges and {Schaerer}, Daniel and {Agienko}, Katerina B. and {Levesque}, Emily M.},
        title = "{The Effects of Stellar Rotation. II. A Comprehensive Set of Starburst99 Models}",
      journal = {\apjs},
         year = 2014,
        month = may,
       volume = {212},
       number = {1},
          eid = {14},
        pages = {14},
          doi = {10.1088/0067-0049/212/1/14},
archivePrefix = {arXiv},
       eprint = {1403.5444},
 primaryClass = {astro-ph.GA},
       adsurl = {https://ui.adsabs.harvard.edu/abs/2014ApJS..212...14L}
}

@ARTICLE{Levesque09,
       author = {{Levesque}, Emily M. and {Massey}, Philip and {Plez}, Bertrand and {Olsen}, Knut A.~G.},
        title = "{The Physical Properties of the Red Supergiant WOH G64: The Largest Star Known?}",
      journal = {\aj},
         year = 2009,
        month = jun,
       volume = {137},
       number = {6},
        pages = {4744-4752},
          doi = {10.1088/0004-6256/137/6/4744},
archivePrefix = {arXiv},
       eprint = {0903.2260},
 primaryClass = {astro-ph.SR},
       adsurl = {https://ui.adsabs.harvard.edu/abs/2009AJ....137.4744L}
}

@ARTICLE{Lewis15,
       author = {{Lewis}, Alexia R. and {Dolphin}, Andrew E. and {Dalcanton}, Julianne J. and {Weisz}, Daniel R. and {Williams}, Benjamin F. and {Bell}, Eric F. and {Seth}, Anil C. and {Simones}, Jacob E. and {Skillman}, Evan D. and {Choi}, Yumi and {Fouesneau}, Morgan and {Guhathakurta}, Puragra and {Johnson}, Lent C. and {Kalirai}, Jason S. and {Leroy}, Adam K. and {Monachesi}, Antonela and {Rix}, Hans-Walter and {Schruba}, Andreas},
        title = "{The Panchromatic Hubble Andromeda Treasury. XI. The Spatially Resolved Recent Star Formation History of M31}",
      journal = {\apj},
         year = 2015,
        month = jun,
       volume = {805},
       number = {2},
          eid = {183},
        pages = {183},
          doi = {10.1088/0004-637X/805/2/183},
archivePrefix = {arXiv},
       eprint = {1504.03338},
 primaryClass = {astro-ph.GA},
       adsurl = {https://ui.adsabs.harvard.edu/abs/2015ApJ...805..183L}
}

@ARTICLE{Li21,
       author = {{Li}, Siyang and {Riess}, Adam G. and {Busch}, Michael P. and {Casertano}, Stefano and {Macri}, Lucas M. and {Yuan}, Wenlong},
        title = "{A Sub-2\% Distance to M31 from Photometrically Homogeneous Near-infrared Cepheid Period-Luminosity Relations Measured with the Hubble Space Telescope}",
      journal = {\apj},
         year = 2021,
        month = oct,
       volume = {920},
       number = {2},
          eid = {84},
        pages = {84},
          doi = {10.3847/1538-4357/ac1597},
archivePrefix = {arXiv},
       eprint = {2107.08029},
 primaryClass = {astro-ph.CO},
       adsurl = {https://ui.adsabs.harvard.edu/abs/2021ApJ...920...84L}
}

@ARTICLE{Limongi18,
   author = {{Limongi}, M. and {Chieffi}, A.},
    title = "{Presupernova Evolution and Explosive Nucleosynthesis of Rotating Massive Stars in the Metallicity Range -3 {\le} [Fe/H] {\le} 0}",
  journal = {\apjs},
archivePrefix = "arXiv",
   eprint = {1805.09640},
 primaryClass = "astro-ph.SR",
     year = 2018,
    month = jul,
   volume = 237,
      eid = {13},
    pages = {13},
      doi = {10.3847/1538-4365/aacb24},
   adsurl = {http://adsabs.harvard.edu/abs/2018ApJS..237...13L}
}

@ARTICLE{LopezSanchez12,
       author = {{L{\'o}pez-S{\'a}nchez}, {\'A}. R. and {Koribalski}, B.~S. and {van Eymeren}, J. and {Esteban}, C. and {Kirby}, E. and {Jerjen}, H. and {Lonsdale}, N.},
        title = "{The intriguing H I gas in NGC 5253: an infall of a diffuse, low-metallicity H I cloud?}",
      journal = {\mnras},
         year = 2012,
        month = jan,
       volume = {419},
       number = {2},
        pages = {1051-1069},
          doi = {10.1111/j.1365-2966.2011.19762.x},
archivePrefix = {arXiv},
       eprint = {1109.0806},
 primaryClass = {astro-ph.CO},
       adsurl = {https://ui.adsabs.harvard.edu/abs/2012MNRAS.419.1051L}
}

@ARTICLE{Lovegrove13,
       author = {{Lovegrove}, Elizabeth and {Woosley}, S.~E.},
        title = "{Very Low Energy Supernovae from Neutrino Mass Loss}",
      journal = {\apj},
         year = 2013,
        month = jun,
       volume = {769},
       number = {2},
          eid = {109},
        pages = {109},
          doi = {10.1088/0004-637X/769/2/109},
archivePrefix = {arXiv},
       eprint = {1303.5055},
 primaryClass = {astro-ph.HE},
       adsurl = {https://ui.adsabs.harvard.edu/abs/2013ApJ...769..109L}
}

@ARTICLE{Maeder00,
   author = {{Maeder}, A. and {Meynet}, G.},
    title = "{Stellar evolution with rotation. VI. The Eddington and Omega -limits, the rotational mass loss for OB and LBV stars}",
  journal = {\aap},
   eprint = {astro-ph/0006405},
     year = 2000,
    month = sep,
   volume = 361,
    pages = {159-166},
   adsurl = {http://adsabs.harvard.edu/abs/2000A%26A...361..159M}
}

@ARTICLE{Martins09,
   author = {{Martins}, F. and {Hillier}, D.~J. and {Bouret}, J.~C. and {Depagne}, E. and 
	{Foellmi}, C. and {Marchenko}, S. and {Moffat}, A.~F.},
    title = "{Properties of WNh stars in the Small Magellanic Cloud: evidence for homogeneous evolution}",
  journal = {\aap},
archivePrefix = "arXiv",
   eprint = {0811.3564},
     year = 2009,
    month = feb,
   volume = 495,
    pages = {257-270},
      doi = {10.1051/0004-6361:200811014},
   adsurl = {http://adsabs.harvard.edu/abs/2009A%26A...495..257M}
}

@ARTICLE{Martins23,
       author = {{Martins}, Fabrice},
        title = "{Surface chemical composition of single WNh stars}",
      journal = {\aap},
         year = 2023,
        month = dec,
       volume = {680},
          eid = {A22},
        pages = {A22},
          doi = {10.1051/0004-6361/202347909},
archivePrefix = {arXiv},
       eprint = {2310.06539},
 primaryClass = {astro-ph.SR},
       adsurl = {https://ui.adsabs.harvard.edu/abs/2023A&A...680A..22M}
}

@ARTICLE{Martin25,
       author = {{Martin}, John C. and {Humphreys}, Roberta M. and {Davidson}, Kris},
        title = "{On the Spatial Distribution of Luminous Blue Variables, B[e] Supergiants, and Wolf─Rayet Stars in the Large Magellanic Cloud}",
      journal = {\apj},
         year = 2025,
        month = dec,
       volume = {994},
       number = {2},
          eid = {159},
        pages = {159},
          doi = {10.3847/1538-4357/ae1021},
archivePrefix = {arXiv},
       eprint = {2508.17114},
 primaryClass = {astro-ph.SR},
       adsurl = {https://ui.adsabs.harvard.edu/abs/2025ApJ...994..159M}
}

@ARTICLE{Massey95,
       author = {{Massey}, Philip and {Lang}, Cornelia C. and
         {Degioia-Eastwood}, Kathleen and {Garmany}, Catharine D.},
        title = "{Massive Stars in the Field and Associations of the Magellanic Clouds: The Upper Mass Limit, the Initial Mass Function, and a Critical Test of Main-Sequence Stellar Evolutionary Theory}",
      journal = {\apj},
         year = "1995",
        month = "Jan",
       volume = {438},
        pages = {188},
          doi = {10.1086/175064},
       adsurl = {https://ui.adsabs.harvard.edu/abs/1995ApJ...438..188M}
}

@ARTICLE{Maiz23,
       author = {{Ma{\'\i}z Apell{\'a}niz}, J. and {Holgado}, G. and {Pantaleoni Gonz{\'a}lez}, M. and {Caballero}, J.~A.},
        title = "{Stellar variability in Gaia DR3. I. Three-band photometric dispersions for 145 million sources}",
      journal = {\aap},
         year = 2023,
        month = sep,
       volume = {677},
          eid = {A137},
        pages = {A137},
          doi = {10.1051/0004-6361/202346759},
archivePrefix = {arXiv},
       eprint = {2304.14249},
 primaryClass = {astro-ph.SR},
       adsurl = {https://ui.adsabs.harvard.edu/abs/2023A&A...677A.137M}
}

@ARTICLE{McDonald22,
       author = {{McDonald}, Sarah L.~E. and {Davies}, Ben and {Beasor}, Emma R.},
        title = "{Red supergiants in M31: the Humphreys-Davidson limit at high metallicity}",
      journal = {\mnras},
         year = 2022,
        month = mar,
       volume = {510},
       number = {3},
        pages = {3132-3144},
          doi = {10.1093/mnras/stab3453},
archivePrefix = {arXiv},
       eprint = {2111.13716},
 primaryClass = {astro-ph.GA},
       adsurl = {https://ui.adsabs.harvard.edu/abs/2022MNRAS.510.3132M}
}

@ARTICLE{McQuinn20,
       author = {{McQuinn}, Kristen. B.~W. and {Berg}, Danielle A. and {Skillman}, Evan D. and {Adams}, Elizabeth A.~K. and {Cannon}, John M. and {Dolphin}, Andrew E. and {Salzer}, John J. and {Giovanelli}, Riccardo and {Haynes}, Martha P. and {Hirschauer}, Alec S. and {Janoweicki}, Steven and {Klapkowski}, Myles and {Rhode}, Katherine L.},
        title = "{The Leoncino Dwarf Galaxy: Exploring the Low-metallicity End of the Luminosity-Metallicity and Mass-Metallicity Relations}",
      journal = {\apj},
         year = 2020,
        month = mar,
       volume = {891},
       number = {2},
          eid = {181},
        pages = {181},
          doi = {10.3847/1538-4357/ab7447},
archivePrefix = {arXiv},
       eprint = {2002.11723},
 primaryClass = {astro-ph.GA},
       adsurl = {https://ui.adsabs.harvard.edu/abs/2020ApJ...891..181M}
}

@ARTICLE{Menon24,
       author = {{Menon}, Athira and {Ercolino}, Andrea and {Urbaneja}, Miguel A. and {Lennon}, Daniel J. and {Herrero}, Artemio and {Hirai}, Ryosuke and {Langer}, Norbert and {Schootemeijer}, Abel and {Chatzopoulos}, Emmanouil and {Frank}, Juhan and {Shiber}, Sagiv},
        title = "{Evidence for Evolved Stellar Binary Mergers in Observed B-type Blue Supergiants}",
      journal = {\apjl},
         year = 2024,
        month = mar,
       volume = {963},
       number = {2},
          eid = {L42},
        pages = {L42},
          doi = {10.3847/2041-8213/ad2074},
       adsurl = {https://ui.adsabs.harvard.edu/abs/2024ApJ...963L..42M}
}

@ARTICLE{Messineo19,
       author = {{Messineo}, M. and {Brown}, A.~G.~A.},
        title = "{A Catalog of Known Galactic K-M Stars of Class I Candidate Red Supergiants in Gaia DR2}",
      journal = {\aj},
         year = 2019,
        month = jul,
       volume = {158},
       number = {1},
          eid = {20},
        pages = {20},
          doi = {10.3847/1538-3881/ab1cbd},
archivePrefix = {arXiv},
       eprint = {1905.03744},
 primaryClass = {astro-ph.GA},
       adsurl = {https://ui.adsabs.harvard.edu/abs/2019AJ....158...20M}
}

@ARTICLE{Meynet94,
   author = {{Meynet}, G. and {Maeder}, A. and {Schaller}, G. and {Schaerer}, D. and 
	{Charbonnel}, C.},
    title = "{Grids of massive stars with high mass loss rates. V. From 12 to 120 M$_{sun}$\_ at Z=0.001, 0.004, 0.008, 0.020 and 0.040}",
  journal = {\aaps},
     year = 1994,
    month = jan,
   volume = 103,
    pages = {97-105},
   adsurl = {http://adsabs.harvard.edu/abs/1994A%26AS..103...97M}
}

@ARTICLE{Micheva22,
       author = {{Micheva}, Genoveva and {Roth}, Martin M. and {Weilbacher}, Peter M. and {Morisset}, Christophe and {Castro}, Norberto and {Monreal Ibero}, Ana and {Soemitro}, Azlizan A. and {Maseda}, Michael V. and {Steinmetz}, Matthias and {Brinchmann}, Jarle},
        title = "{MUSE crowded field 3D spectroscopy in NGC 300. III. Characterizing extremely faint HII regions and diffuse ionized gas}",
      journal = {\aap},
         year = 2022,
        month = dec,
       volume = {668},
          eid = {A74},
        pages = {A74},
          doi = {10.1051/0004-6361/202244017},
archivePrefix = {arXiv},
       eprint = {2210.04786},
 primaryClass = {astro-ph.GA},
       adsurl = {https://ui.adsabs.harvard.edu/abs/2022A&A...668A..74M}
}

@ARTICLE{Mingozzi22,
       author = {{Mingozzi}, Matilde and {James}, Bethan L. and {Arellano-C{\'o}rdova}, Karla Z. and {Berg}, Danielle A. and {Senchyna}, Peter and {Chisholm}, John and {Brinchmann}, Jarle and {Aloisi}, Alessandra and {Amor{\'\i}n}, Ricardo O. and {Charlot}, St{\'e}phane and {Feltre}, Anna and {Hayes}, Matthew and {Heckman}, Timothy and {Henry}, Alaina and {Hernandez}, Svea and {Kumari}, Nimisha and {Leitherer}, Claus and {Llerena}, Mario and {Martin}, Crystal L. and {Nanayakkara}, Themiya and {Ravindranath}, Swara and {Skillman}, Evan D. and {Sugahara}, Yuma and {Wofford}, Aida and {Xu}, Xinfeng},
        title = "{CLASSY IV. Exploring UV Diagnostics of the Interstellar Medium in Local High-z Analogs at the Dawn of the JWST Era}",
      journal = {\apj},
         year = 2022,
        month = nov,
       volume = {939},
       number = {2},
          eid = {110},
        pages = {110},
          doi = {10.3847/1538-4357/ac952c},
archivePrefix = {arXiv},
       eprint = {2209.09047},
 primaryClass = {astro-ph.GA},
       adsurl = {https://ui.adsabs.harvard.edu/abs/2022ApJ...939..110M}
}

@ARTICLE{Mingozzi24,
       author = {{Mingozzi}, Matilde and {James}, Bethan L. and {Berg}, Danielle A. and {Arellano-C{\'o}rdova}, Karla Z. and {Plat}, Adele and {Scarlata}, Claudia and {Aloisi}, Alessandra and {Amor{\'\i}n}, Ricardo O. and {Brinchmann}, Jarle and {Charlot}, St{\'e}phane and {Chisholm}, John and {Feltre}, Anna and {Gazagnes}, Simon and {Hayes}, Matthew and {Heckman}, Timothy and {Hernandez}, Svea and {Kewley}, Lisa J. and {Kumari}, Nimisha and {Leitherer}, Claus and {Martin}, Crystal L. and {Maseda}, Michael and {Nanayakkara}, Themiya and {Ravindranath}, Swara and {Rigby}, Jane R. and {Senchyna}, Peter and {Skillman}, Evan D. and {Sugahara}, Yuma and {Wilkins}, Stephen M. and {Wofford}, Aida and {Xu}, Xinfeng},
        title = "{CLASSY. VIII. Exploring the Source of Ionization with UV Interstellar Medium Diagnostics in Local High-z Analogs}",
      journal = {\apj},
         year = 2024,
        month = feb,
       volume = {962},
       number = {1},
          eid = {95},
        pages = {95},
          doi = {10.3847/1538-4357/ad1033},
archivePrefix = {arXiv},
       eprint = {2306.15062},
 primaryClass = {astro-ph.GA},
       adsurl = {https://ui.adsabs.harvard.edu/abs/2024ApJ...962...95M}
}

@ARTICLE{Mokiem07,
   author = {{Mokiem}, M.~R. and {de Koter}, A. and {Vink}, J.~S. and {Puls}, J. and 
	{Evans}, C.~J. and {Smartt}, S.~J. and {Crowther}, P.~A. and 
	{Herrero}, A. and {Langer}, N. and {Lennon}, D.~J. and {Najarro}, F. and 
	{Villamariz}, M.~R.},
    title = "{The empirical metallicity dependence of the mass-loss rate of O- and early B-type stars}",
  journal = {\aap},
archivePrefix = "arXiv",
   eprint = {0708.2042},
     year = 2007,
    month = oct,
   volume = 473,
    pages = {603-614},
      doi = {10.1051/0004-6361:20077545},
   adsurl = {http://adsabs.harvard.edu/abs/2007A%26A...473..603M}
}

@ARTICLE{Monreal-Ibero12,
       author = {{Monreal-Ibero}, A. and {Walsh}, J.~R. and {V{\'\i}lchez}, J.~M.},
        title = "{The ionized gas in the central region of NGC 5253. 2D mapping of the physical and chemical properties}",
      journal = {\aap},
         year = 2012,
        month = aug,
       volume = {544},
          eid = {A60},
        pages = {A60},
          doi = {10.1051/0004-6361/201219543},
archivePrefix = {arXiv},
       eprint = {1206.2275},
 primaryClass = {astro-ph.CO},
       adsurl = {https://ui.adsabs.harvard.edu/abs/2012A&A...544A..60M}
}

@ARTICLE{MunozSanchez24,
       author = {{Munoz-Sanchez}, G. and {Kalitsounaki}, M. and {de Wit}, S. and {Antoniadis}, K. and {Bonanos}, A.~Z. and {Zapartas}, E. and {Boutsia}, K. and {Christodoulou}, E. and {Maravelias}, G. and {Soszynski}, I. and {Udalski}, A.},
        title = "{The dramatic transition of the extreme Red Supergiant WOH G64 to a Yellow Hypergiant}",
      journal = {arXiv e-prints},
         year = 2024,
        month = nov,
          eid = {arXiv:2411.19329},
        pages = {arXiv:2411.19329},
          doi = {10.48550/arXiv.2411.19329},
archivePrefix = {arXiv},
       eprint = {2411.19329},
 primaryClass = {astro-ph.SR},
       adsurl = {https://ui.adsabs.harvard.edu/abs/2024arXiv241119329M}
}

@ARTICLE{Nally24,
       author = {{Nally}, Conor and {Jones}, Olivia C. and {Lenki{\'c}}, Laura and {Habel}, Nolan and {Hirschauer}, Alec S. and {Meixner}, Margaret and {Kavanagh}, P.~J. and {Boyer}, Martha L. and {Ferguson}, Annette M.~N. and {Sargent}, B.~A. and {Nayak}, Omnarayani and {Temim}, Tea},
        title = "{JWST MIRI and NIRCam unveil previously unseen infrared stellar populations in NGC 6822}",
      journal = {\mnras},
         year = 2024,
        month = jun,
       volume = {531},
       number = {1},
        pages = {183-198},
          doi = {10.1093/mnras/stae1163},
archivePrefix = {arXiv},
       eprint = {2309.13521},
 primaryClass = {astro-ph.GA},
       adsurl = {https://ui.adsabs.harvard.edu/abs/2024MNRAS.531..183N}
}

@ARTICLE{Nandal24a,
       author = {{Nandal}, Devesh and {Regan}, John A. and {Woods}, Tyrone E. and {Farrell}, Eoin and {Ekstr{\"o}m}, Sylvia and {Meynet}, Georges},
        title = "{Explaining the high nitrogen abundances observed in high-z galaxies via population III stars of a few thousand solar masses}",
      journal = {\aap},
         year = 2024,
        month = mar,
       volume = {683},
          eid = {A156},
        pages = {A156},
          doi = {10.1051/0004-6361/202348035},
archivePrefix = {arXiv},
       eprint = {2402.03428},
 primaryClass = {astro-ph.GA},
       adsurl = {https://ui.adsabs.harvard.edu/abs/2024A&A...683A.156N}
}

@ARTICLE{Nandal24b,
       author = {{Nandal}, Devesh and {Sibony}, Yves and {Tsiatsiou}, Sophie},
        title = "{Fast-rotating massive Population III stars as possible sources of extreme N enrichment in high-redshift galaxies}",
      journal = {\aap},
         year = 2024,
        month = aug,
       volume = {688},
          eid = {A142},
        pages = {A142},
          doi = {10.1051/0004-6361/202348866},
archivePrefix = {arXiv},
       eprint = {2405.11235},
 primaryClass = {astro-ph.GA},
       adsurl = {https://ui.adsabs.harvard.edu/abs/2024A&A...688A.142N}
}

@ARTICLE{Nandi23,
       author = {{Nandi}, Payel and {Stalin}, C.~S. and {Saikia}, D.~J. and {Muneer}, S. and {Mountrichas}, George and {Wylezalek}, Dominika and {Sagar}, R. and {Kissler-Patig}, Markus},
        title = "{Star Formation in the Dwarf Seyfert Galaxy NGC 4395: Evidence for Both AGN and SN Feedback?}",
      journal = {\apj},
         year = 2023,
        month = jun,
       volume = {950},
       number = {2},
          eid = {81},
        pages = {81},
          doi = {10.3847/1538-4357/accf1e},
archivePrefix = {arXiv},
       eprint = {2304.08986},
 primaryClass = {astro-ph.GA},
       adsurl = {https://ui.adsabs.harvard.edu/abs/2023ApJ...950...81N}
}

@ARTICLE{Neugent10,
   author = {{Neugent}, K.~F. and {Massey}, P. and {Skiff}, B. and {Drout}, M.~R. and 
	{Meynet}, G. and {Olsen}, K.~A.~G.},
    title = "{Yellow Supergiants in the Small Magellanic Cloud: Putting Current Evolutionary Theory to the Test}",
  journal = {\apj},
archivePrefix = "arXiv",
   eprint = {1006.5742},
 primaryClass = "astro-ph.SR",
     year = 2010,
    month = aug,
   volume = 719,
    pages = {1784-1795},
      doi = {10.1088/0004-637X/719/2/1784},
   adsurl = {http://adsabs.harvard.edu/abs/2010ApJ...719.1784N}
}

@ARTICLE{Neugent23,
       author = {{Neugent}, Kathryn F. and {Massey}, Philip},
        title = "{Newly Discovered Wolf-Rayet Stars in M31}",
      journal = {\aj},
         year = 2023,
        month = aug,
       volume = {166},
       number = {2},
          eid = {68},
        pages = {68},
          doi = {10.3847/1538-3881/ace25f},
archivePrefix = {arXiv},
       eprint = {2306.11949},
 primaryClass = {astro-ph.GA},
       adsurl = {https://ui.adsabs.harvard.edu/abs/2023AJ....166...68N}
}

@ARTICLE{Nieuwenhuijzen90,
   author = {{Nieuwenhuijzen}, H. and {de Jager}, C.},
    title = "{Parametrization of stellar rates of mass loss as functions of the fundamental stellar parameters M, L, and R}",
  journal = {\aap},
     year = 1990,
    month = may,
   volume = 231,
    pages = {134-136},
   adsurl = {http://adsabs.harvard.edu/abs/1990A%26A...231..134N}
}

@ARTICLE{Ohnaka24,
       author = {{Ohnaka}, K. and {Hofmann}, K. -H. and {Weigelt}, G. and {van Loon}, J. Th. and {Schertl}, D. and {Goldman}, S.~R.},
        title = "{Imaging the innermost circumstellar environment of the red supergiant WOH G64 in the Large Magellanic Cloud}",
      journal = {\aap},
         year = 2024,
        month = nov,
       volume = {691},
          eid = {L15},
        pages = {L15},
          doi = {10.1051/0004-6361/202451820},
       adsurl = {https://ui.adsabs.harvard.edu/abs/2024A&A...691L..15O}
}
\bibliographystyle{aa}

\newpage

\appendix


\section{Numbers}
\subsection{Numbers of stars and estimated star-formation rates \label{app:numbers}}

\begin{table*}[]
\caption{\label{tab:gal_sfrs_etc}
Numbers of certain types of stars and SFRs in different galaxies.}
\small
\centering
\begin{tabular}{llllllll}
\hline
\hline
Galaxy & SFR (lit.) & $N_\mathrm{HeB,\,lit.\,SFR}$ & $N_\mathrm{cSG,\,obs}$ & SFR$_\mathrm{min}$ & $N_\mathrm{cSG,\,pred}$ & $N_\mathrm{cSG,\,obs}$ & $N_\mathrm{WR,\,obs}$ \\
 & [M$_\odot$\,yr$^{-1}$] & {\tiny (5$<$logL$<$5.6)} & {\tiny (5$<$logL$<$5.6)} & [M$_\odot$\,yr$^{-1}$] & {\tiny (logL$>$5.6)} & {\tiny (logL$>$5.6)} & {\tiny (logL$>$5.6)} \\ 
\hline
I\,Zw\,18 & 0.1-1 & 150-1500 & 28 & 0.019 & 90 (10$^{a)}$) & 0 & "tens" (estimated, no $L$ provided) \\
SMC & 0.05 & 75 & 41$^{b)}$ & 0.027 & 32 & 0 & 11 (3 can be H-burning$^{c)}$)\\
NGC\,5253 & 0.1, 0.2 & 150, 300 & 71 & 0.047 & 52 & 1 & 30 to 40 (estimated, no $L$ provided)\\
LMC & 0.2 & 300 & 72$^{b)}$ & 0.048 & 21 & 1 & $\sim$80 (half can be H-burning$^{c)}$)\\
NGC\,300 & 0.08-0.16 & 120-240 & 86 & 0.057 & 21 & 0 & 30 to 40 (estimated, no $L$ provided) \\
NGC\,4395 & 0.03, 0.5 & 45, 750 & 86 & 0.057 & 19 & 0 & "WR features" \\
M\,31 & 0.7 & 1050 & 117$^{d)}$ & 0.078 & & & \\
\hline
\end{tabular}
\tablefoot{The text in Appendix\,\ref{app:numbers} describes how numbers of cool SG stars and the minimum SFRs are obtained. Explanation of notes in the table: $^{a)}$: if, instead, a constant cool SG lifetime fraction of 7\% of the stellar lifetime is assumed -- $^{b)}$: from \cite{Davies18} -- $^{c)}$: these stars can be hydrogen burning because they have temperatures lower than the zero-age main-sequence temperature. -- $^{d)}$: from \cite{McDonald22}.
Literature SFR references are the following. I\,Zw\,18: table\,1 of \cite{Hirschauer24}, \cite{Bortolini24}; SMC: \cite{Harris04}; N5253: \cite{LopezSanchez12};  LMC: \cite{Harris09}; N300: \cite{Helou04}, \cite{Kang25}; N4395: \cite{Smirnova20}, \cite{Nandi23}; M31: \cite{Lewis15}.
References for $N_\mathrm{WR,\,obs}$: Sect.\,\ref{sec:sfh}, and \cite{Hainich14, Hainich15} for LMC and SMC, respectively.
}
\end{table*}

In Table\,\ref{tab:gal_sfrs_etc}, we compile literature SFRs of the galaxies discussed in this study.
From there, we can estimate the theoretically expected number of helium-burning stars based on the literature SFR ($N_\mathrm{HeB,\,lit.\,SFR}$) in the luminosity range $5.0 <\log (L / L_\odot) < 5.6$  as follows. We first notice that during helium burning this luminosity range corresponds to initial masses in the range $16 \leq M_\mathrm{ini} / M_\odot \leq 32$. Then we follow the method described in appendix\,C of \cite{Schootemeijer21}. There, the IMF of \cite{Kroupa01} is integrated to calculate the fraction of stars born in the initial mass range of interest, and their average mass. Then, we look up average lifetimes for stellar models \citep{Schootemeijer19} in that mass range, and the typical fraction of their lifetime they spend burning helium (7\%). With this information we are able to roughly estimate the relation between the number of helium burning stars in the range $5.0 <\log (L / L_\odot) < 5.6$ and the SFR their host galaxy as

\begin{equation}
N_\mathrm{HeB,\,lit.\,SFR} \approx \frac{1500}{ (\mathrm{M}_\odot \, / \, \mathrm{yr})}.
\label{eq:sfr_to_n}
\end{equation}

For comparison, per galaxy we also provide the observed numbers of cool SGs, $N_\mathrm{cSG,\,obs}$, in the same luminosity range (Sect.\,\ref{sec:ldists}). In addition to this, we use Eq.\,\ref{eq:sfr_to_n} to calculate the minimum constant SFR, SFR$_\mathrm{min}$, associated with the observed numbers of cool SGs. We emphasize that these are minimum values, because stars can also burn helium as hotter objects (e.g., warm SGs and helium stars), and therefore comparisons with literature SFRs should be performed prudently.

Finally, for the highest-luminosity end ($\log (L / L_\odot) > 5.6$), Table\,\ref{tab:gal_sfrs_etc} provides: i), the number of cool SGs we predicted in Sect\,\ref{sec:ldists} based on BoOST models and the number of dimmer cool SGs, ii), the observed number of cool SGs (same references as the other $N_\mathrm{cSG,\,obs}$ column), and, iii), a brief note on the number of WR stars from literature.

While this table is mainly meant for future reference, we briefly describe a few noticeable features in it. Our values for SFR$_\mathrm{min}$ are usually somewhat below the literature SFRs, as is to be expected. However, in some cases the values of SFR$_\mathrm{min}$ are rather far below the literature SFRs, in particular for I\,Zw\,18 and M\,31. For I\,Zw, this difference can be attributed to uncertainties in cool SG lifetimes at extremely low metallicity (see the last paragraph of Sect.\,\ref{sec:izw18_iotons}). Also, because we inferred the presence of about as much warm SGs as cool SGs in the luminosity range $5.0 <\log (L / L_\odot) < 5.6$ in I\,Zw\,18, it is likely that it is forming stars at a rate of at least $\sim$0.04\,M$_\odot$\,yr$^{-1}$.
For M\,31, the difference could be caused by relatively short cool SG lifetimes, for example because stars evolve into low-luminosity WR stars ($5.0 <\log (L / L_\odot) < 5.6$), warm SGs, or intermediate-temperature objects. According to \cite{McDonald22}, their sample is rather complete, which would disqualify completeness issues as a cause for the low number of cool SGs observed in M\,31. Finally, based on the number of cool SGs at ($5.0 <\log (L / L_\odot) < 5.6$) in the SMC, we would expect as much as 32 cool SGs or other helium-burning objects at $\log (L / L_\odot) > 5.6$. However, in this luminosity range there are at most 11 WR stars, and objects with temperatures between those of cool SGs and main-sequence stars seem to be no more than a handful in number \citep{Blaha89, Massey95, Schootemeijer21}. As such, about half of the expected helium-burning stars are missing at $\log (L / L_\odot) > 5.6$. Possible explanations could have to do with one or more of the following: a recent decrease in SFR on the order of 50\%; a steeper IMF; or stars burning helium at temperatures that overlap with the main sequence.

\subsection{He$^+$-ionizing photon rates \label{app:rates}}

To estimate the He$^+$ ionizing photon production rate in the window of opportunity indicated by the purple shading in Fig\,\ref{fig:iotons}, we consider three different metallicities. For the example below we take $\log (Z / Z_\odot) = -1.75$. For our calculation we take the following steps:
\begin{itemize}
    \item At $\log (Z / Z_\odot) = -1.75$, the window of opportunity has a luminosity range of $5.6 \leq \log (L / L_\odot) \leq 6.4$. This corresponds to helium-burning models in the initial mass range $30 \lesssim M_\mathrm{ini}/M_\odot \lesssim 100$ \citep[estimated using models from][]{Schootemeijer19}. Their initial-mass-function-weighted average mass is $\sim$50\,M$_\odot$.
    \item Repeating the procedure described in the first paragraph of Appendix\,\ref{app:numbers} (integrating the IMF, looking up lifetimes of helium-burning models), we estimate that a galaxy with an SFR of 1\,M$_\odot$/yr hosts 375 helium-burning stars from this initial mass range.
    \item We assume that for evolved massive stars born with at least 30\,M$_\odot$, the helium core makes up half of the initial mass of the star \citep[in reasonable agreement with models from][]{Georgy13, Schootemeijer19}. Then, the total helium-star mass from the initial mass range $30 \leq M_\mathrm{ini} \leq 100$ is $ 375 \cdot 50 \cdot 0.5 = 9375$\,M$_\odot$ for an SFR of 1\,M$_\odot$/yr.
    \item We notice that in the helium-star models from \cite{Sander20} that are more massive than 15\,M$_\odot$, the He$^+$-ionizing flux per unit mass is rather constant: $Q(\mathrm{He}^+) \approx 10^{47.35}\,\mathrm{s}^{-1}\, \mathrm{M}_\odot^{-1}$.
    \item Multiplying that with the number of 9375\,M$_\odot$ from the previous step, at $\log (Z / Z_\odot) = -1.75$ we obtain $Q(\mathrm{He^+}) \approx 2.1 \cdot 10^{51} \ \mathrm{s}^{-1} / \ (\mathrm{M}_\odot  \, / \, \mathrm{yr})$.
\end{itemize}
We repeat this exercise for $\log (Z / Z_\odot) = -1.50$ ($-1.25$). There, the window of opportunity extends to $\log (L / L_\odot) = 6.2$ (6.0). \
As such, it extends to an initial mass range of $M_\mathrm{ini} \approx 70$\,M$_\odot$ ($50$\,M$_\odot$), with an average initial mass of $\sim$45\,M$_\odot$ ($\sim$40\,M$_\odot$). At an SFR of 1\,M$_\odot$/yr, this corresponds to 336 (263) stars, with a total helium star mass of 7728\,M$_\odot$ (5250\,M$_\odot$). The final result at $\log (Z / Z_\odot) = -1.50$ (-1.25) that follows is then $Q(\mathrm{He^+}) \approx 1.7 \cdot 10^{51} \ \mathrm{s}^{-1} / \ (\mathrm{M}_\odot  \, / \, \mathrm{yr})$ (at $\log (Z / Z_\odot) = -1.25$: $Q(\mathrm{He^+}) \approx 1.2 \cdot 10^{51} \ \mathrm{s}^{-1} / \ (\mathrm{M}_\odot  \, / \, \mathrm{yr})$).

As such, the correlation between $\log (Z / Z_\odot)$ and $Q(\mathrm{He^+})$ is not far from linear in the range $-1.75 \leq \log (Z / Z_\odot) \leq -0.75$. Since it is meant to serve as a rough estimate, we present a linear relation in Eq.\,\ref{eq:woo}. 

\end{document}